\documentclass[useAMS,usenatbib]{mn2e} 
\usepackage{epsfig} 
\usepackage{natbib} 
 
\DeclareMathVersion{bold}

\title{Comparison of filters for the detection of point sources in 
Planck simulations} 
 
\author[L\'opez-Caniego et al.]{M. L\'opez-Caniego$^{1,2}$\footnotemark, 
D. Herranz$^{2}$, J. Gonz\'alez-Nuevo$^{3}$, 
J. L. Sanz$^{1}$,\newauthor R. B. Barreiro $^{1,2}$, P. Vielva$^{1}$, 
F. Arg\"ueso$^{4}$, L. Toffolatti$^{5}$ \\ $^{1}$ Instituto de 
F\'\i{sica} de Cantabria (CSIC-UC), Avda. los Castros s/n, 39005 
Santander, Spain \\ $^{2}$ Departamento de F\'\i sica Moderna, 
Universidad de Cantabria, Avda. los Castros, s/n, 39005 Santander, 
Spain \\ $^{3}$ SISSA-I.S.A.S, via Beirut 4, I-34014 Trieste, Italy \\ 
$^{4}$ Departamento de Matem\'aticas, Universidad de Oviedo, 
Avda. Calvo Sotelo s/n, 33007 Oviedo, Spain \\ $^{5}$ Departamento de 
F\'\i{sica}, Universidad de Oviedo, Avda. Calvo Sotelo s/n, 33007 
Oviedo, Spain }

\begin{document} 
 
\maketitle 
 
\begin{abstract} 
 
 We study the detection of extragalactic point sources in
two-dimensional flat simulations for all the frequencies of the
forthcoming ESA's Planck mission. In this work we have used the most
recent available templates of the microwave sky: as for the diffuse
Galactic components and the Sunyaev-Zel'dovich clusters we have used
the ``Plank Reference Sky Model''; as for the extragalactic point
sources, our simulations - which comprise all the source populations
relevant in this frequency interval - are based on up-to-date
cosmological evolution models for sources. To consistently compare the
capabilities of different filters for the compilation of the -
hopefully - most complete blind catalogue of point sources, we have
obtained three catalogues by filtering the simulated sky maps with:
the Matched Filter (MF), the Mexican Hat Wavelet (MHW1) and the
Mexican Hat Wavelet 2 (MHW2), the first two members of the MHW
Family. For the nine Planck frequencies we show the number of real and
spurious detections and the percentage of spurious detections at
different flux detection limits as well as the completeness level of
the catalogues and the average errors in the estimation of the flux
density of detected sources. Allowing a $5\%$ os spurious detections,
we obtain the following number of detections by filtering with the
MHW2 an area equivalent to half of the sky: 580 (30 GHz), 342 (44
GHz), 341 (70 GHz), 730 (100 GHz), 1130 (143 GHz), 1233 (217 GHz), 990
(353 GHz), 1025 (545 GHz) and 3183 (857 GHz). Our current results
indicate that the MF and the MHW2 yield similar results, whereas the
MHW1 performs worse in some cases and especially at very low
fluxes. This is a relevant result, because we are able to obtain
comparable results with the well known Matched Filter and with this
specific wavelet, the MHW2, which is much easier to implement and use.
 
\end{abstract}

 
\section{Introduction} \label{sec:intro} 
\footnotetext{E-mail: caniego@ifca.unican.es} In two years from now, 
the ESA's Planck\footnote{http://www.rssd.esa.int/index.php?project=Planck} 
satellite~\citep{tauber} will inaugurate a new era in the studies of 
the Cosmic Microwave Background (CMB) radiation. Planck will observe 
the microwave sky with unprecedented angular resolution and 
sensitivity in nine frequency channels ranging from 30 to 857 GHz. 
Besides CMB anisotropies, Planck will also yield all-sky maps of all 
the major sources of microwave emission, including a large number of 
extragalactic sources that have not yet been observed at these 
frequencies. The study of these sources of microwave emission --often 
referred to as \emph{``contaminants''} or \emph{``foregrounds''}-- is 
twofold: on one hand, it is necessary to remove them in order to have 
the cleanest possible view of the CMB and, on the other hand, a better 
knowledge of the different foregrounds is a scientific goal in itself. 
As the launch date approaches, a big effort is being made to develop 
and to test state-of-the-art data processing techniques that will 
optimise the scientific exploitation of the forthcoming Planck data. 
 
The case of extragalactic foregrounds deserves to be considered in 
detail. Radio and infrared selected galaxies will be seen by Planck 
as point-like objects due to their very small projected angular size 
as compared to the experiment resolution (see Table 1). Hence, they 
are usually referred to as extragalactic point sources or just as 
\emph{point sources} (as we do hereafter). Point sources are 
expected to be a major contaminant for Planck at multipoles 
$\ell\geq 500-1000$ \citep{tof98,dz99,hob99,max00} and it is 
necessary to detect and remove as many of them as possible in order 
to clean CMB maps. As a very important astrophysics by-product, the 
detection process will yield all-sky catalogues   of extragalactic 
point sources in a frequency interval in which they are lacking and 
that will prove very useful to constrain models of galaxy formation 
and evolution. 
 
Unfortunately, many of the component separation techniques that are 
generally used to separate diffuse Galactic foregrounds are not well 
suited to deal with point sources. This is mainly due to the fact that 
each galaxy is an independent source, different, in principle, to any 
other source in the sky. Albeit average energy spectra can be defined 
for different source populations, the spectral emission law is 
different for each galaxy and it makes impossible to apply separation 
methods which exploits a single energy spectrum for this foreground 
component.

If the lack of knowledge on their spectral emission properties can be
troublesome, their projected angular shape is basically the same for
all of them: a Dirac's $\delta$-like response convolved with the
instrument beam response function.  Thus, techniques that take into
account the specific angular shape of point sources, such as wavelets
and band-pass filters, are particularly useful for detecting them. In
the last few years a number of techniques have been proposed for the
specific case of point source detection in CMB maps, including the
Mexican Hat Wavelet \citep[]{cay00,patri01a,patri01b,patri03}, the
Matched filter (MF) \citep[]{max98}, the Scale-Adaptive filter
\citep{sanz01,yo02a,yo02b}, the Adaptive Top Hat filter \citep{chi02},
the Biparametric Scale-Adaptive filter \citep{can05} and the recently
introduced Mexican Hat Wavelet Family \citep[]{jgn06}.  All the
previously mentioned methods belong to the class of wavelet and linear
filter techniques.  Additionally, non-linear techniques such as
Bayesian detection \citep{hob03} have been successfully applied to
point source detection in the CMB context, but since the use of those
techniques implies a totally different methodological approach we will
focus on the above mentioned filters and wavelets. Whereas some works
have made an effort to compare the performances of some of them both
theoretically and by using almost-ideal simulations
\citep{bar03,vio04,can05}, an attempt to compare the existent
techniques under {\it (almost)} ``real life'' conditions for the
future Planck mission has not been done, yet.
 
In this work we intend to reproduce the conditions of a blind point 
source survey as it will be carried out by Planck. We will use the 
Planck Reference Sky and the nominal Planck instrument characteristics 
and goal performance to simulate realistic sky emission as it should 
be observed at the nine frequencies covered by the satellite. Using 
such realistic simulations we will compare the performance of 
different filters.  The ultimate goal is to decide which is the best 
tool of choice for the Planck case. The `goodness' of a filter will 
be evaluated according to the following criteria: 
 
\renewcommand{\labelenumi}{(\roman{enumi})} 
 
\begin{enumerate} 
\item \label{launo} The filter must be well suited to conduct a blind 
  survey, that is, it must be able to work well with a minimum number 
  of a priori assumptions about the data. 
\item \label{launo_bis} Besides, it must be robust against the effect 
  of possible systematics. 
\item \label{lados} It must yield, after the detection process, a high 
  number of positive detections. 
\item \label{latres} It must yield, after the detection process, a low
  number of false detections (not higher that, let us say, $5\%$ of
  the integral, or total, number of detections above the corresponding
  detection threshold).
\item \label{lacuatro} Moreover, additional factors, such as the flux 
  detection limit of the output catalogue, its completeness and the 
  accuracy with which the positions and the flux densities of the 
  sources are estimated, will be taken into account. 
\end{enumerate} 
 
We would like to remind that previous criteria are similar to those
required for the Early Release Compact Source Catalogue, ERCSC, a
blind catalogue of point sources that is one of the objectives of the
Planck collaboration. In the compilation of this kind of catalogue,
factors such as quickness and accuracy of the estimation of the flux
of the sources have the priority over a high absolute number of
detections. Keeping this in mind, criterion~\ref{launo} eliminates
from the competition tools such as the Biparametric Scale-Adaptive
filter that, even if they are potentially very powerful, require a
detailed knowledge of the probability distribution of the
foregrounds\footnote{Such tools could be very useful for exhaustive
data mining the sky down to very low flux limits.}. The Adaptive Top
Hat filter is known to produce strong ringing artifacts around the
sources which can lead to a high number of false detections --against
criterion~\ref{latres}--, in particular in the vicinity of bright
sources. The Scale-Adaptive filter seems to perform similarly or
slightly worse than the MF \citep{yo02c}.  Therefore, in this work we
will focus on the comparison of two tools: the Mexican Hat Wavelet
Family (MHWF), which includes the standard Mexican Hat Wavelet, and
the Matched Filter\footnote{In particular, \citet{jgn06} have shown
that the second member of the Mexican Hat Wavelet Family (to be
introduced in section~\ref{sec:method}) is the one that gives the best
results for the case of the Planck Low Frequency Instrument (LFI)
channels, and therefore we will limit the discussion to this wavelet,
comparing it with the standard Mexican Hat (the first member of the
family), that has been widely used in the literature with good
results, and the matched filter.}.
 
At first glance, the MF should be the obvious winner in the 
comparison. By definition, it is the best linear operator that can 
be applied to the data in order to maximise the signal to noise 
ratio of a signal with a known profile embedded in additive noise. 
But, in practise, the use of MFs does not lack of subtleties that 
must be considered here. 
 
In Fourier space the MF is proportional to the inverse of 
the power spectrum of the noise. That means that the power spectrum of 
the noise must either be known a priori or be estimated from the data 
in order to construct the filter. If point sources are scarce, the 
power spectrum of the noise can be reasonably approximated by the 
power spectrum of the observed data, that is easy to \emph{estimate} 
by means of any of the power spectra toolboxes available in scientific 
softwares. But any estimated power spectrum, as good as it may be, is 
just a good guess of the real thing. This leads to a number of 
problems: 
 
\begin{itemize} 
\item Firstly, it is necessary to estimate the value of the power 
  spectrum for all the Fourier modes present in the image. This 
  implies the estimation of several hundreds of numbers for a typical 
  CMB image. For the typical image size of the sky patches we work 
  with, each Fourier mode must be estimated from a small number of data 
  samples. Therefore, the estimated power spectrum is \emph{noisy}, 
  especially at low Fourier modes. On the contrary, to use the 
  various members of the Mexican Hat Wavelet Family it is only necessary 
  to determine one single number, the optimal scale of the wavelet 
  and, thus, it is much less sensitive to noise estimation. 
\item Since the estimated power spectrum is noisy, if it is directly
  used to construct the MF it very often happens that the resulting
  filter is so full of discontinuities and `jumps' in Fourier space
  that strong ringing effects appear in the filtered image. Therefore,
  it is necessary to smooth the power spectrum before constructing the
  filter. The different possible choices used in the smoothing
  procedure introduce some degree of arbitrarity in the use of MFs:
  one can, for example, apply some binning and interpolation scheme, or
  use polynomial fitting to the power spectrum, etc.
\item An additional problem appears when it is not possible to
  properly estimate some Fourier modes. This is the case when the
  image which has to be filtered is not complete (for example, if a
  mask is applied to the data in order to cut bad pixels, or in areas
  of the sky where there is incomplete coverage by the instrument). In
  that case the missing modes must be somehow guessed in order to
  build the MF. This problem is much less relevant for the case of
  wavelets.
\item Moreover, if we make considerations in the sphere we will have 
  to deal with important problems when using a Matched filter as 
  compared with a wavelet. These problems arise from the fact that the 
  foregrounds are very different in different regions of the sky and 
  therefore we need to use the appropriate filter for every 
  region. The approach followed by \citet{patri03} using the Spherical 
  MHW (SMHW) was to divide the sky in a number N of regions and obtain 
  the optimal scale for all of them. Then they determined that many of 
  these scales gave similar results and that they could be divided in 
  a small number of different groups. This allowed them to construct 
  just a few filters in the sphere with their corresponding scales and 
  filter the maps a small number of times instead of N. This is 
  important because depending on the resolution, the filtering process 
  may need a lot of CPU time. The problem with the MF is that instead 
  of calculating N optimal scales we would need to calculate N 
  filters, calculating the power spectra from the N regions, and 
  filtering the maps N times, once for every filter. This process will 
  require enormous amounts of CPU time, especially when dealing with 
  high resolution images, and therefore will make it unfeasible in 
  practise.

\end{itemize} 
 
Therefore, and as any experienced practitioner perfectly knows, the 
use of the MF is not the same as filtering by the inverse of the 
squared Fourier transform of the data. On the contrary, it requires 
a non negligible effort of handmade tuning that is not free from 
arbitrarities. Besides, all the previous effects lead to an 
unavoidable degradation of the performance of the MF under realistic 
conditions. 
 
Furthermore, all the theoretical superiority of the MF with respect to
other linear filters comes from the fact that it maximises the signal
to noise ratio, that is, it minimises the variance of the filtered
noise. But if the noise is not Gaussian, as it is the case in CMB maps
due to the emission contributed by Galactic foregrounds, \textit{the
fact that the variance is minimum does not guarantee that the number
of false detections be minimum}. There may be outliers that are not
removed. In that case, it is not by any means clear that the MF should
be better than any other filter.
 
Taking all the previous points into account, the comparison between 
the MF and the wavelets is still necessary. The performance of 
wavelets will degrade as well when going from ideal to realistic 
conditions. Wavelets, however, are much less sensitive than MFs to 
the problems described in the paragraphs above. Therefore, it is 
expected that the performance degradation will be less severe. If we 
can find a wavelet that performs nearly as well as the MF, but 
without having to resort to handmade tuning, we will have a 
detection tool that is as good as the MF regarding 
criteria~\ref{lados},~\ref{latres} and~\ref{lacuatro} but is better 
regarding criterion~\ref{launo} and~\ref{launo_bis}. Such a wavelet 
would be preferable to MF for the compilation of a Planck blind 
point source catalogue. 
 
In Section~\ref{sec:method} we briefly review the tools to be used in 
this paper: the MF and the wavelets belonging to the Mexican Hat 
Wavelet Family. In Section~\ref{sec:sims} we describe the realistic 
Planck simulations we use. The results are summarised in 
Section~\ref{sec:results}.  In Section~\ref{sec:discussion} we discuss 
some additional issues regarding point source detection in microwave 
satellite missions.  Finally, we describe our main conclusions in 
Section~\ref{sec:Conclusions}.

\section{Methodology}   \label{sec:method} 
 
In this Section we consider the detection of compact (point-like) 
sources where $A$ is the intrinsic flux density of the source and 
$s(x)=A\tau(x)$ is the observed flux, filtered by a circular Gaussian 
profile $\tau (x) = \exp (- x^2/2R^2)$, $x=|\vec{x}|$, whose Fourier 
transform is $\tau (q) = R^2 \exp (-(qR)^2/2)$, 
$q\equiv|\vec{q}|$. Such profile represents a commonly used useful 
approximation to the real profile of the beam response function, which 
can be more complicated and not analytically simple to describe. Of 
course, the MF can be easily designed for a non-axysimmetric beam, but 
also elliptical extension of the MHWF can be done (for instance, the 
elliptical MHW has been used in \citet{mac05} and \citet{cru06}). In 
general, the elliptical approximation is good enough for the typical 
beam shapes of the CMB experiments. In any case, dealing with 
non-axysimmetric tools implies increasing the CPU time, since 
different orientation must be considered. Fortunately, as it was shown 
by \citet{patri03}, data with non-axysimmetric but well known main 
beams can be still analysed with isotropic filters by characterising 
the bias introduced for the particular beam shape. 
 
Given the above, the filtered sources in the map appear as circularly 
symmetric objects and, with the assumption that the background is 
statistically homogeneous and isotropic, the most natural thing to do 
- for detecting point-like sources - is the application of circularly 
symmetric filters to the map. More specifically, let us consider a 2D 
filter $\Psi (\vec{x}; R, \vec{b})$, where $R$ and $\vec{b}$ define a 
scaling and a translation respectively, then 
\begin{equation} \label{eq:psi} 
  \Psi(\vec{x}; R, \vec{b}) \equiv \frac{1}{R^2} \psi \left( 
  \frac{|\vec{x} - \vec{b}|}{R} \right) . 
\end{equation} 
If we filter our field $f(\vec{x})$ with  $\Psi (\vec{x}; R, 
\vec{b})$, the filtered map will be 
\begin{equation} \label{eq:wav} 
  w(R, \vec{b}) = \int d\vec{x}\,f(\vec{x})\Psi (\vec{x}; R, \vec{b}). 
\end{equation} 
\noindent The filter is normalised such that the flux of the 
source in the position of the source ($b=\vec{0}$) is the same after 
filtering: 
\begin{equation} 
  \int d\vec{x} \, \tau(\vec{x}) \Psi(\vec{x}; R, \vec{0}) = 1. 
\end{equation} 
\noindent 
For the filtered map the moment of order-n is defined as 
\begin{equation} 
  \sigma_n^2 \equiv 2 \pi \int_0^{\infty} dq \ q^{1+2n} P(q) 
  \psi^2(q). 
\end{equation} 
\noindent where $P(q)$ is the power spectrum of the unfiltered 
map. Note that the zeroth-order moment, $\sigma_0$, is the 
dispersion of the filtered map. 
 
As a result, the source profile described above is characterised by 
the ``natural scale'' $R$. This parameter appears in all the filters 
that we will consider in the following. For a given ``optimal scale'', 
$R_0$, the amplification $(A/\sigma _{0})$ of the source flux has a 
maximum value. By working at this particular scale -- at which a 
filter maximise the flux of a compact source with respect to the 
average surrounding background value -- it is possible to 
significantly improve the performance of the chosen filter in terms of 
detecting compact objects \citep[]{patri01a,patri03,can04,can05}.

\subsection{The Matched Filter (MF)} 
 
\begin{figure*} 
\begin{center} 
\includegraphics[width=5.8cm]{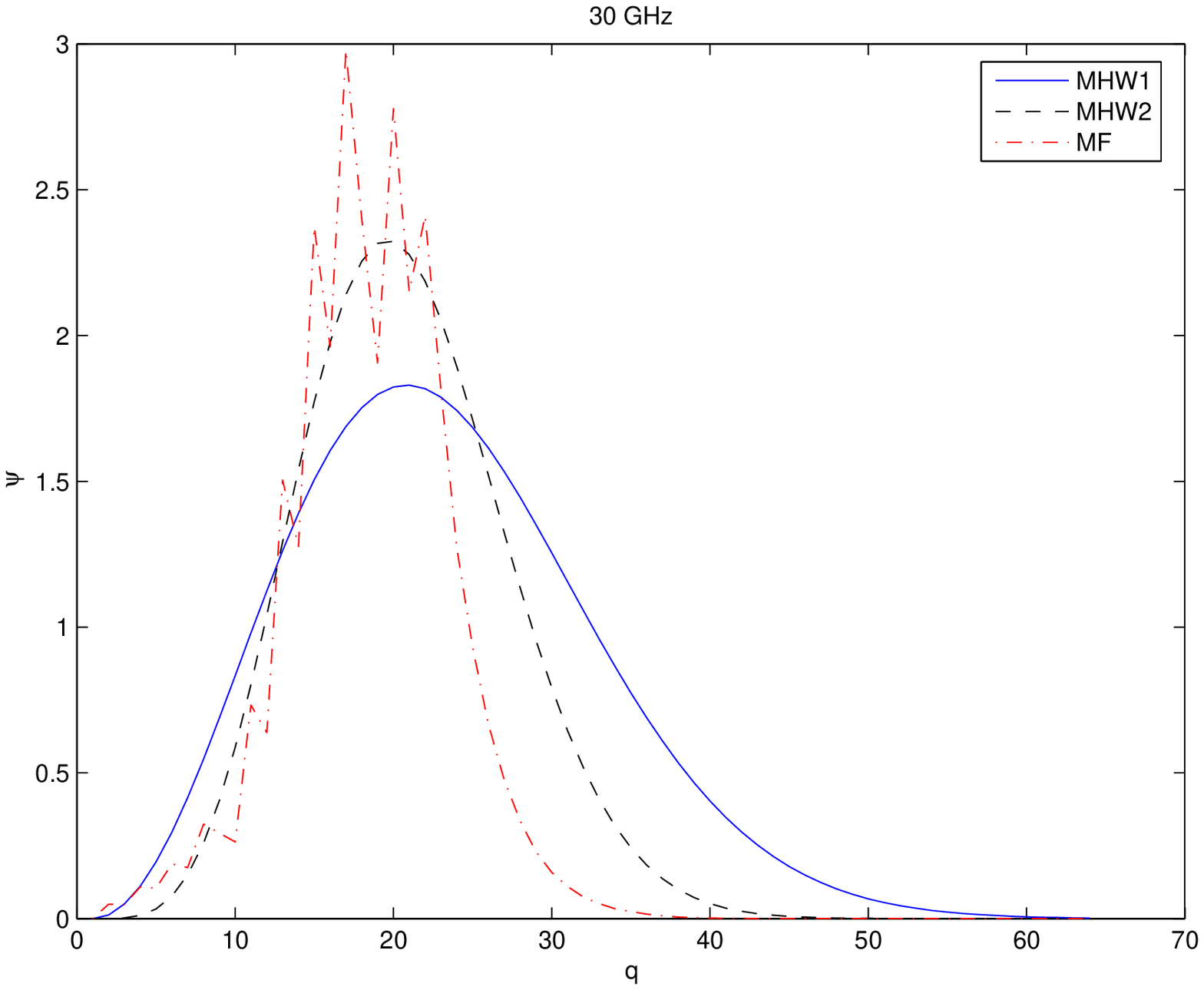} 
\includegraphics[width=5.8cm]{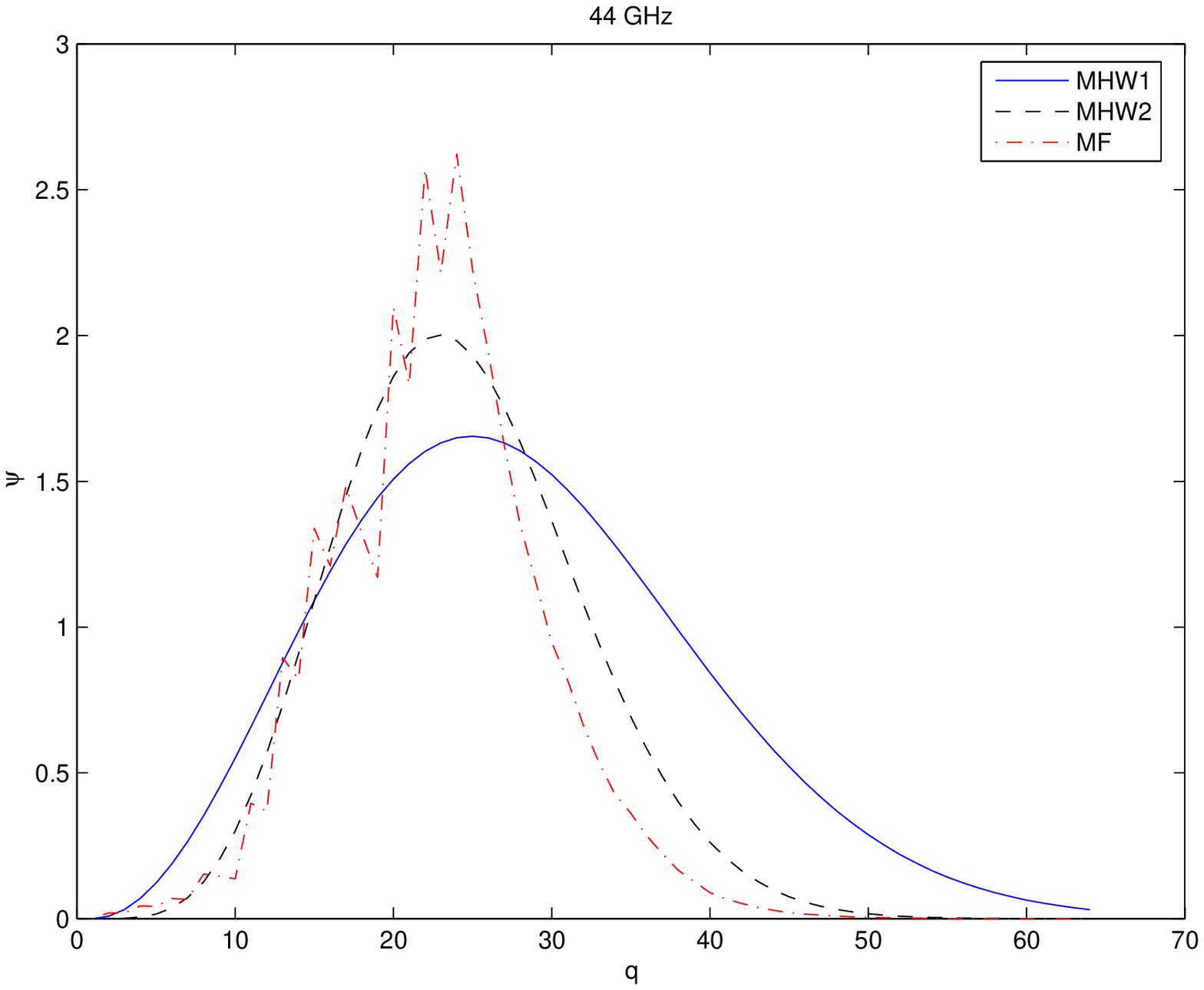} 
\includegraphics[width=5.8cm]{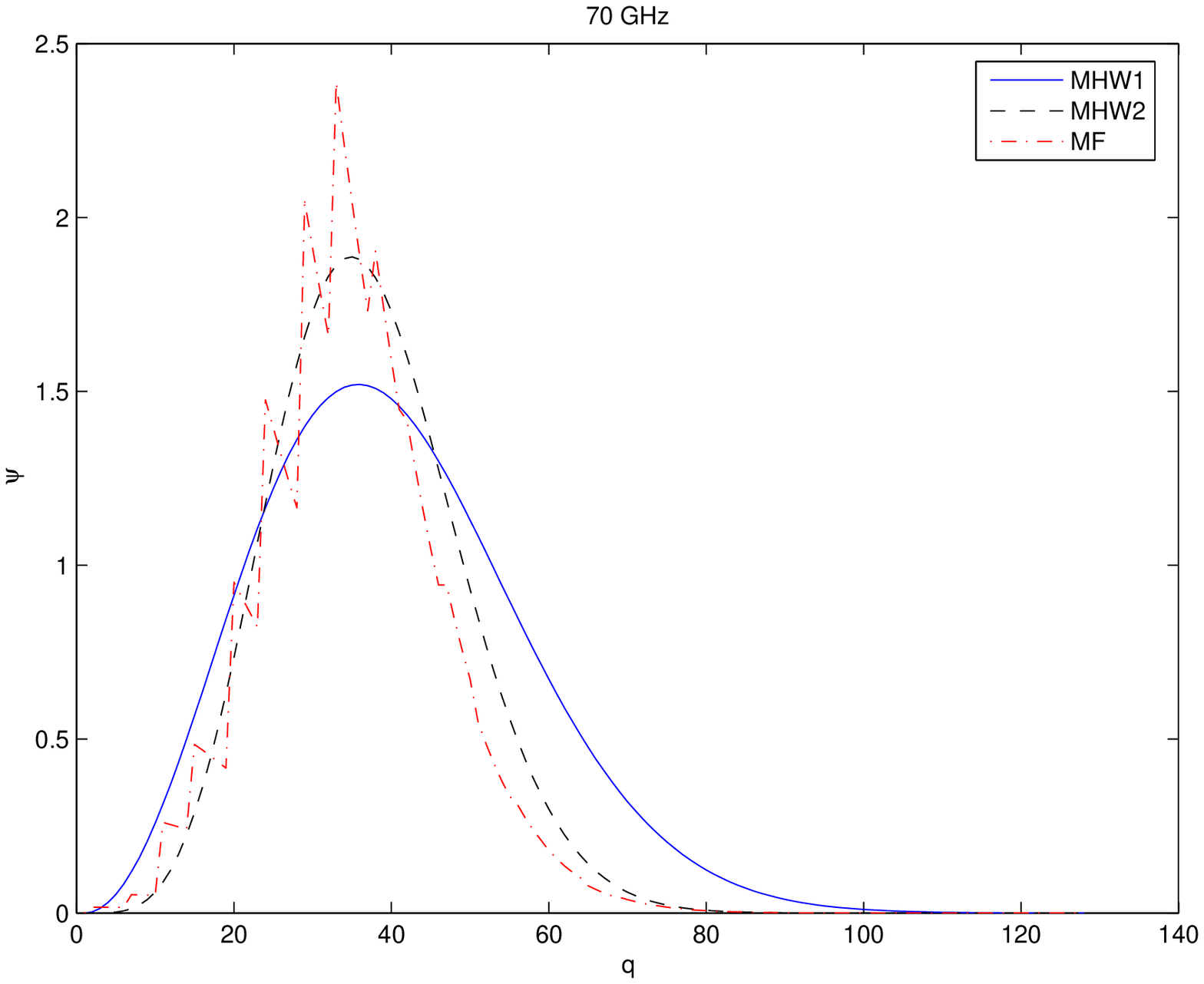} 
~~ 
\includegraphics[width=5.8cm]{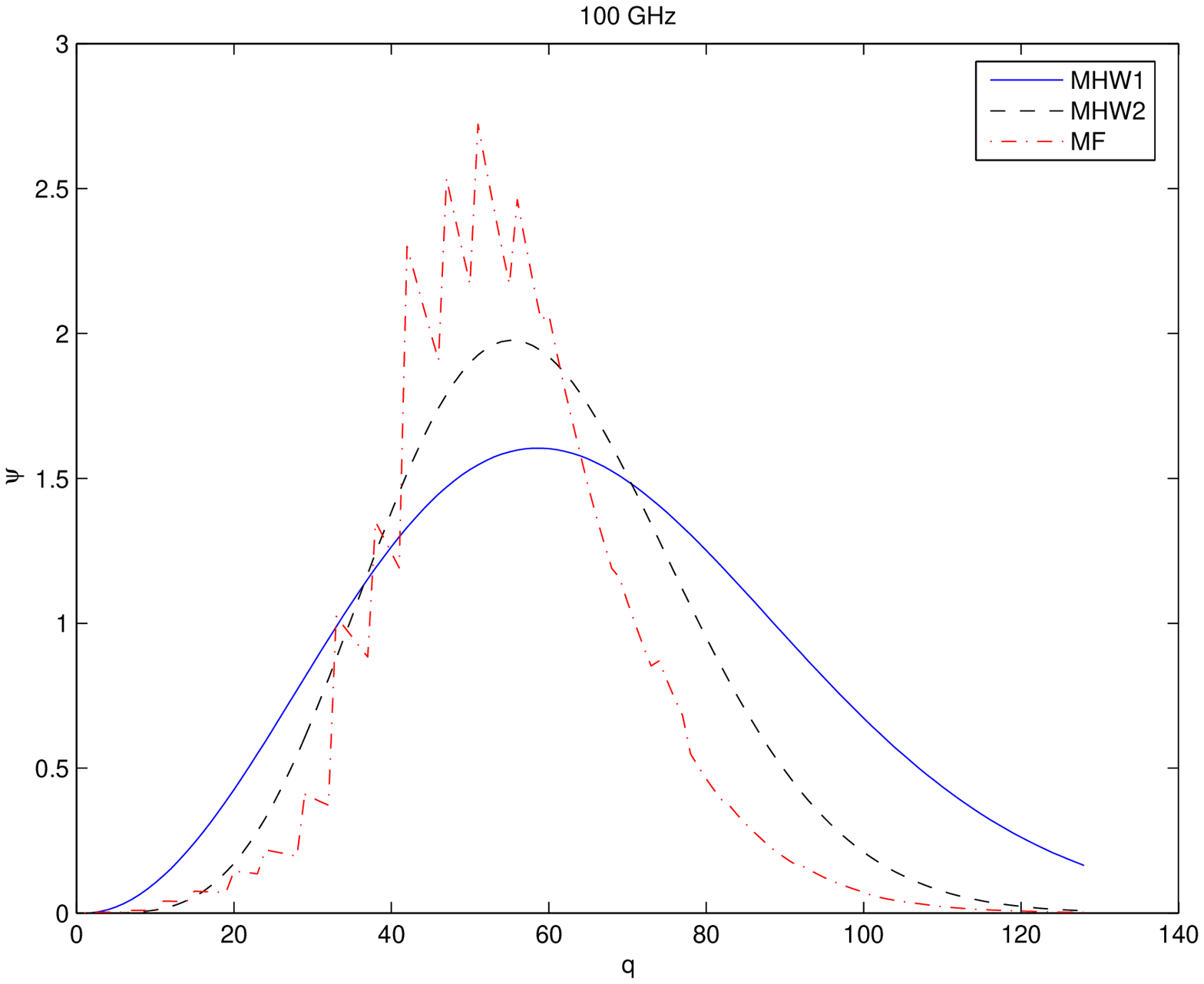} 
\includegraphics[width=5.8cm]{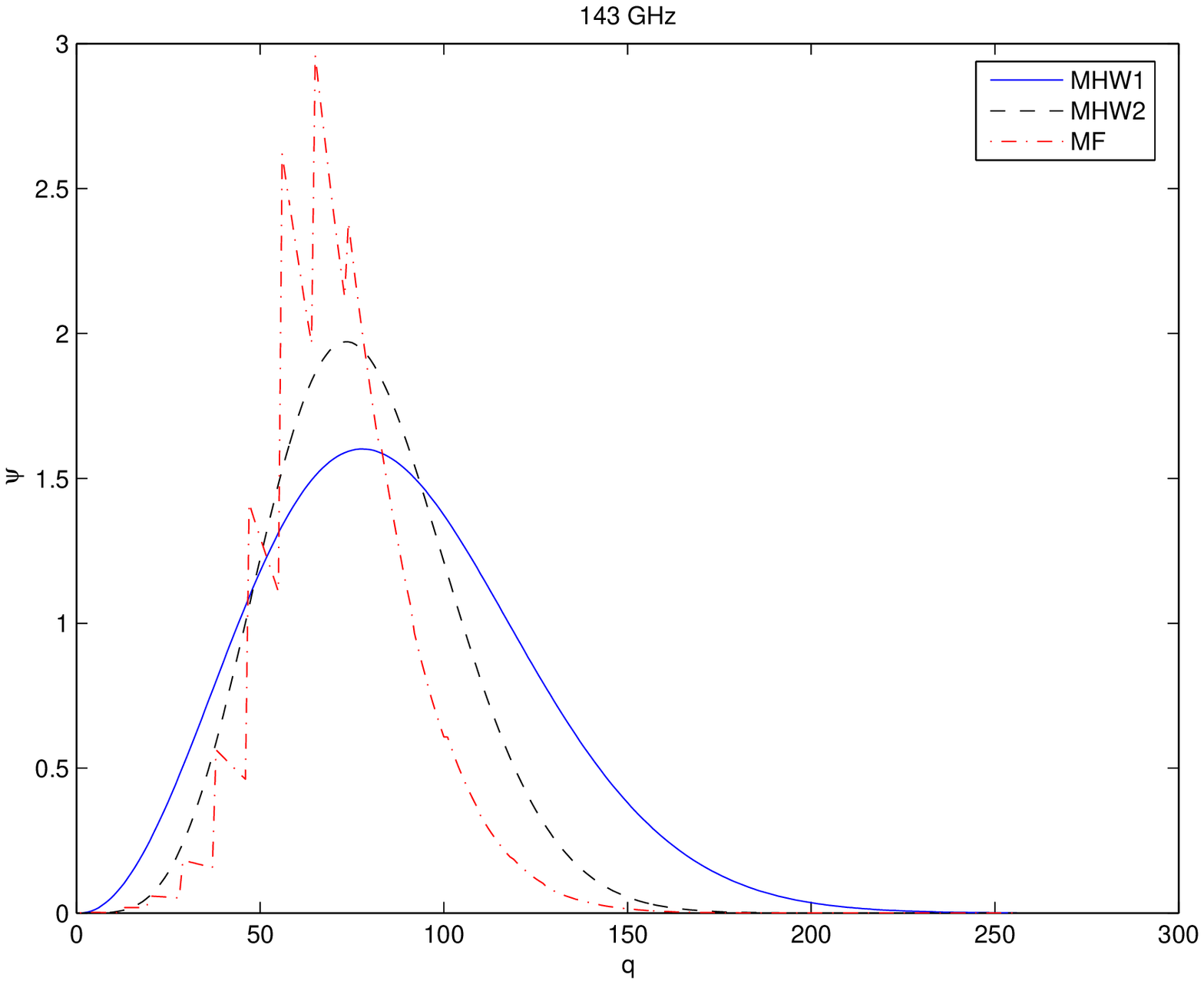} 
\includegraphics[width=5.8cm]{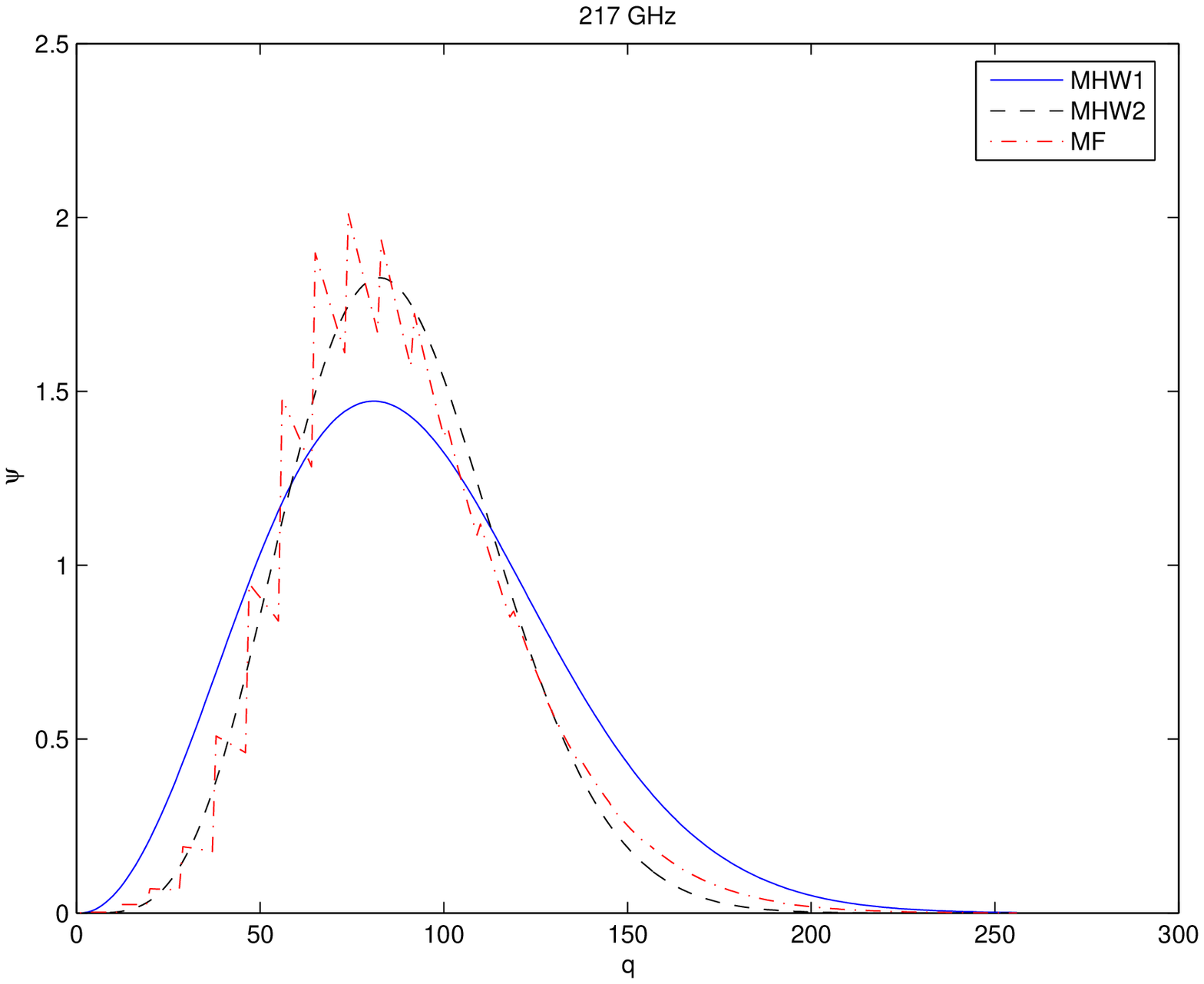} 
~~ 
\includegraphics[width=5.8cm]{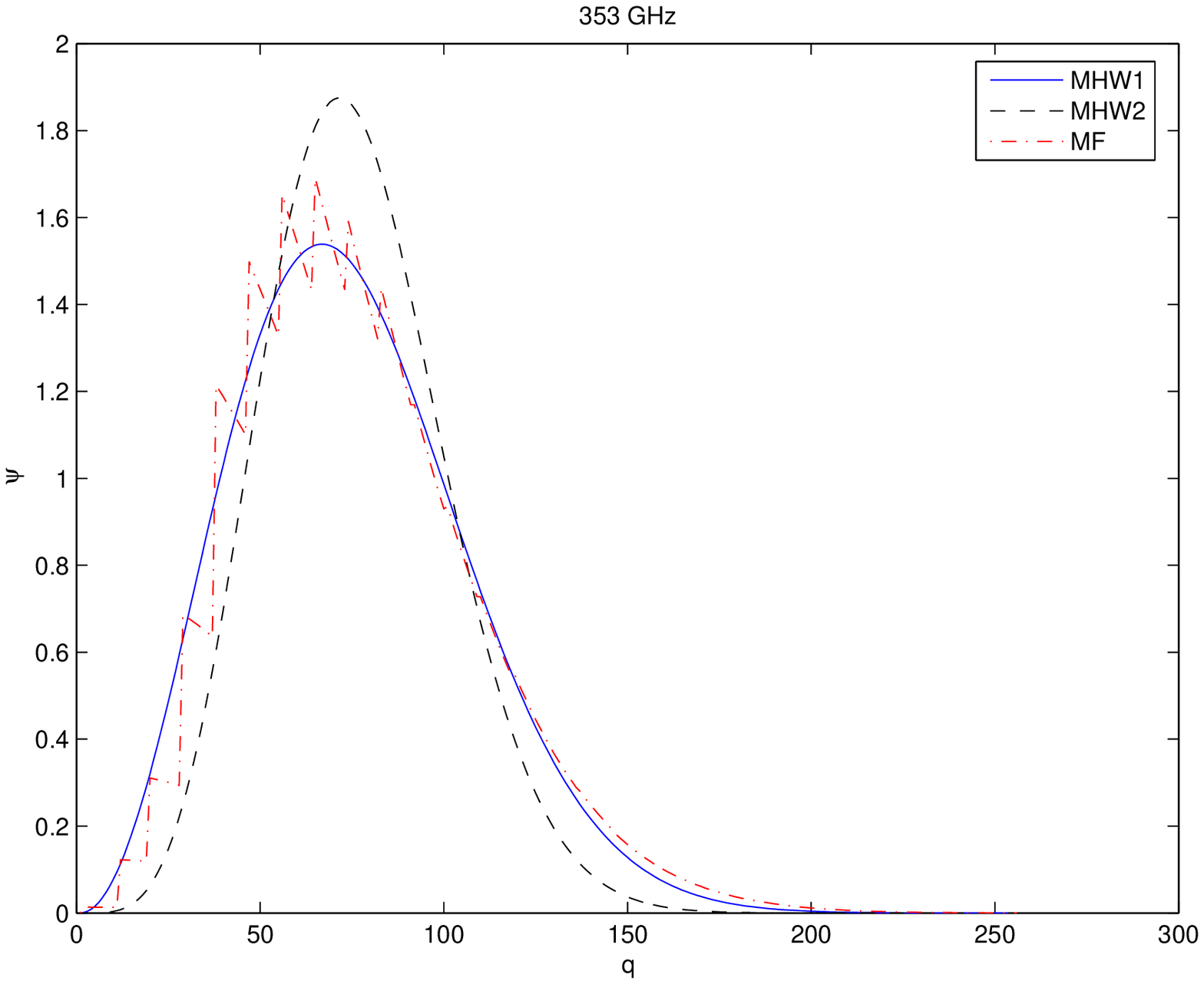} 
\includegraphics[width=5.8cm]{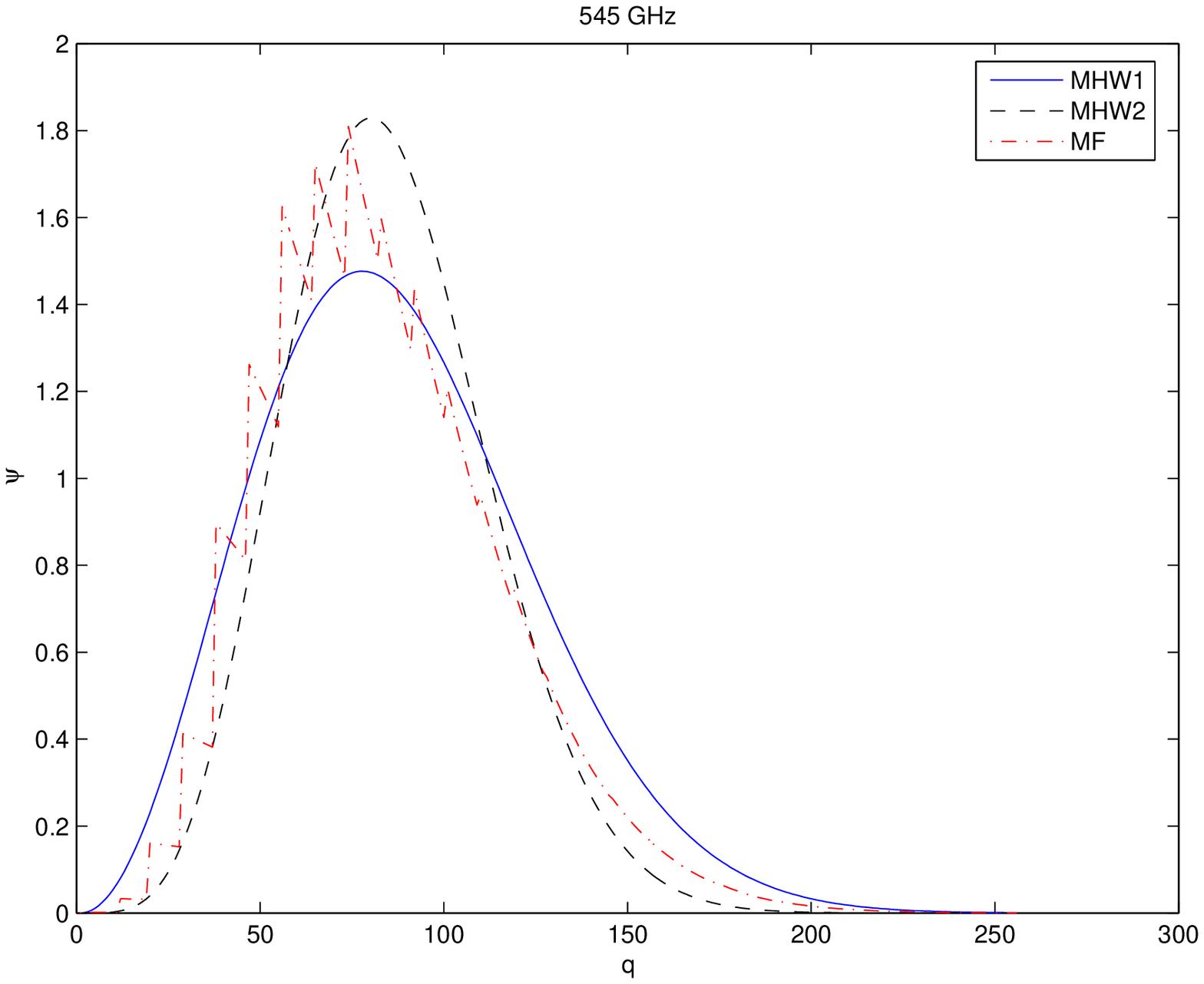} 
\includegraphics[width=5.8cm]{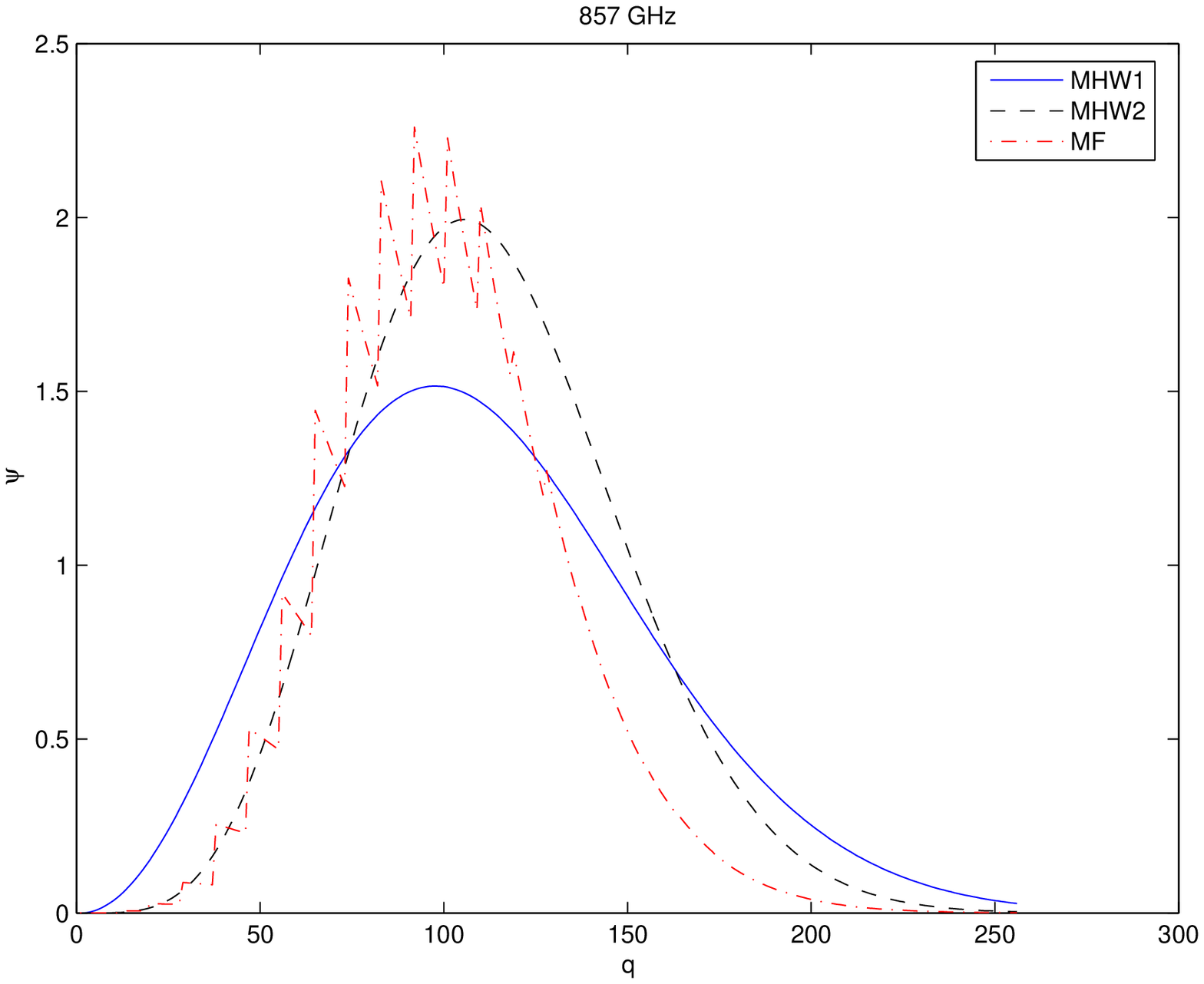} 
\caption{In each panel we compare the three chosen filters in Fourier 
  space as they result at each Planck frequency for one 
  simulation. The solid line corresponds to the MHW1, the dashed line 
  to the MHW2 and the dot-dashed one to the MF.  The MHW1 and MHW2 
  operate at their optimal scale, a scale that yields the maximum 
  average amplification of the source flux. The plots show that the MF 
  is noisy as compared to the two adopted wavelets. As a remarkable 
  result, we want to point out that the MF and the MHW2 have very 
  similar behaviours in most of the plots. \label{fig:mfs}} 
\end{center} 
\end{figure*}

The MF  can be obtained by introducing a circularly-symmetric 
filter, $\Psi(x;R,b)$, and imposing the following two conditions: 
$(1) \ \langle w(R_0, 0)\rangle = s(0) \equiv A$, i. e. $w(R, 0)$ 
is an \emph{unbiased} estimator of the flux density of the source; 
$(2)$ the variance of $w(R, b)$ has a minimum at the scale $R_0$, 
i. e. it is an \emph{efficient} estimator 
\begin{equation} \label{eq:mf} 
\hat{\psi}_{MF} = \frac{1}{a}\frac{\tau (q)}{P(q)}, \ \ \ 
a = 2 \pi \int dq q \frac{\tau^2 (q)}{P(q)}. 
\end{equation} 
As mentioned in the previous section, the practical implementation of
the MF requires some additional work. In equation (\ref{eq:mf}) the
source profile $\tau(q)$ is known, but the power spectrum $P(q)$ must
be estimated from the data. Moreover, the normalisation $a$ must be
calculated by integrating $P(q)$.  The estimation of the power
spectrum from the data can be easily done, but the estimates are
necessarily noisy.  In our case we use flat patches with sizes varying
from $128 \times 128$ to $512 \times 512$ pixels. For small patches
like the ones we use, we obtain a extremely noisy power spectrum and
we need to smooth it before it can be used. We have found that a
smoothing procedure that consists in binning the power spectrum modes
and performing linear interpolation for the power spectrum values
lying between the bin centres works well. Our tests suggest that the
results vary slightly for any number of bins between 25 and 50, but
the performance degrades quickly outside this interval. As for the
results discussed in this paper, we bin the Fourier modes in 40
uniformly spaced bins. Even so, the filter that we construct is noisy,
as seen in Figure \ref{fig:mfs}. In this figure we have compared the
three chosen filters at each Planck frequency and for one of the
simulations: the MF and the first two members of the MHWF (that will
be introduced in the next section). It is easy to appreciate that the
resulting MF is very noisy if compared to the smooth curves of the two
wavelets, that only depend on the optimal scale, a quantity which can
be easily estimated (see below).

\subsection{The Mexican Hat Wavelet Family}

The Mexican Hat Wavelet Family (MHWF) has been discussed by 
\citet{jgn06} and it is the extension of the MHW,  obtained by 
applying iteratively the Laplacian operator to the Gaussian 
function. 
 
\noindent The Fourier transform of $\psi_{n}(x)$ is 
 
\begin{equation} 
\label{cc2} 
 \hat\psi_{n} (q) = \int_0^{\infty}dx\,xJ_0(qx)\psi_{n} (x), 
\end{equation} 
 
\noindent where $\vec{q}$ is the wave number, $q\equiv |\vec{q}|$ and 
$J_0$ is the Bessel function of the first kind. Any member of this 
family can be written in Fourier space as 
\begin{equation} 
\hat\psi_{n}(q) \propto q^{2n}e^{-\frac{q^{2}}{2}}. 
\end{equation} 
 
The expression in real space for these wavelets is 
\begin{equation} 
\label{cc1} \psi_{n}(x) \propto \triangle^{n}{\varphi(x)}, 
\end{equation} 
\noindent where $\varphi$ is the 2D Gaussian $\varphi(x)=\frac{1}{2 
\pi}e^{-x^2/2}$. 
 
Note that $\psi_{1}(x)$ is the usual MHW (hereinafter MHW1), a wavelet 
that has been extensively tested in the field of point detection with 
excellent results. Note that we call MHWn the member of the MHWF with 
index n. 
 
The goal in the work by \citet{jgn06} was to compare the 
performance of some members of the MHWF and the main conclusion was 
that the second level wavelet $\psi_{2}(x)$ did perform better than 
$\psi_{1}(x)$ for the channels of the low frequency instrument of Planck. 
 
In this work we are going to concentrate on these two wavelets, 
$\psi_{1}(x)$ and $\psi_{2}(x)$, and compare them with the Matched 
filter, as described in the previous subsection, for both the LFI and 
HFI channels. 
 
Now let us consider a field $f(\vec{x})$ on the plane $R^2$, where 
$\vec{x}$ is an arbitrary point. One can define the wavelet 
coefficient at scale $R$ at the point $\vec{b}$ in the form given by 
equations (\ref{eq:psi}) and (\ref{eq:wav}). 
 
 The wavelet coefficients $w_n(\vec{b},R)$ for the members of the MHWF 
can be obtained in the following form 
 
\begin{equation} 
\label{cc3} 
 w_n(\vec{b},R) = \int d\vec{q}e^{-i\vec{q}\cdot 
\vec{b}}f(\vec{q})\hat\psi_{n} (qR), 
\end{equation} 
 
 \noindent assuming the appropriate differential and boundary 
 conditions for the field this expression can be rewritten as 
 
\begin{equation} 
\label{cc4} 
 w_n(\vec{b}, R) \propto \int d\vec{x}\,[{\triangle^n}f(\vec{x})]\varphi\Big(\frac{|\vec{x}-\vec{b}|}{R}\Big). 
\end{equation} 
 
\noindent%
 and the wavelet coefficient at point $\vec{b}$ can be 
interpreted as the filtering by a Gaussian window of the invariant 
$(2n)^{th}$-order differences of the field $f$.

\section{Simulations}  \label{sec:sims} 
 
We want to compare the capabilities of the MHW1, the MHW2 and of 
the MF when dealing with the detection of compact sources in 
realistic CMB maps. 
 
To test these tools in the most realistic conditions so far, we used 
the latest ``Planck Reference Sky Model'' provided by the Working 
Group 2 (WG2, ``Component separation") of the Planck Consortium. All 
the available astrophysics foregrounds components are provided in 
all-sky maps in the HEALPix format \citep{heal05} and include the 
following components: 
\begin{itemize} 
\item Thermal dust emission: has been created using a combination of 
two grey bodies with mean emissivity parameters $\alpha_1=1.67$ and 
$\alpha_2=2.70$ and mean temperatures $T_1=9.4 K$ and $T_2=16.2 K$ 
based on the \citet{fink99} model. 
 
\item Synchrotron emission: is a cleaned version of the 408 GHz 
\citet{has82} map made by \citet{gia02}. 
 
\item Free-Free radiation: it adopts the free-free model of 
\citet{dick03} which has been obtained by different $H\alpha$ 
surveys (e.g. WHAM \citep{haf99}, SHASSA \citep{gau01}). 
 
\item S-Z Clusters: follow-up of the works by 
\citet[]{col97,dz99,dz05} and by P. Mazzotta. 
 
\end{itemize} 
 
\begin{figure} 
\begin{center} 
        \includegraphics[width=8.0cm]{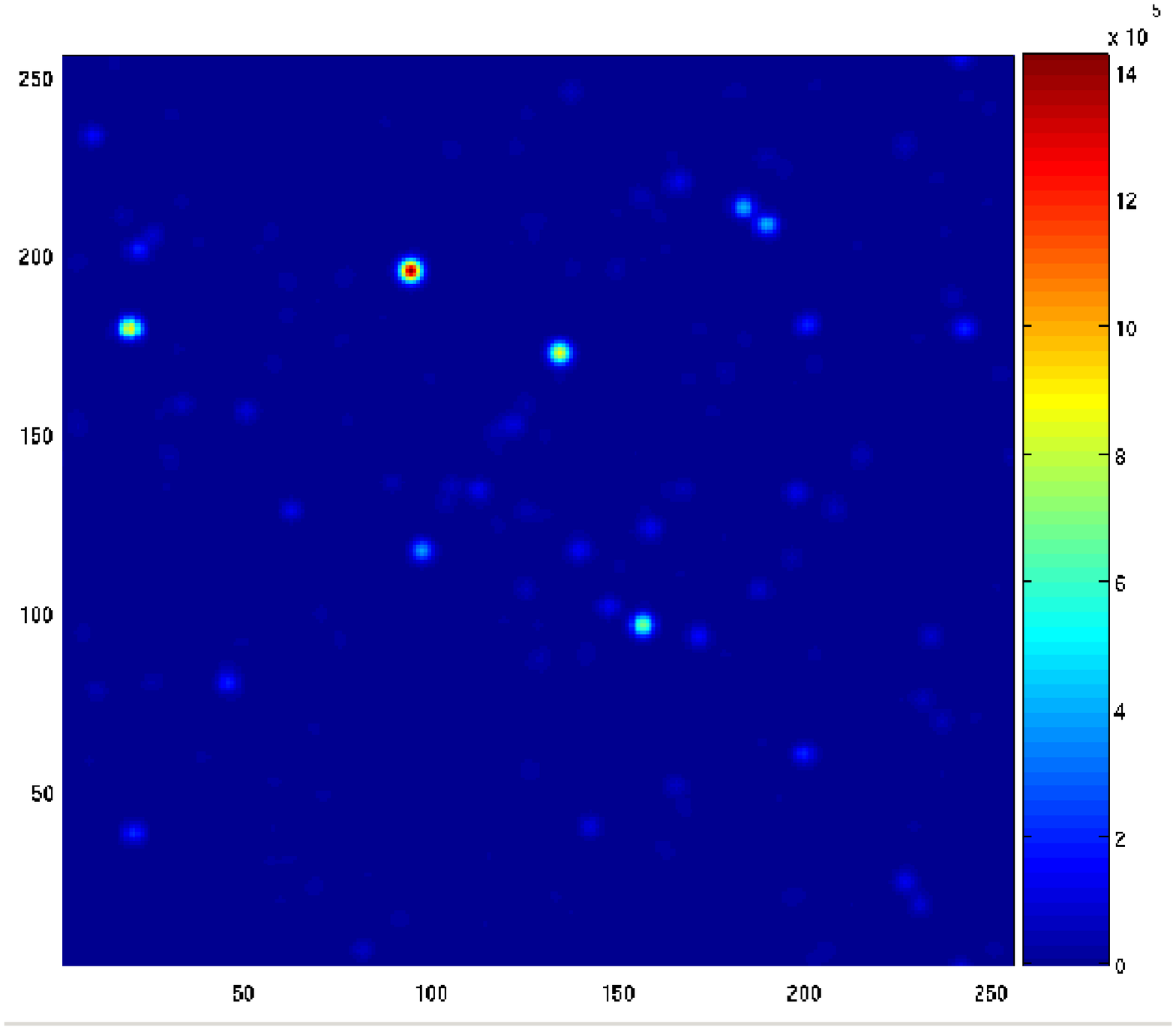} 
~~ 
        \includegraphics[width=8.0cm]{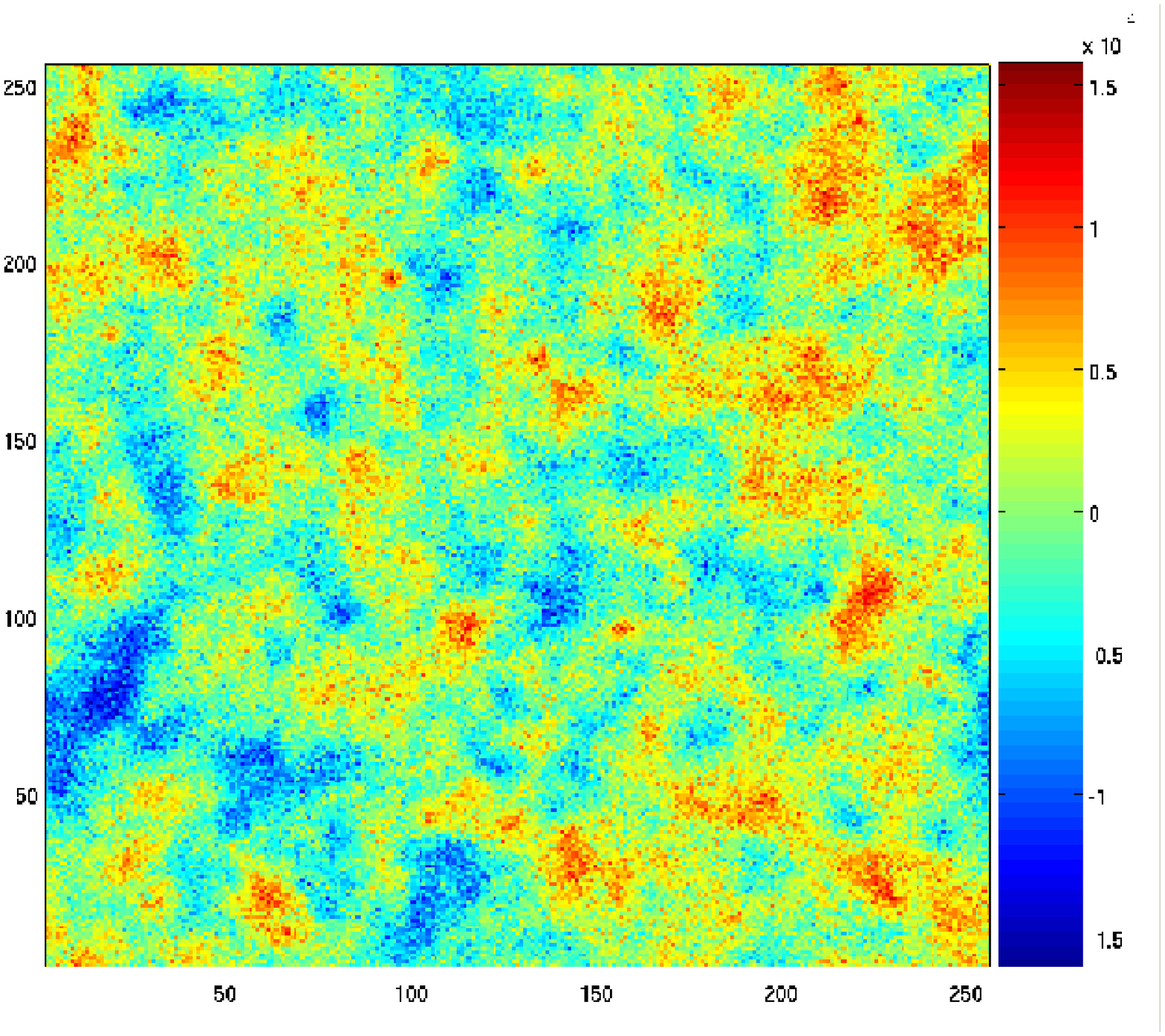} 
~~ 
        \includegraphics[width=8.0cm]{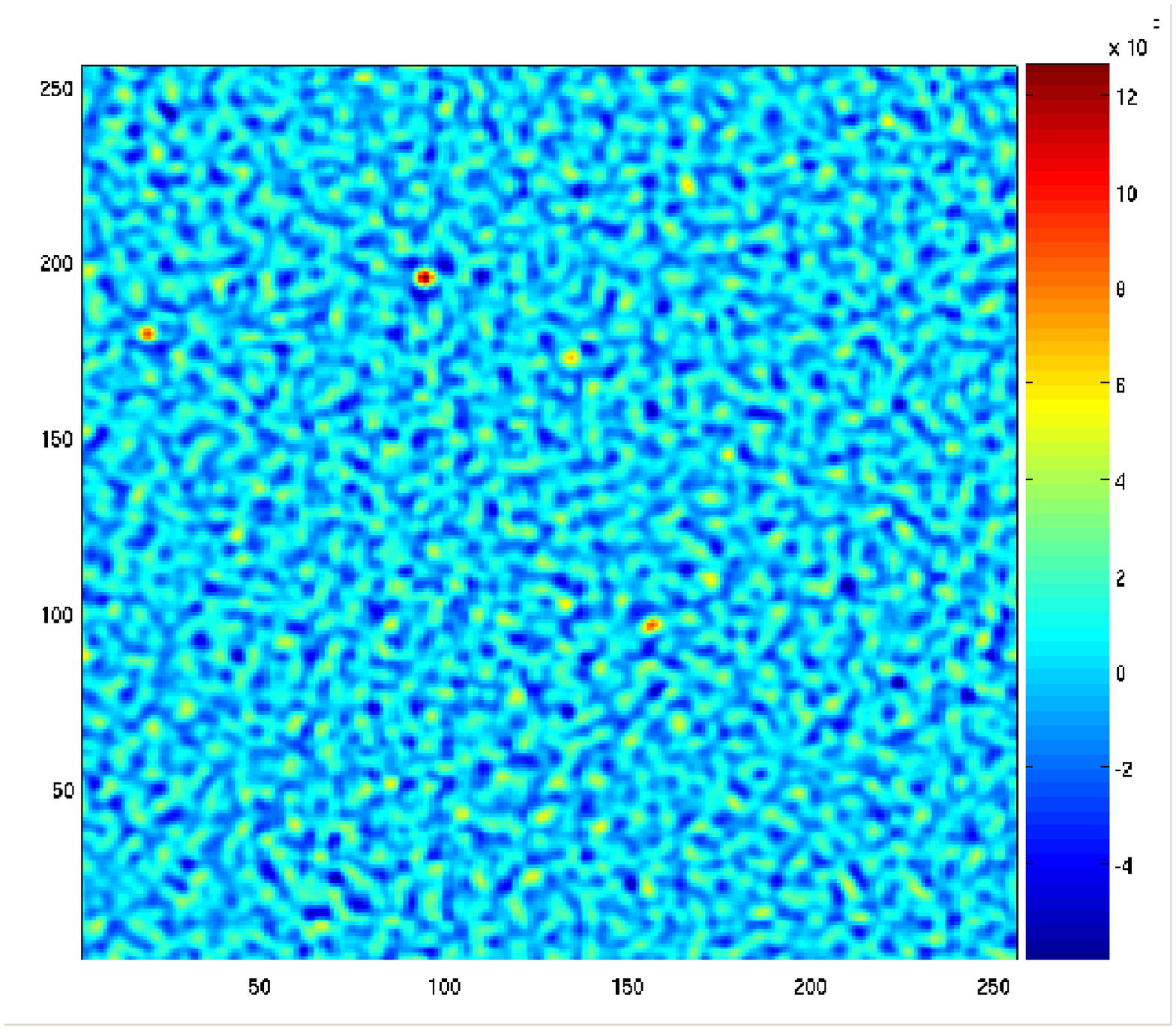} 
~~ 
\caption{In this figure we show one of the simulations at 70 GHz. In 
the upper panel we show a simulation of point sources after they have 
been convolved with the beam of the instrument. In the middle and 
lower panels we have the the combination of the different components 
before and after filtering with the MHW2.\label{fig:ejemplo_patches}} 
\end{center} 
\end{figure} 
 
As for point sources (not contained in the ``Planck Reference Sky 
model'', yet) we have used the simulation software (EPSS-2D) presented 
by \citet{jgn05} which is a very fast and efficient tool for 
simulating maps of extragalactic point sources. For this first 
comparison among the capabilities of different filters for source 
detection, we have adopted a simple Poisson distribution of point 
sources in the sky, not considering the clustering properties of the 
different source populations. 
 
The input number counts are the ones foreseen by the \citet{dz05}
evolution model for radio-selected sources and by the
\citet[]{gr01,gr04} model for high-$z$ proto-spheroidal galaxies.
These two models have proven very successful in fitting the observed
source number counts recently obtained by the VSA, ATCA, WMAP and CBI
surveys (at $\sim$15-30 GHz), and by the SCUBA and MAMBO surveys (in
the far-IR/sub-mm domain), respectively. As for late-type far-IR
galaxies we followed the same phenomenological approach as in
\citet{ngr04}.
 
At each frequency, 145 patches of 12.8x12.8 square degrees have been 
simulated, covering a total effective area of $2\pi$ sr (half of the 
sky). Each patch has contribution from: a CMB realisation following 
the Concordance Model \citep{spe03}, a point source realisation given 
by the model described above and the sum of all the Galactic 
components. Galactic foregrounds have been selected from 12 different 
regions of the sky (at Galactic latitude $¦b¦ > 30^\circ$). These 12 
regions are representatives (in terms of its dispersion) of the 
Galactic emission outside the above mentioned Galactic cut. Each one 
of these 12 Galactic simulations has been used accordantly to their 
representativeness. This is a similar procedure to the one followed in 
\citet{patri01a} (in that case 10 Galactic patches were considered). 
 
Once the CMB, the point sources and the Galactic components have been 
added, the resultant map is convolved using the instrumental Planck 
specifications (see Table \ref{tab:table1}). Finally a Gaussian white noise 
realisation is added, with a dispersion per pixel according to the 
Planck specifications (see Table 1). 
 
Each image is filtered with the MF, the MHW1 and the MHW2. To avoid 
spurious artifacts due to border effects, five pixels on the edge of 
the image have not been considered.  The MHW1 and MHW2 operates at 
their respective optimal scales. This optimal scale can be easily 
obtained by filtering the image with a series of scales and by finding 
the particular scale that produces an image with maximum 
amplification. Note that this procedure does not affect the flux 
of the sources, since the filters are normalised in such a way, 
eq. (3), that the flux of the source is always preserved. 
 
We then compare the list of source candidates (our recovered
catalogue) with the reference (input) source catalogue, by trying to
match the detected sources with the simulated ones. A source candidate
is considered as a positive detection if it is found within a radius
$r\leq \mathrm{FWHM}/2$ from the corresponding source in the reference
catalogue. Only those objects in the candidate catalogue with a
positive matching and with an error in the estimate of their flux
density $\leq 100\%$ (with respect to the flux of the simulated
source) are considered as real (positive) detections. Every other
candidate source with flux above a fixed detection threshold in the
filtered map (see Figures ) is considered as a false or {\it spurious
detection} and it has not been included in the final catalogue,
corresponding to that particular Planck channel.

\begin{table} 
  \centering 
  \begin{tabular}{ccccc} 
   Freq.  & Im. Size & FWHM & Pix. Size & $\sigma_{noise}$  \\ 
   (GHz) & (pixels)& (arcmin) & (arcmin) & ($10^{-6}$)\\ 
  \hline 
    857 & 512 & 5.0 & 1.5 &  6700.0 \\ 
    545 & 512 & 5.0 & 1.5 &  147.0 \\ 
    353 & 512 & 5.0 & 1.5 &  14.7 \\ 
    217 & 512 & 5.0 & 1.5 &  4.8 \\ 
    143 & 512 & 7.1 & 1.5 &  2.2 \\ 
    100 & 256 & 9.5 & 3.0 & 2.5\\ 
    70 & 256 & 14.0 & 3.0 &  4.7 \\ 
    44 & 128 & 24.0 & 6.0 &  2.7 \\ 
    30 & 128 & 33.0 & 6.0 &  2.0 \\ 
  \hline 
\end{tabular} 
  \caption{Instrument performance goals for all the Planck satellite 
channels. The antenna is assumed to be a circular Gaussian one and the 
instrumental noise level is in units of $\Delta T/T$ per $FWHM$ (e.g 
in a square whose side is the FWHM of the beam).\label{tab:table1}} 
\end{table} 
 
\section{Results and discussion} \label{sec:results} 
 
In this Section we compare the performances of the filters 
described and discussed in the previous sections (MHW1, MHW2, MF). 
 
\begin{figure*} 
\begin{center} 
 
        \includegraphics[width=5.8cm]{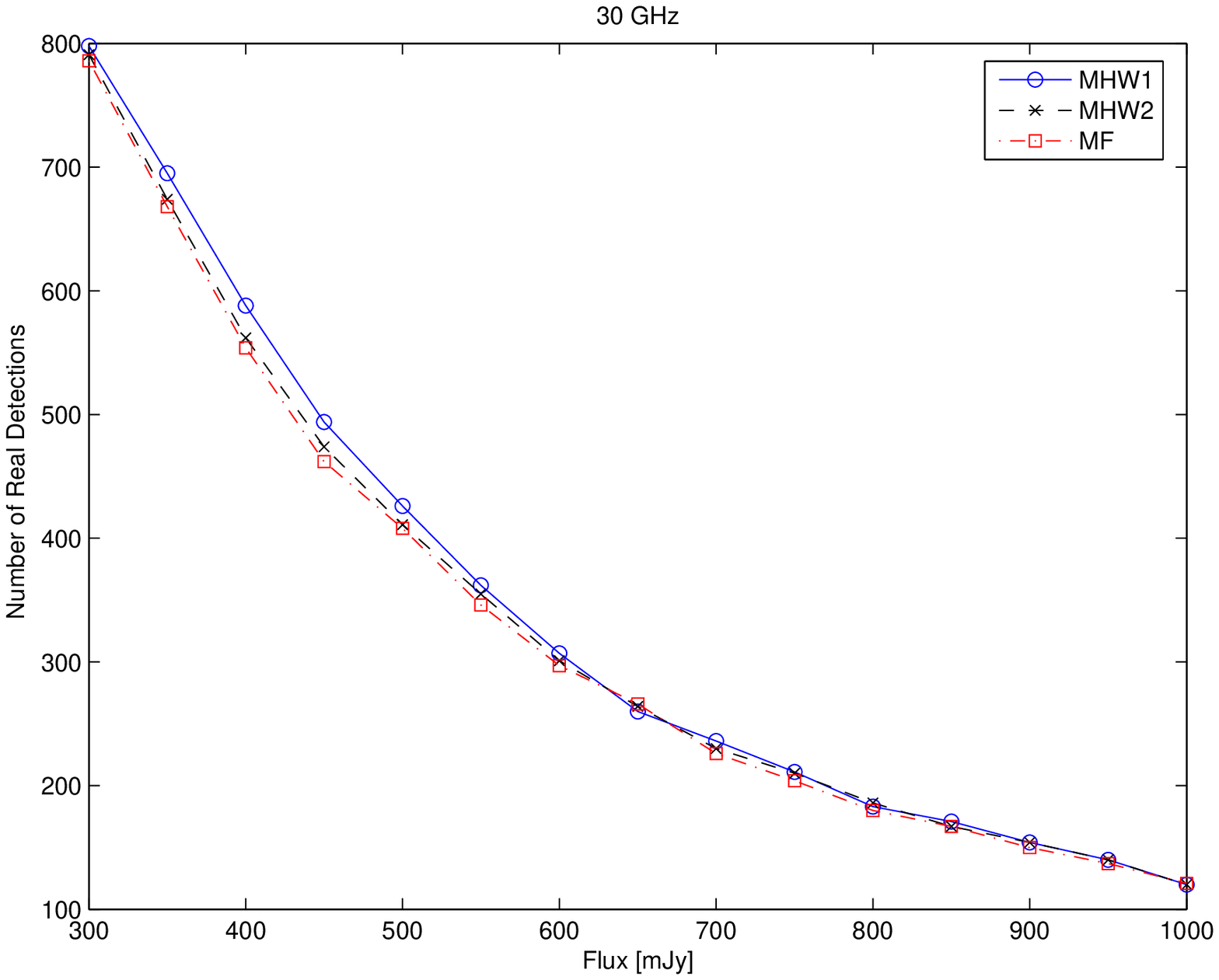} 
        \includegraphics[width=5.8cm]{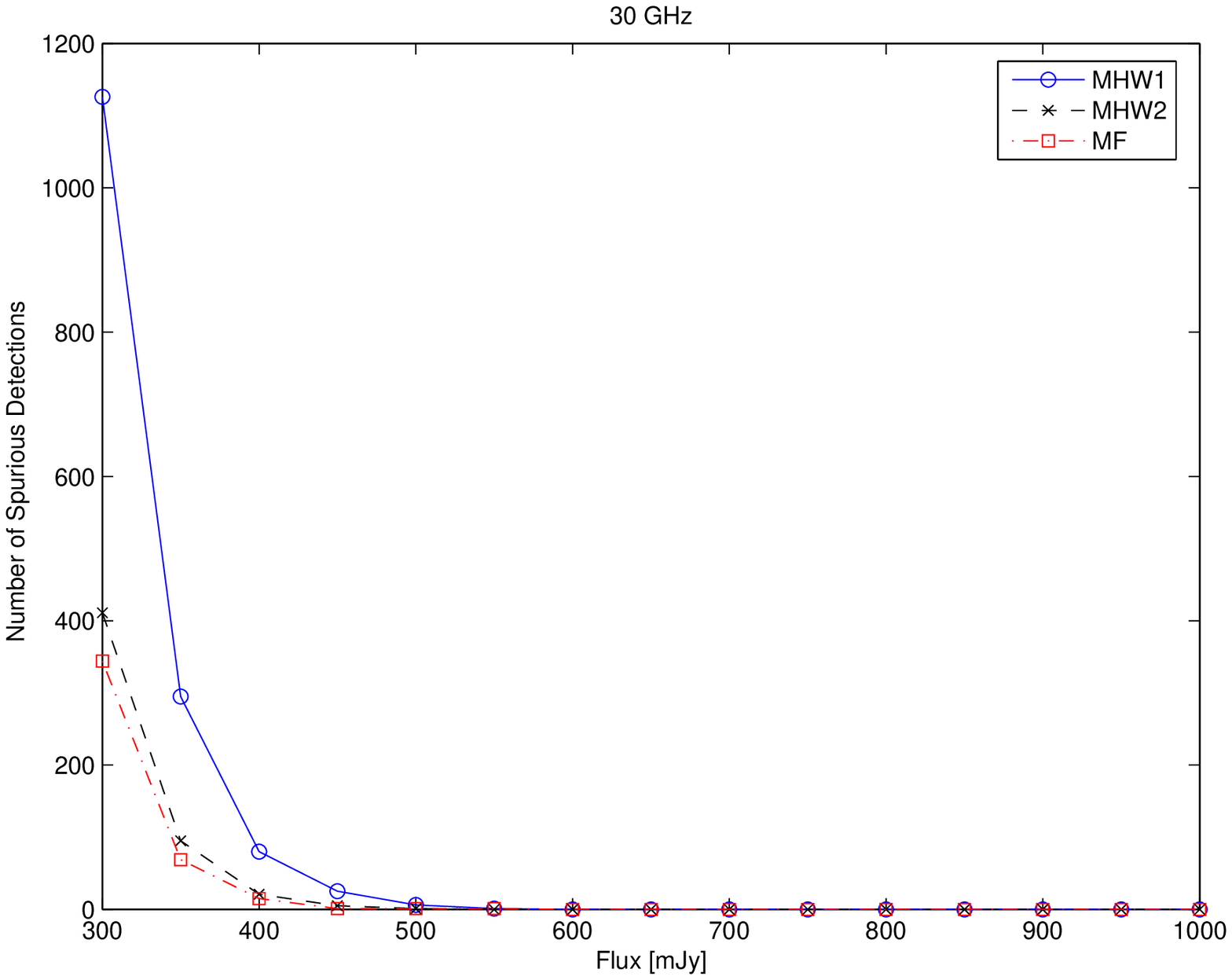} 
~~ 
        \includegraphics[width=5.8cm]{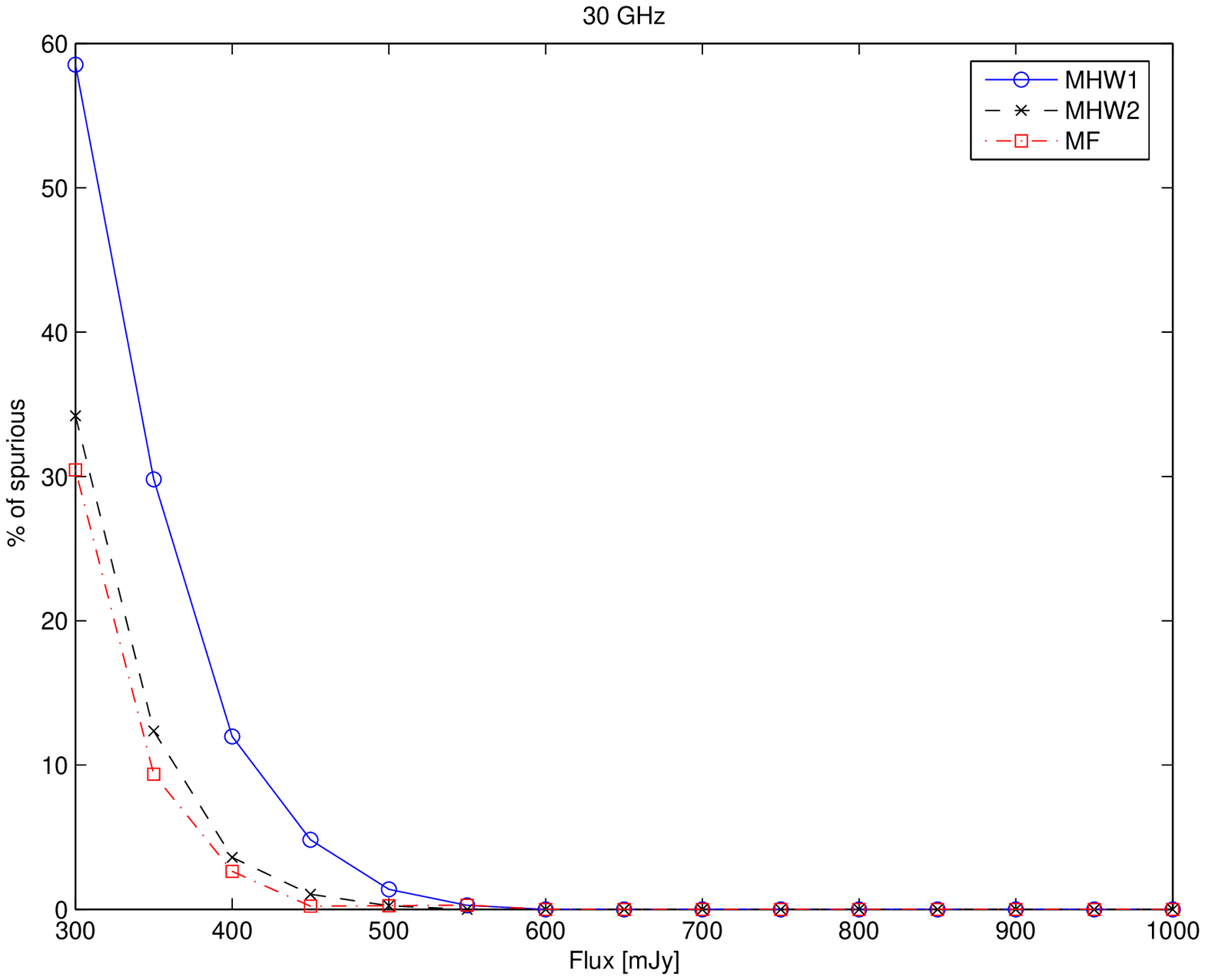} 
        \includegraphics[width=5.8cm]{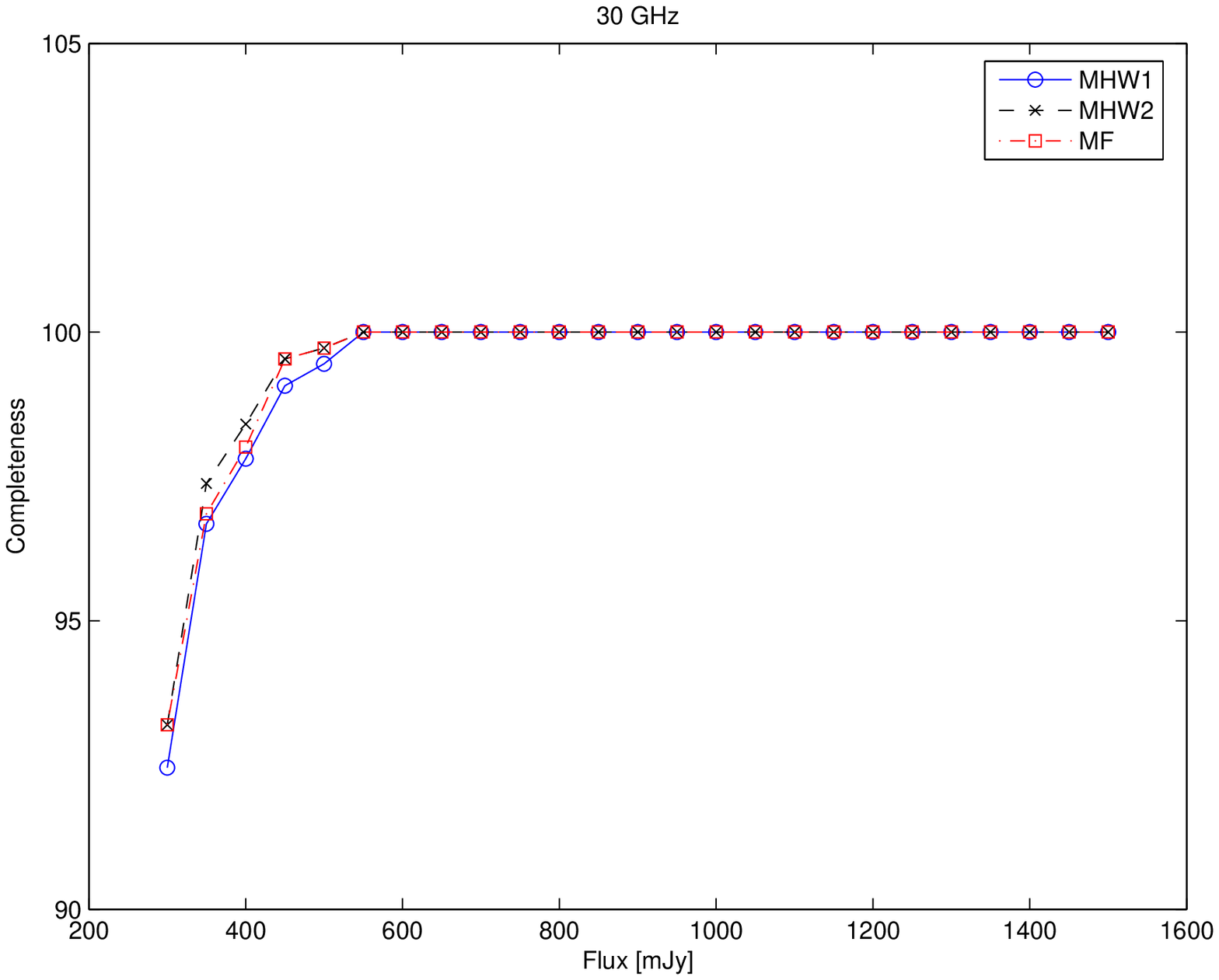} 
    \caption{30 GHz. In this figure we compare the results 
      obtained for the three considered filters, the MHW1 and MHW2 
      at their optimal scales and the MF. In the upper left 
      figure we show the number of detected sources above a given 
      flux detection limit. In the upper right figure we present the number of 
      spurious or false detections above a given flux detection limit. The 
      lower left figure shows the percentage of spurious sources 
      (defined as the number of spurious divided by the sum of 
      spurious and real detections) above a given flux detection limit. The 
      lower right figure shows the completeness level of the 
      catalogues above a given flux detection limit. The solid line 
      corresponds to the MHW1, the dashed line corresponds to the 
      MHW2 and the dot-dashed line corresponds to the 
      MF. Similarly, the circle, cross and squares correspond to 
      the MHW, MHW2 and the MF 
      respectively. \label{fig:fig_ps_30}} 
 
\end{center} 
\end{figure*}

\begin{figure*} 
\begin{center} 
        \includegraphics[width=6.2cm]{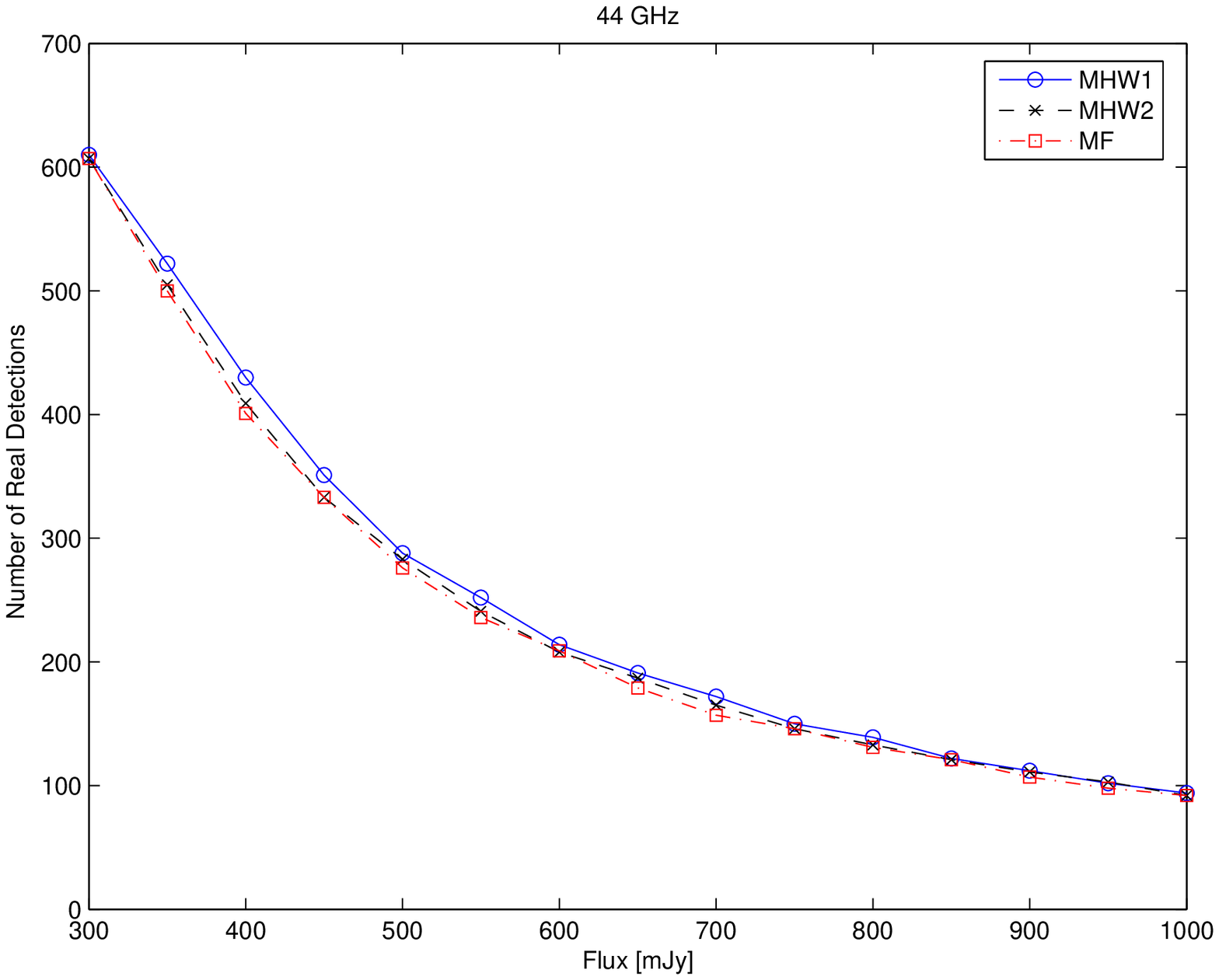} 
        \includegraphics[width=6.2cm]{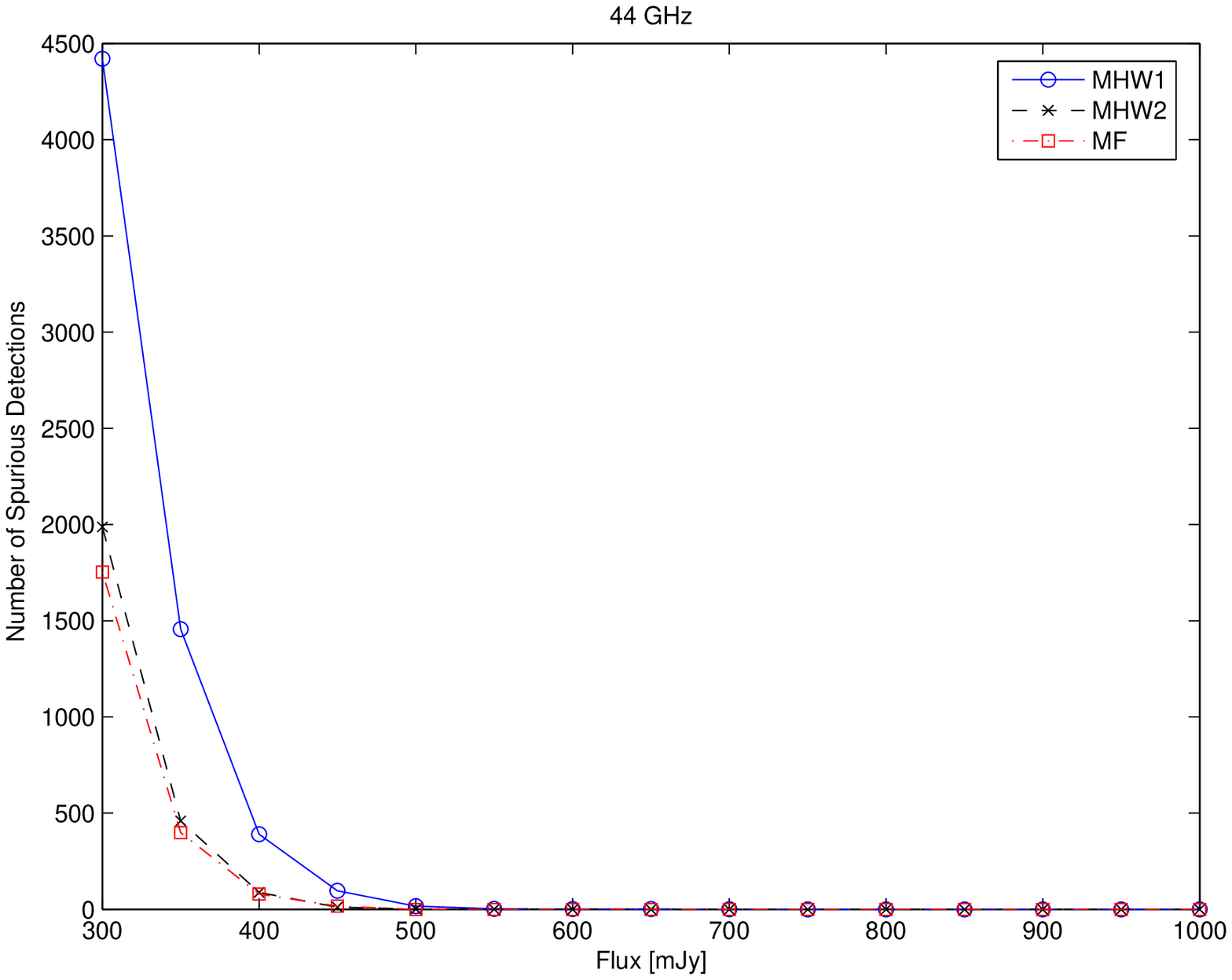} 
~~ 
        \includegraphics[width=6.2cm]{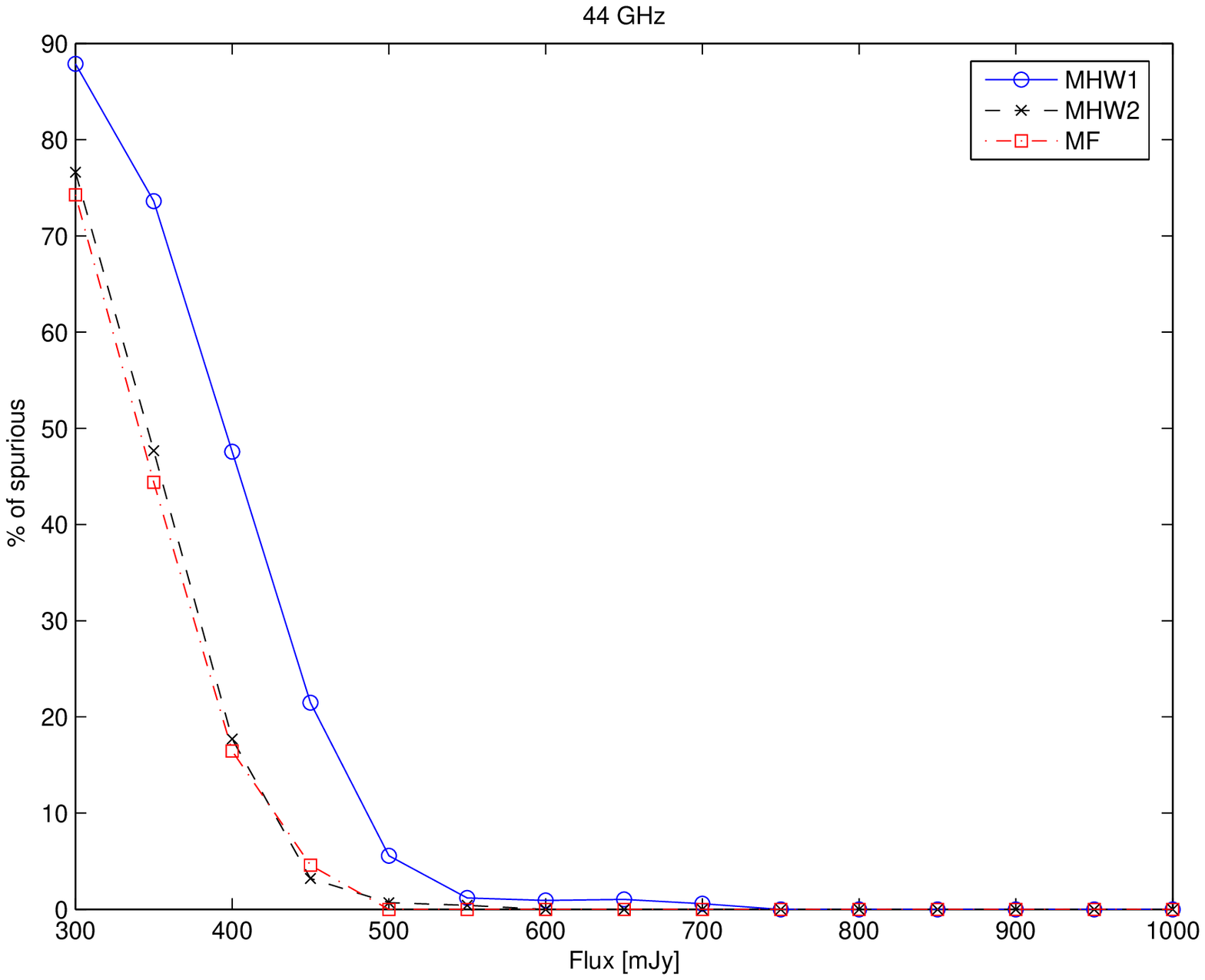} 
        \includegraphics[width=6.2cm]{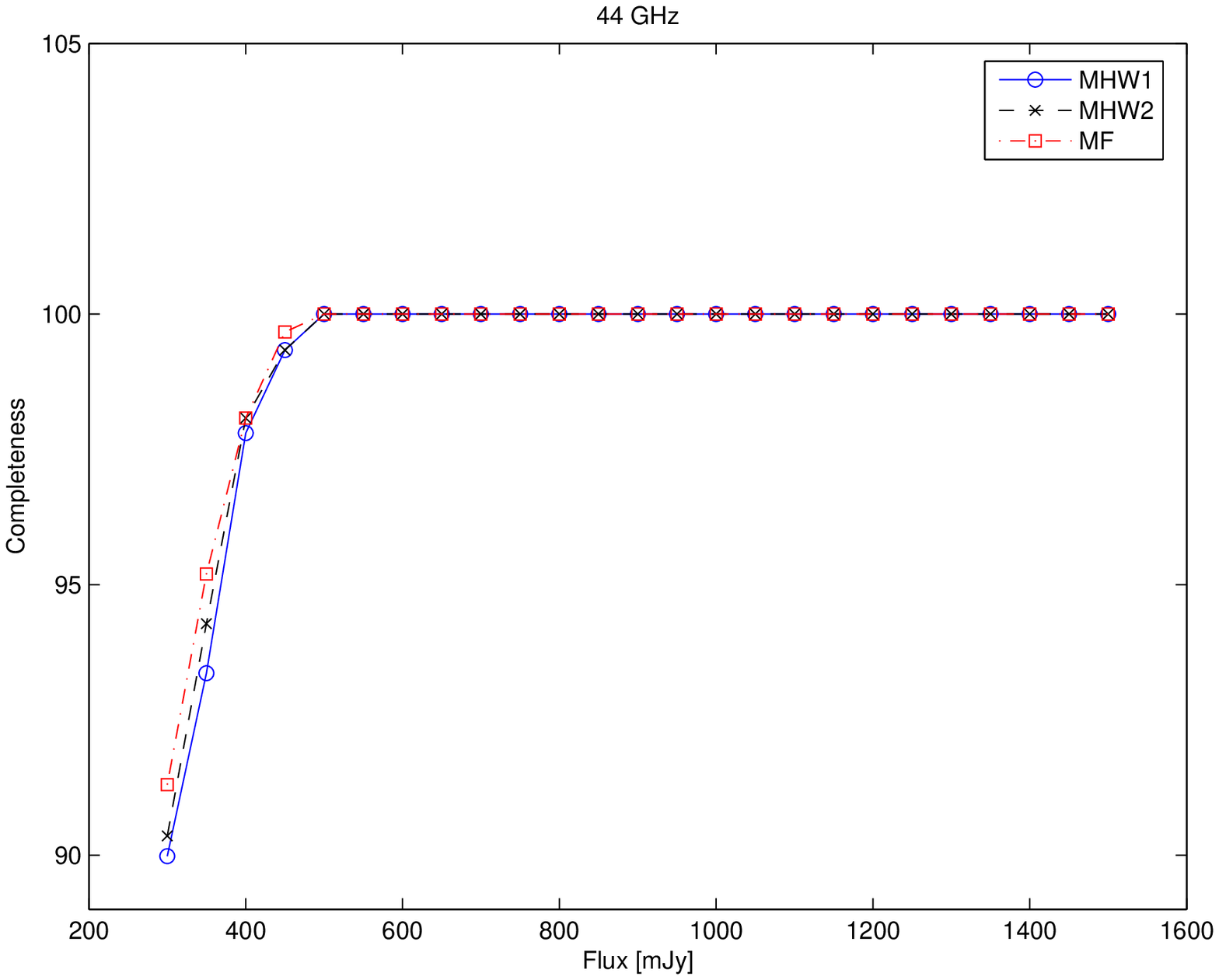} 
\caption{The same as in Figure \ref{fig:fig_ps_30} but at 44 GHz. \label{fig:fig_ps_44}}
\end{center} 
\end{figure*}

\begin{figure*} 
\begin{center} 
 
        \includegraphics[width=6.2cm]{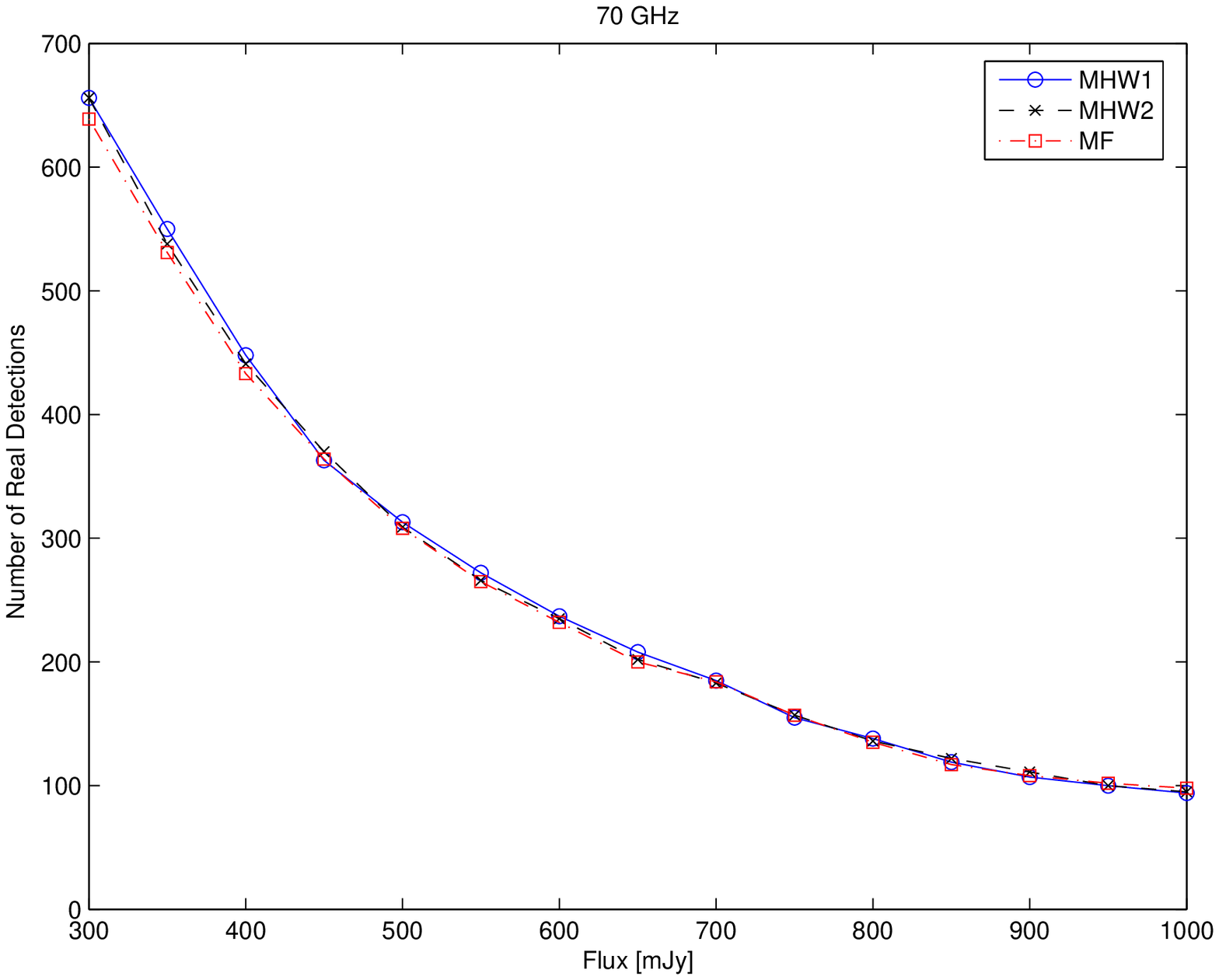} 
        \includegraphics[width=6.2cm]{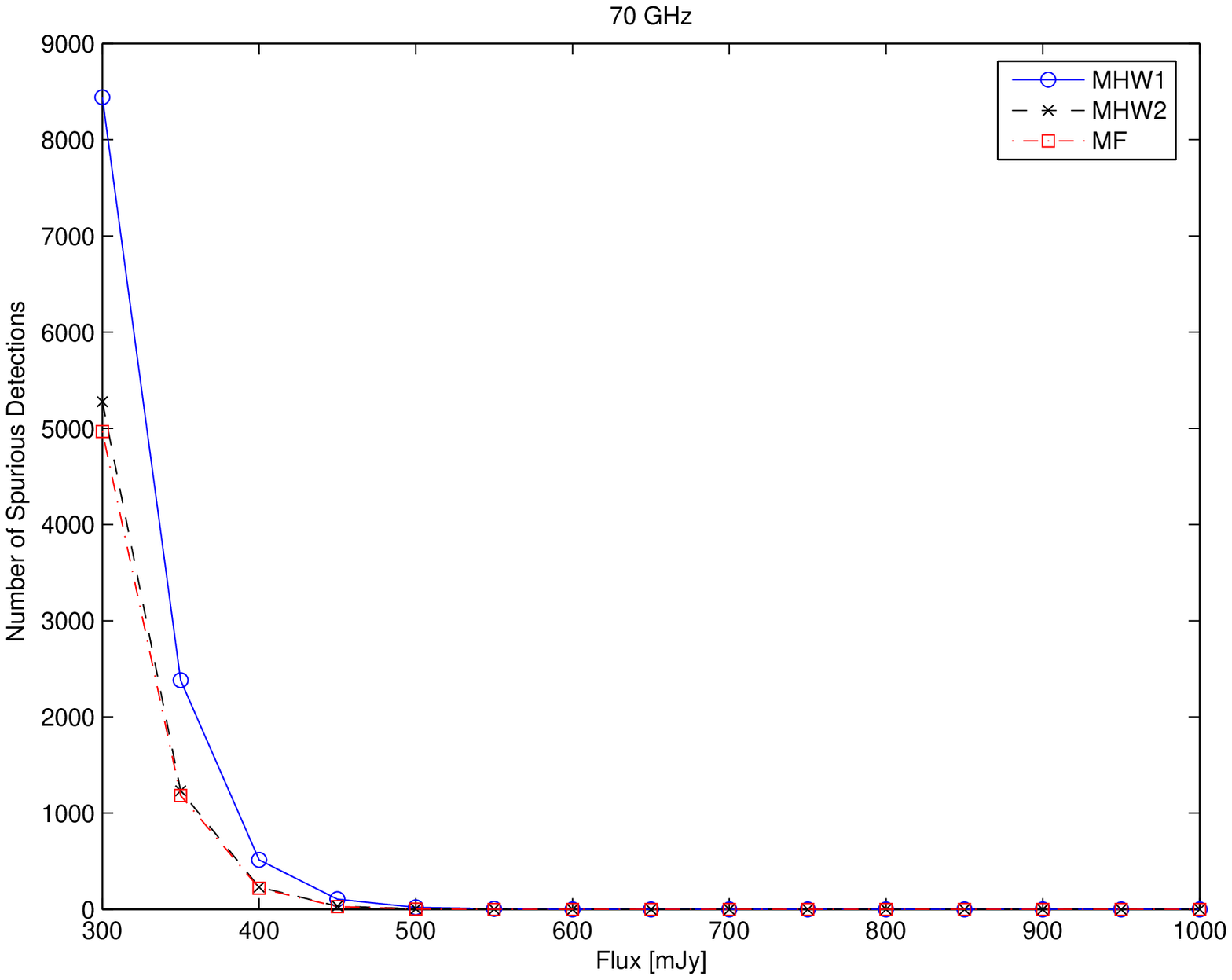} 
~~ 
        \includegraphics[width=6.2cm]{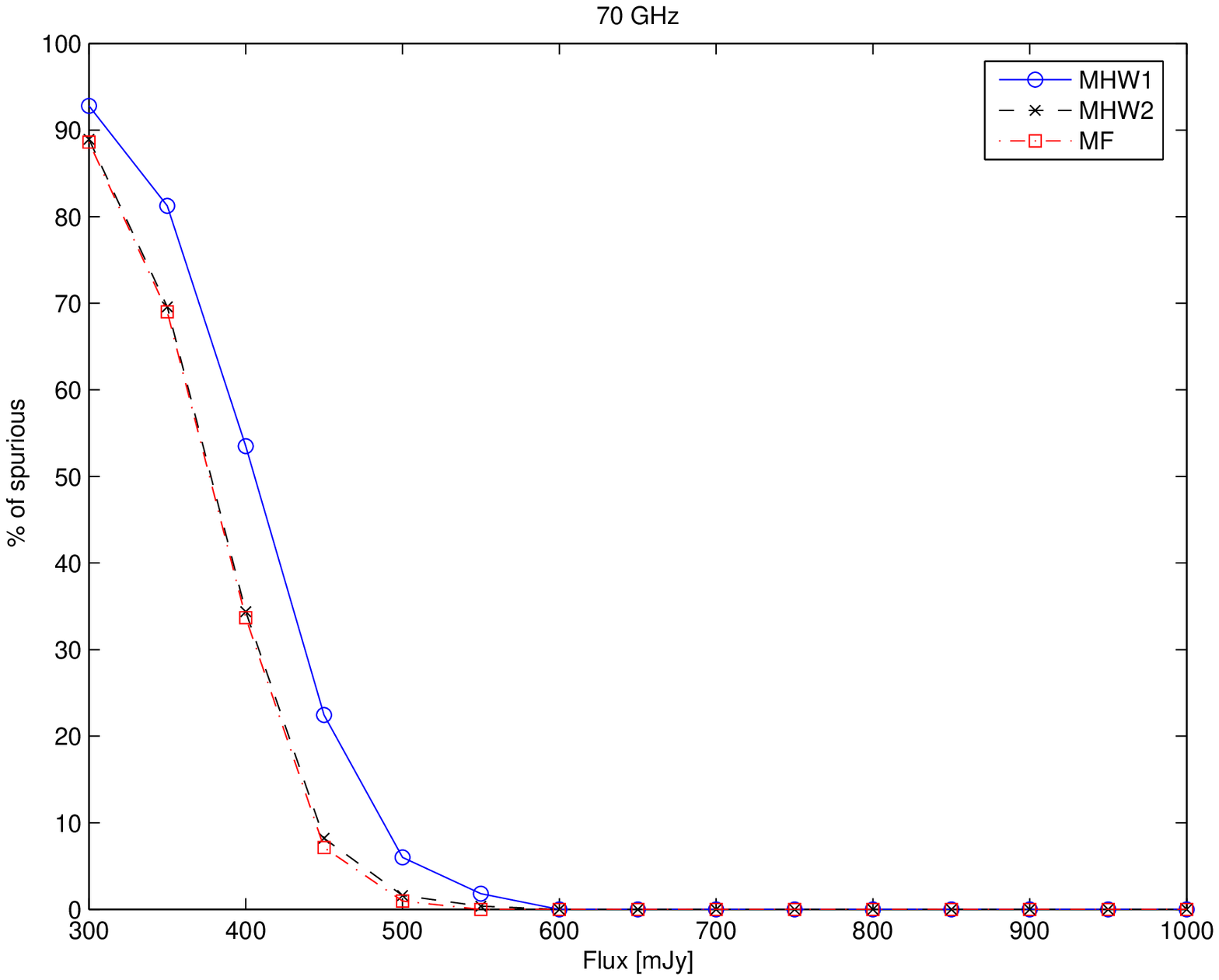} 
        \includegraphics[width=6.2cm]{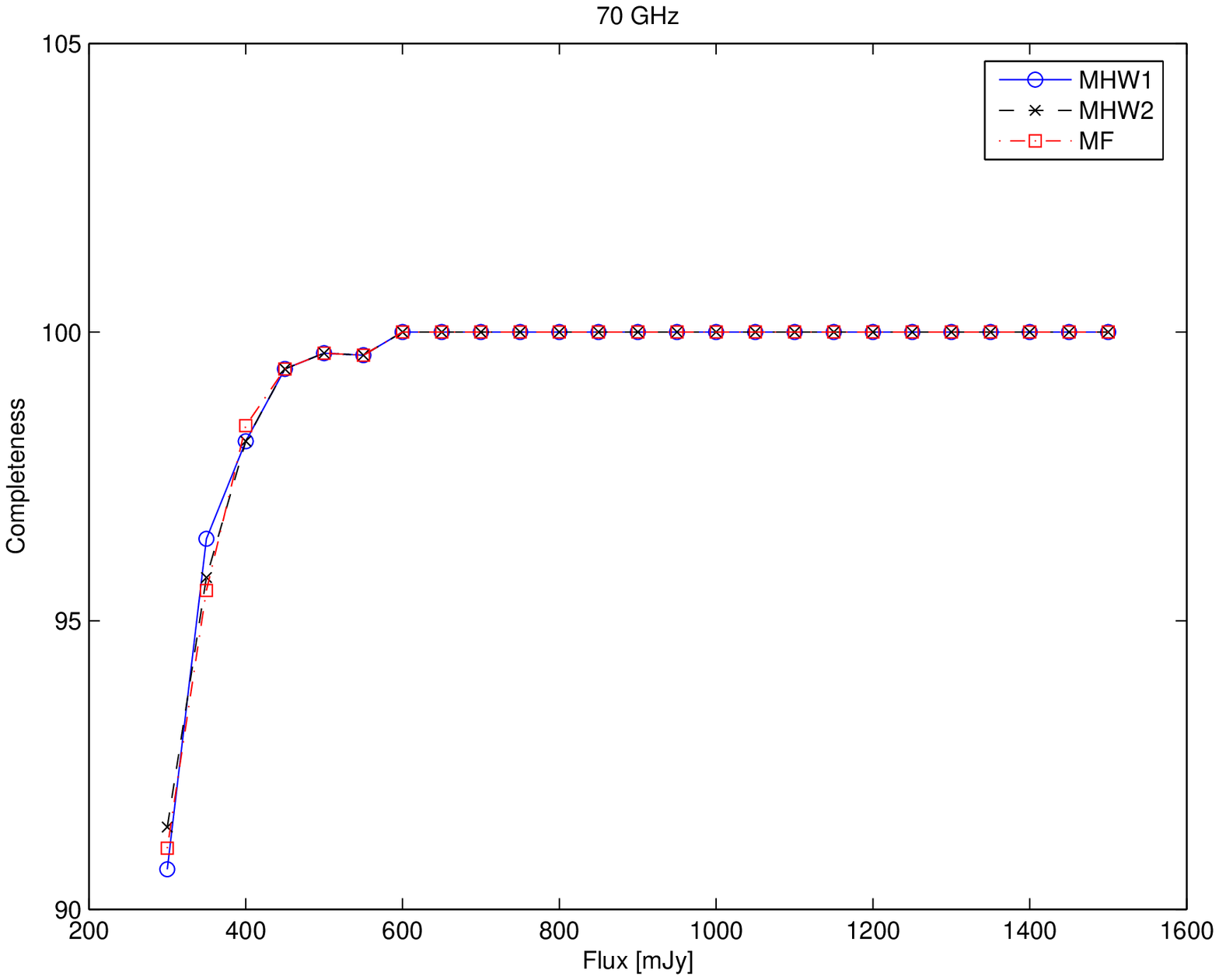} 
 
\caption{The same as in Figure \ref{fig:fig_ps_30} but at 70 GHz. \label{fig:fig_ps_70}}
\end{center} 
\end{figure*}

\begin{figure*} 
\begin{center} 
        \includegraphics[width=6.2cm]{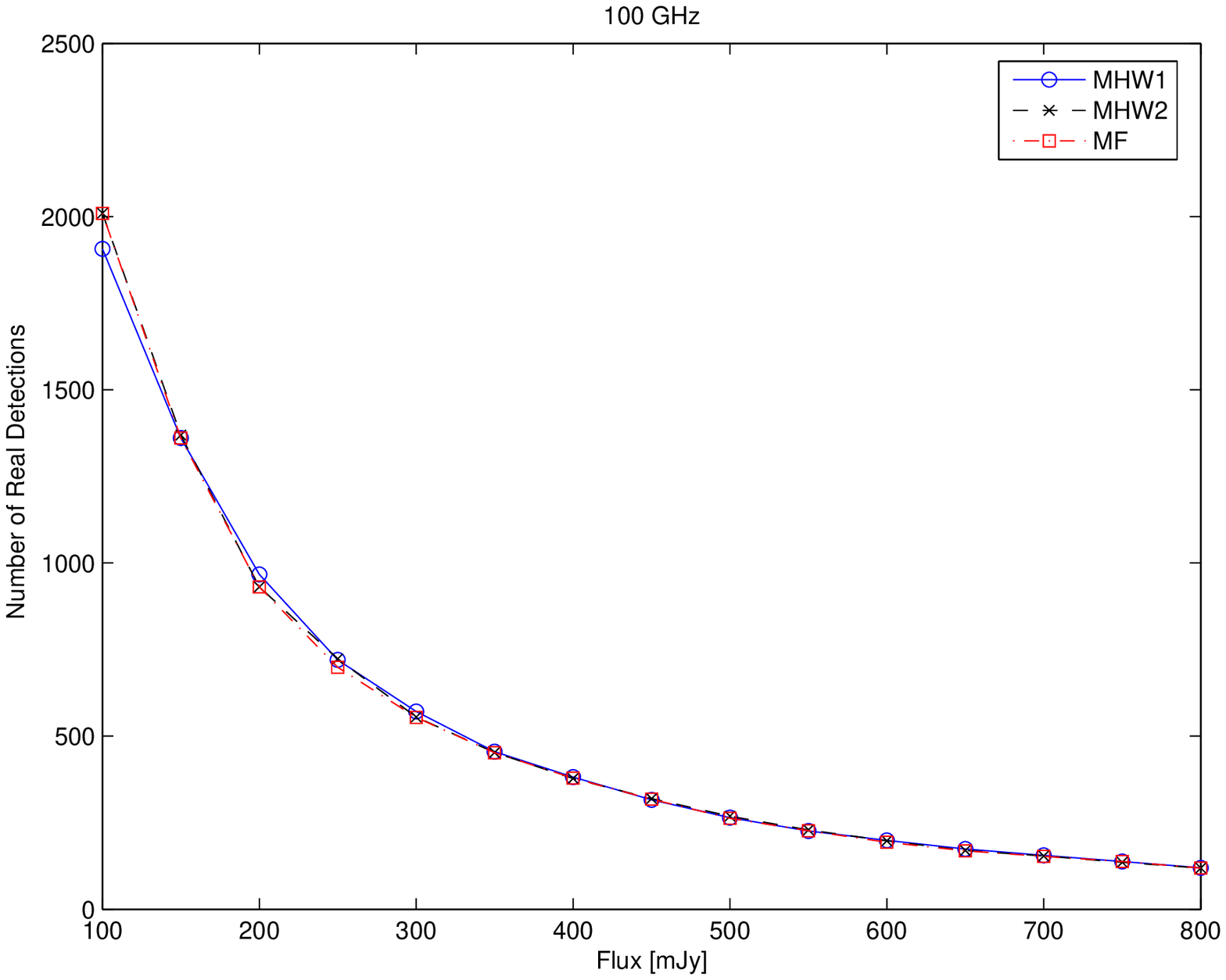} 
        \includegraphics[width=6.2cm]{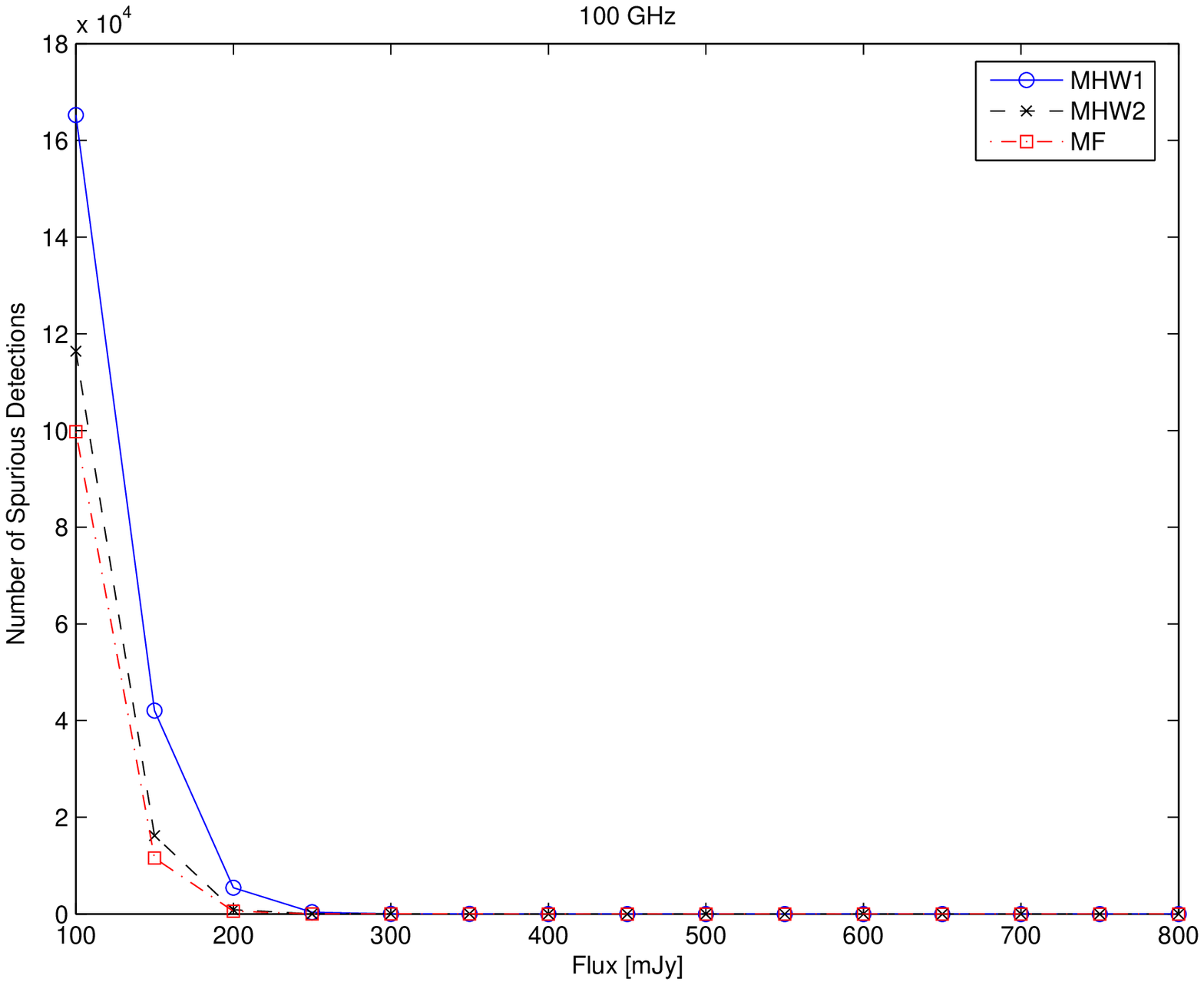} 
~~ 
        \includegraphics[width=6.2cm]{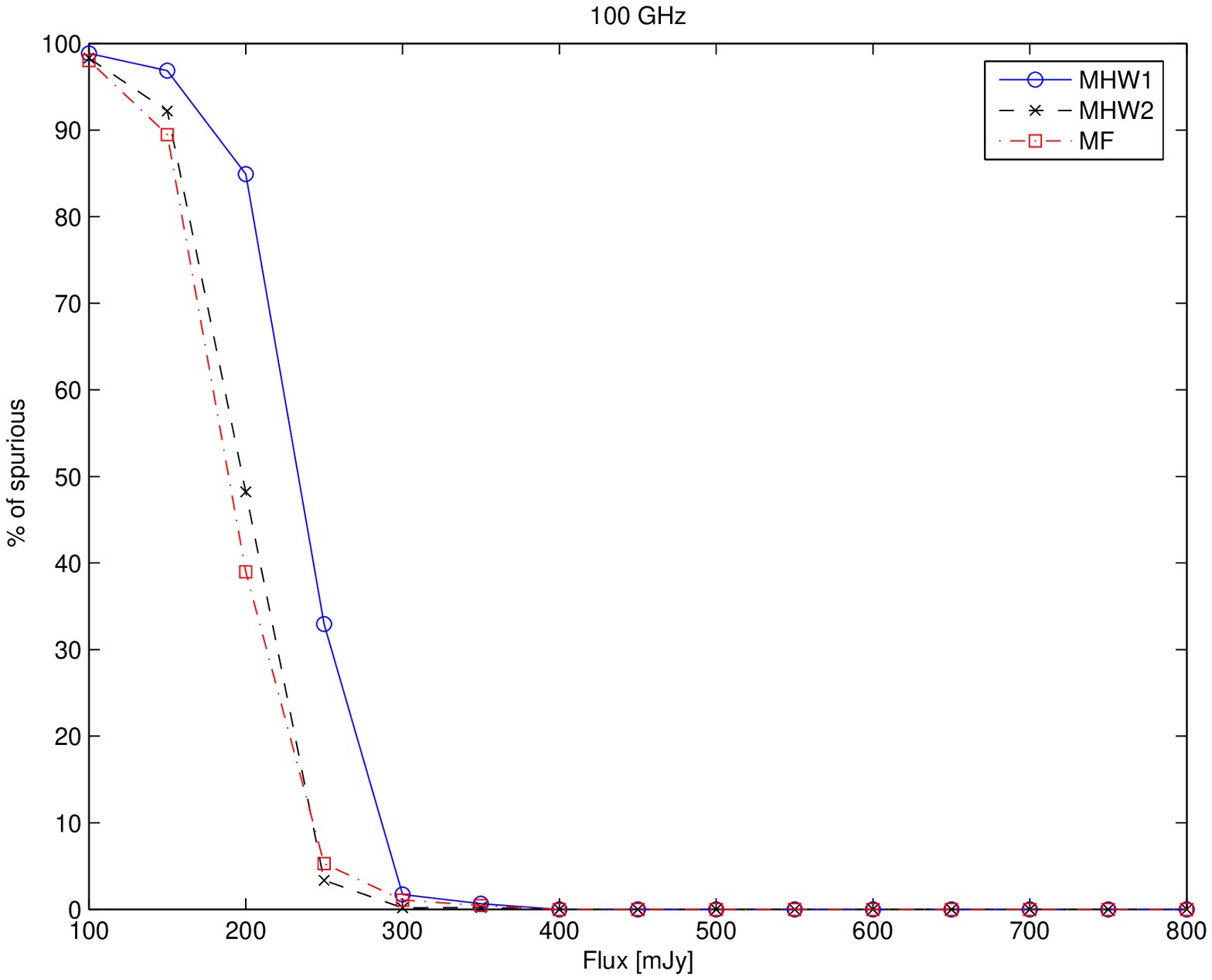} 
        \includegraphics[width=6.2cm]{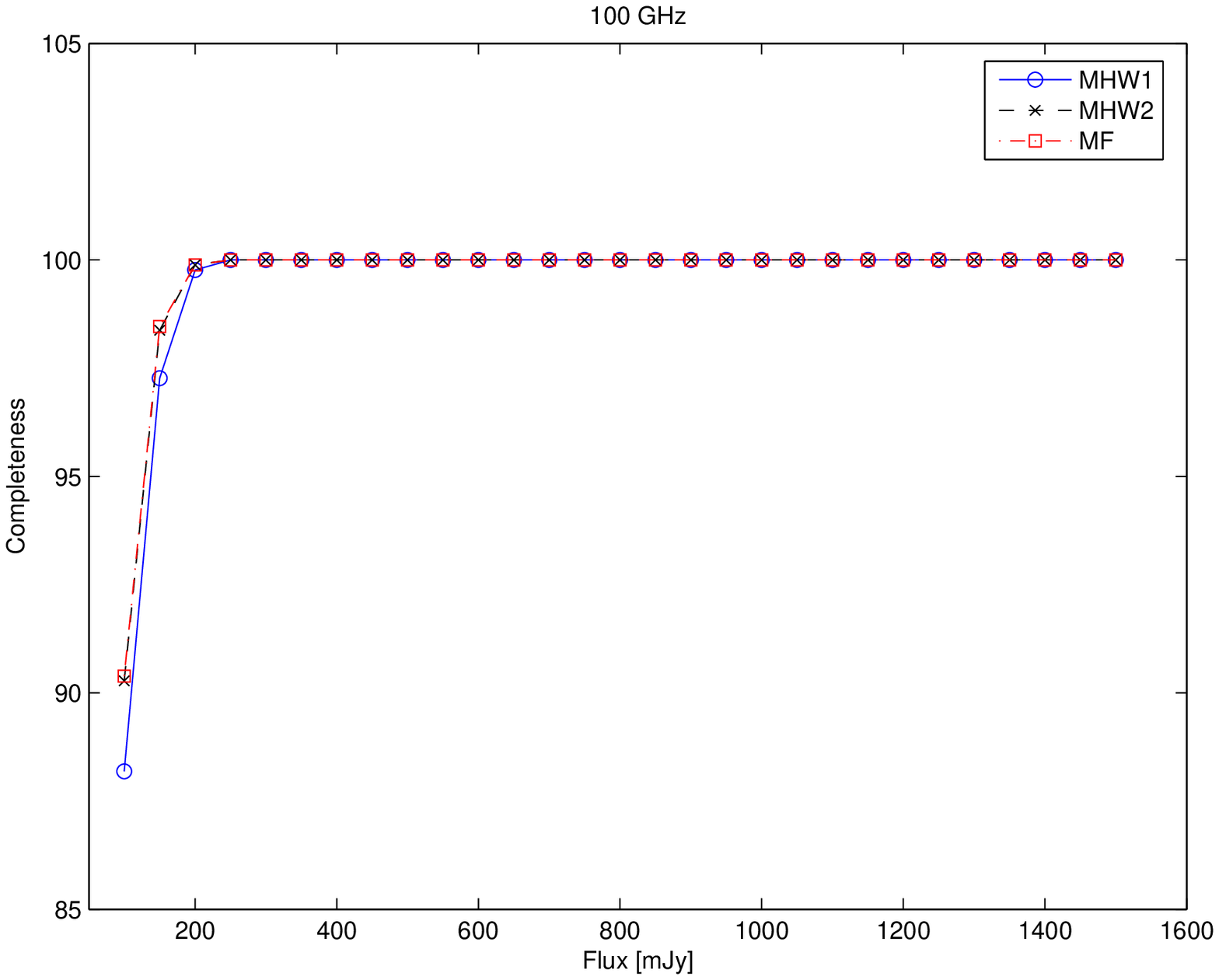} 
\caption{The same as in Figure \ref{fig:fig_ps_30} but at 100 GHz. \label{fig:fig_ps_100}}
\end{center} 
\end{figure*}

\begin{figure*} 
\begin{center} 
        \includegraphics[width=6.2cm]{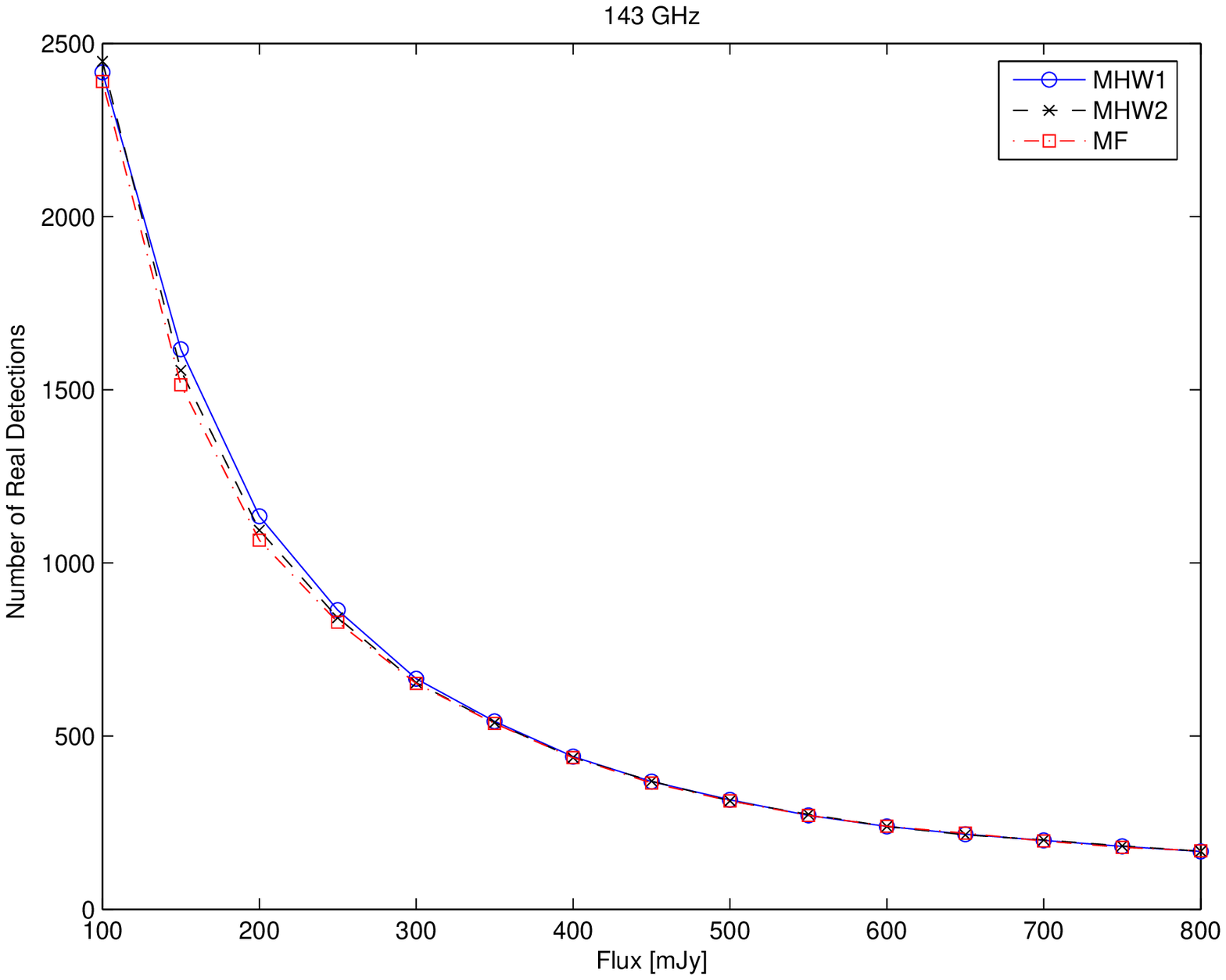} 
        \includegraphics[width=6.2cm]{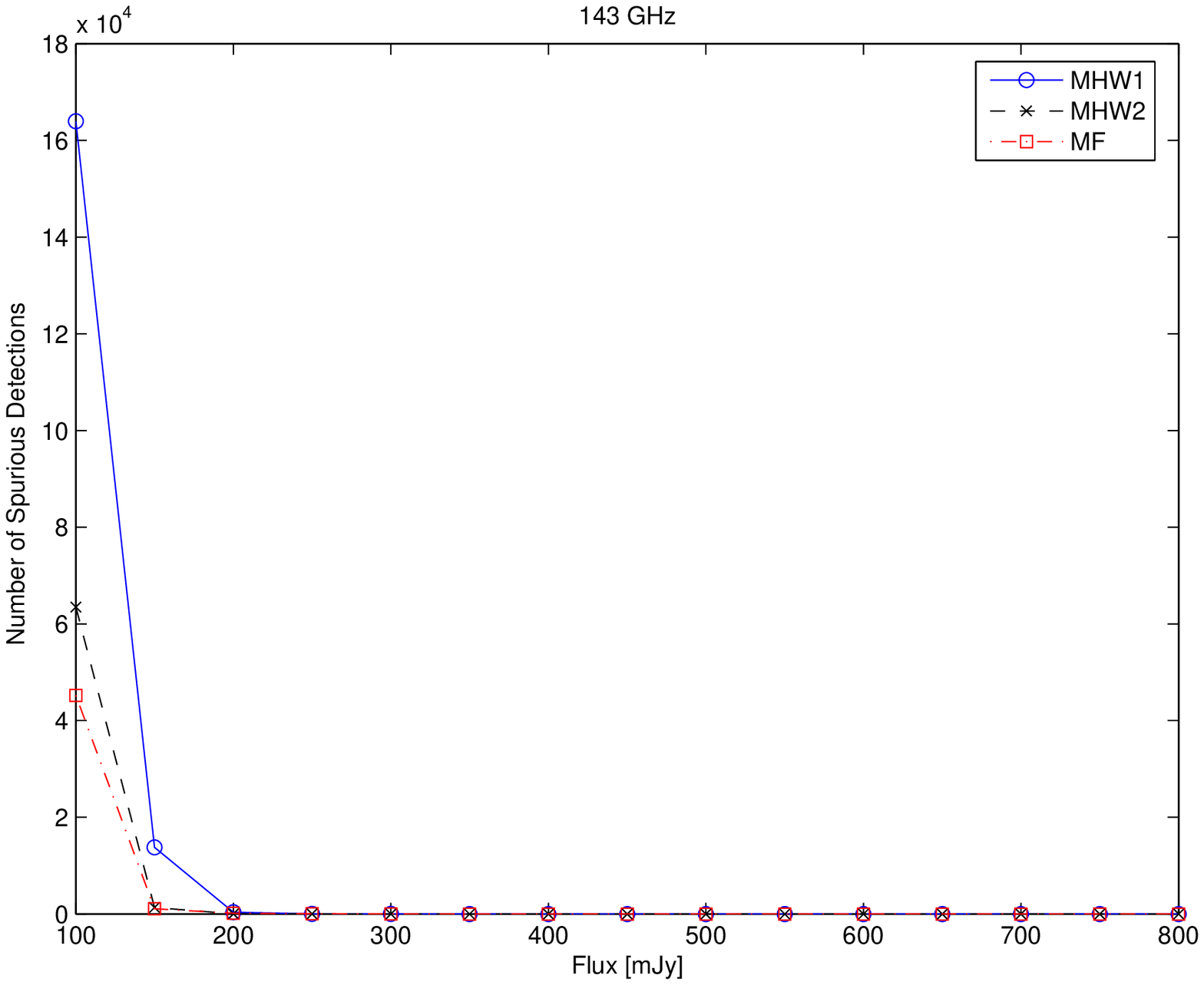} 
~~ 
        \includegraphics[width=6.2cm]{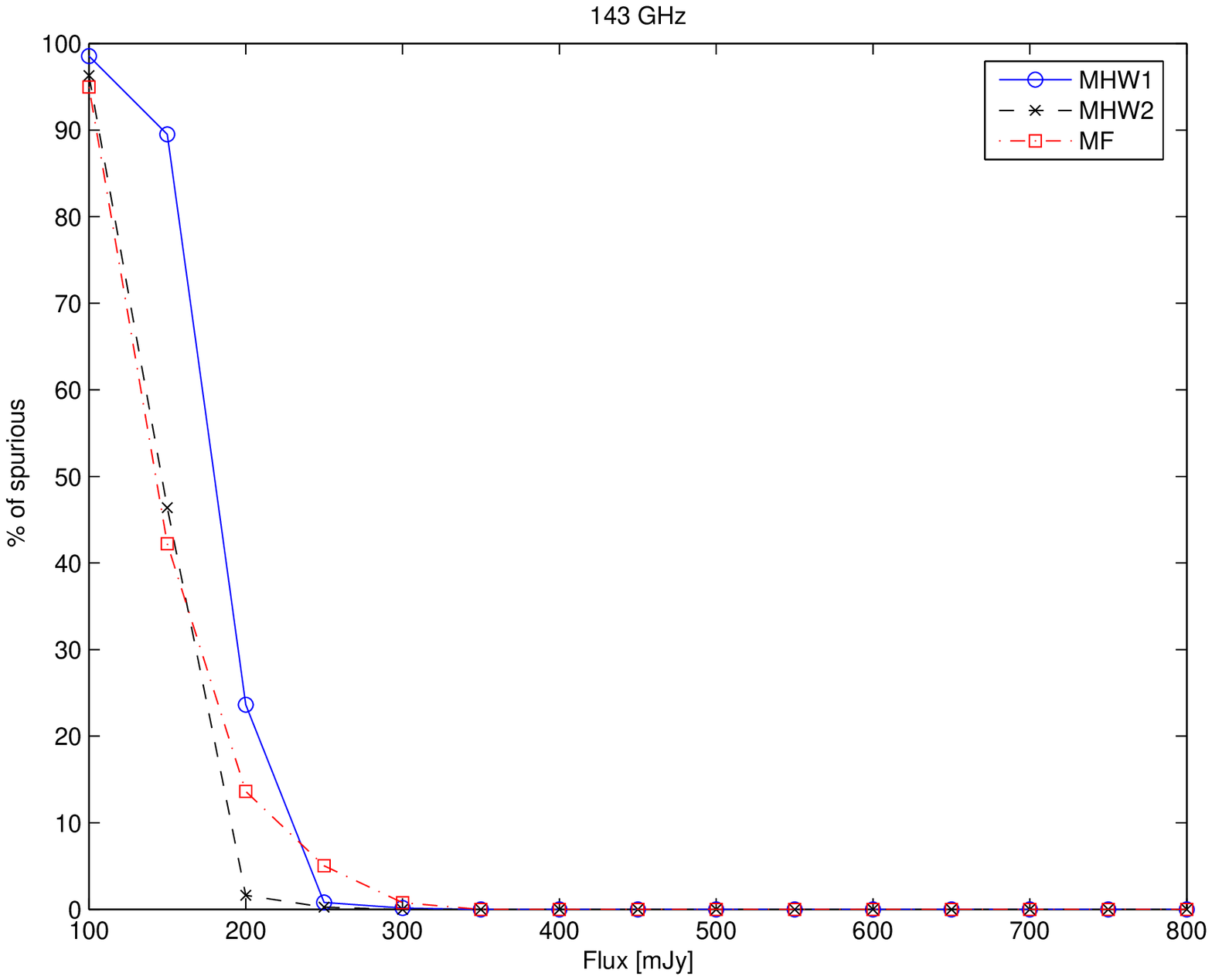} 
        \includegraphics[width=6.2cm]{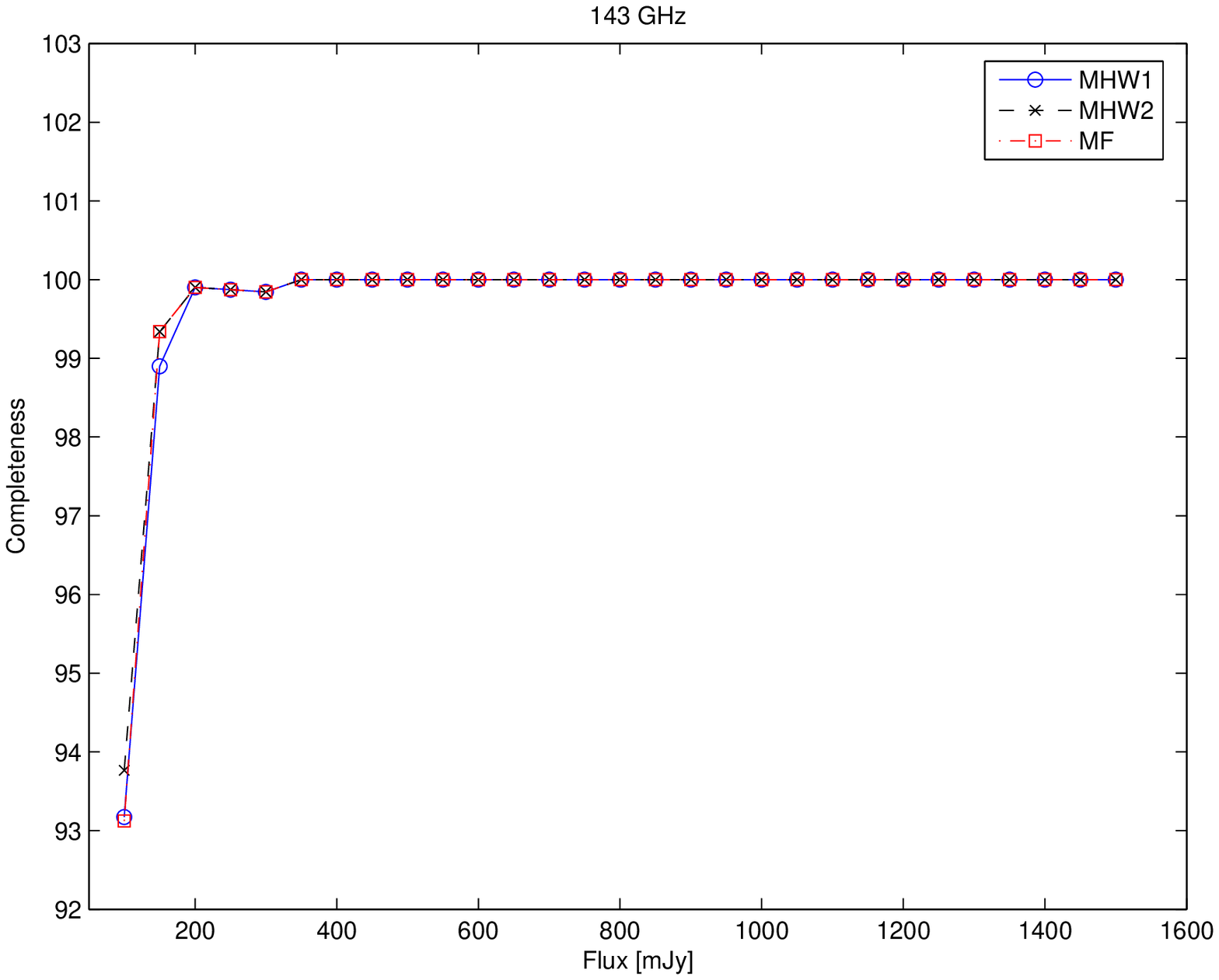} 
\caption{The same as in Figure \ref{fig:fig_ps_30} but at 143 GHz.\label{fig:fig_ps_143} }
\end{center} 
\end{figure*}

\begin{figure*} 
\begin{center} 
        \includegraphics[width=6.2cm]{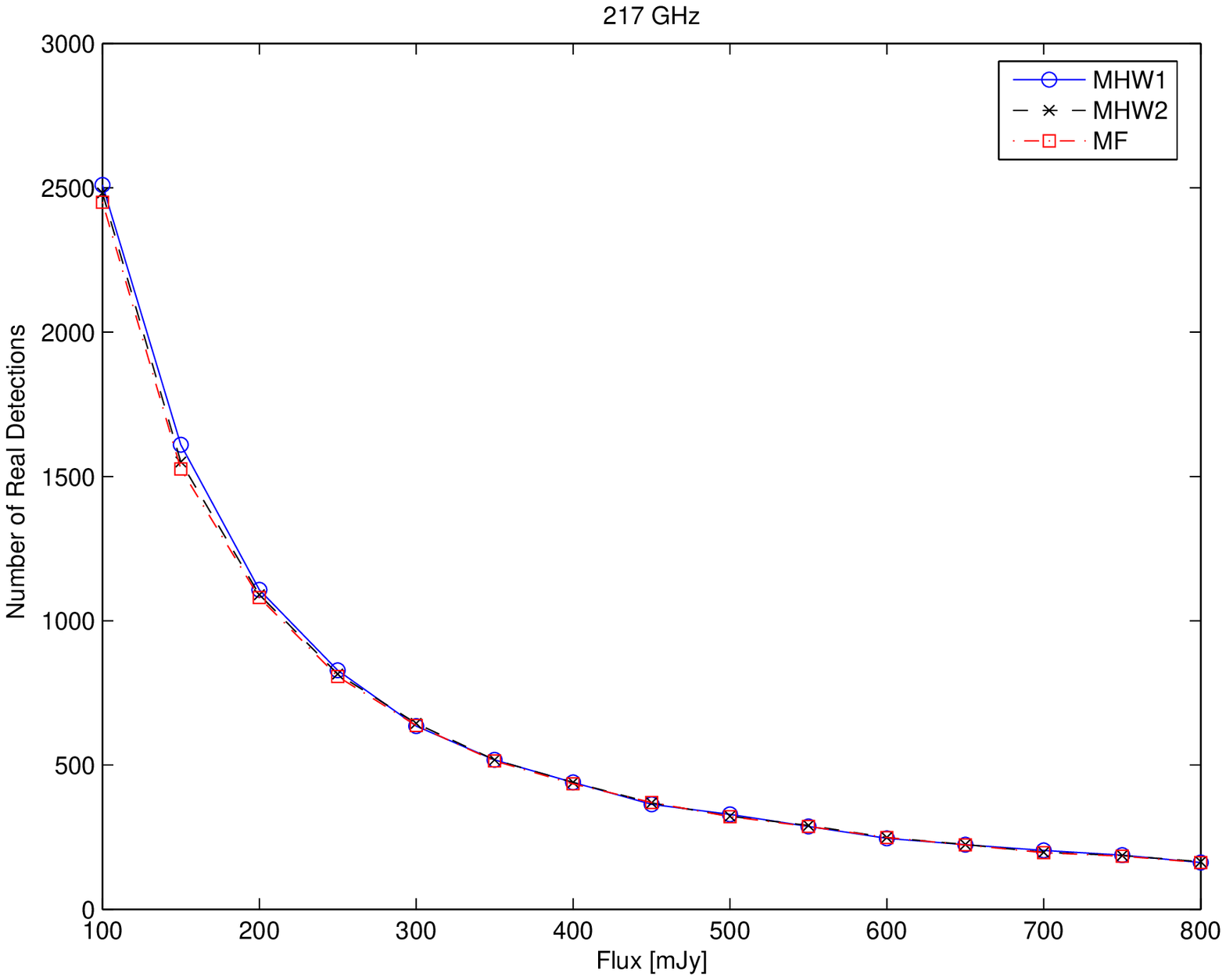} 
        \includegraphics[width=6.2cm]{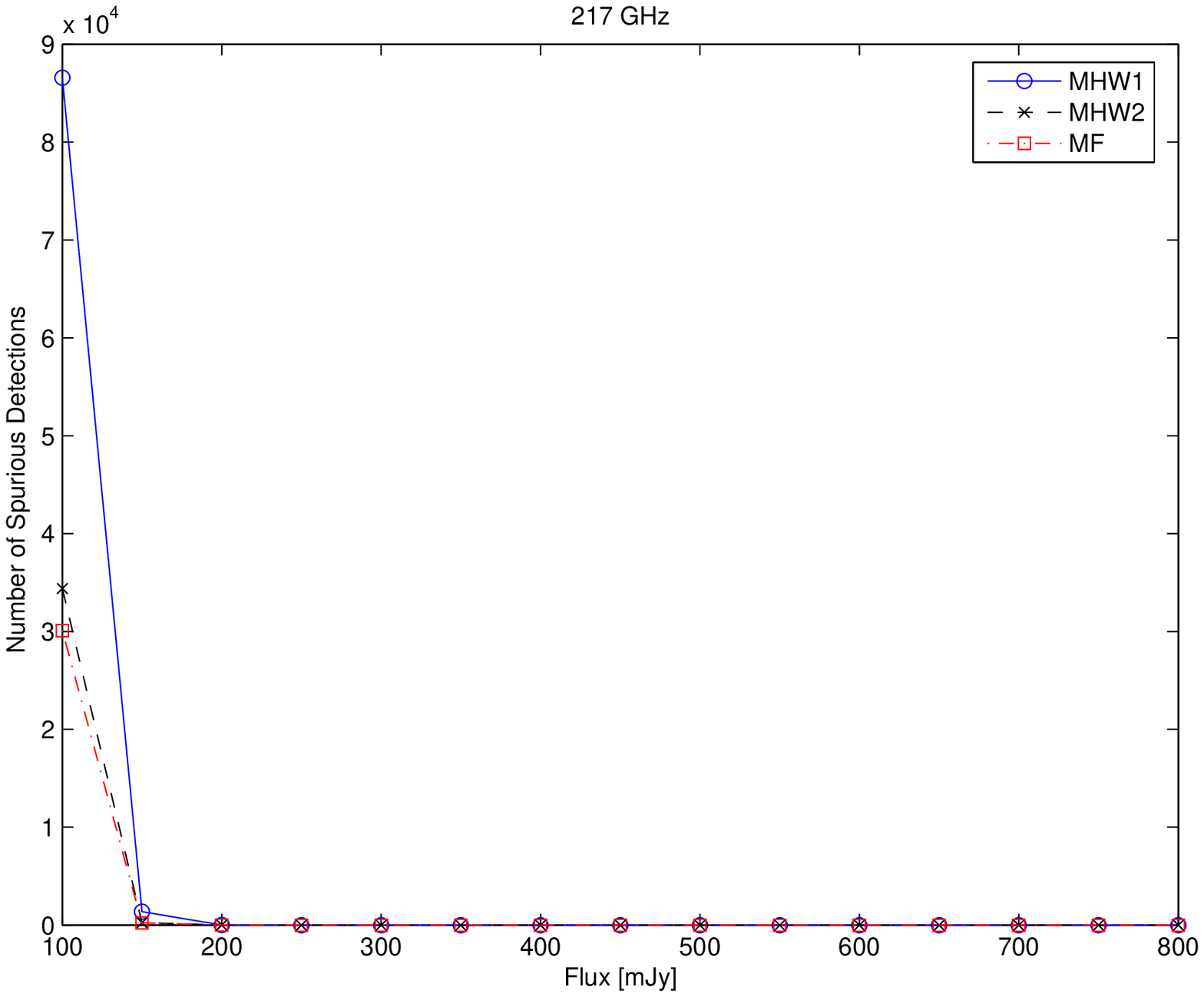} 
~~ 
        \includegraphics[width=6.2cm]{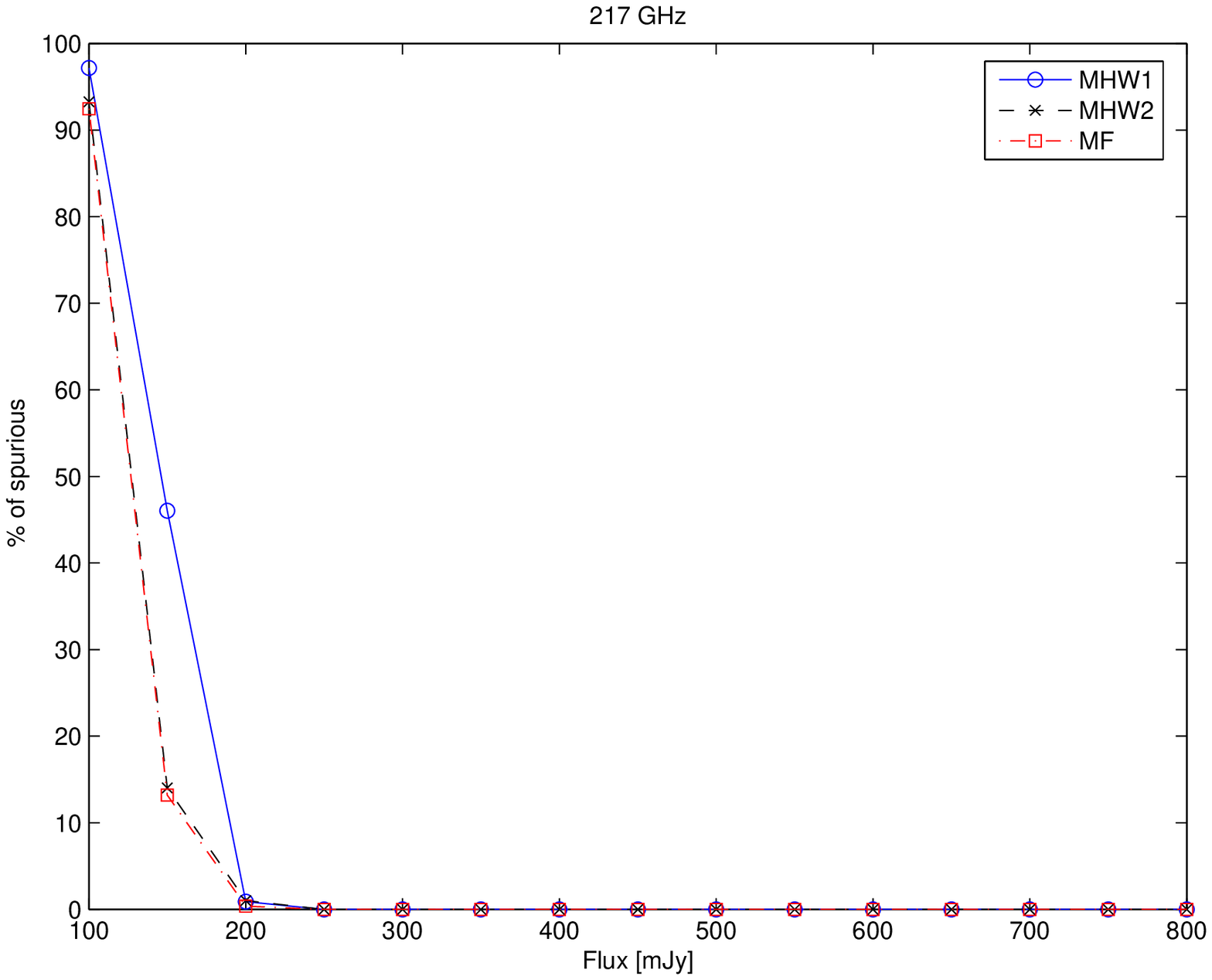} 
        \includegraphics[width=6.2cm]{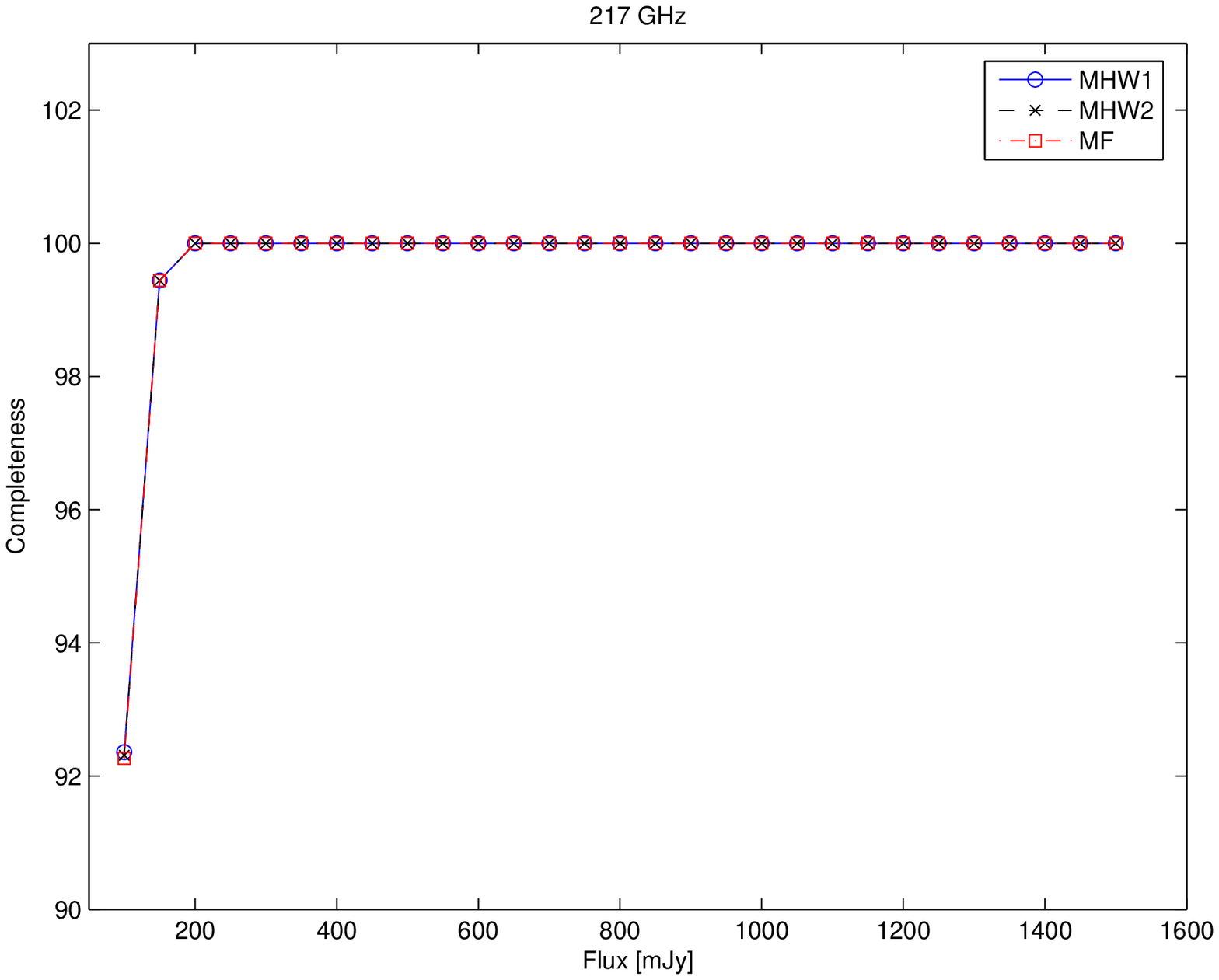} 
\caption{The same as in Figure \ref{fig:fig_ps_30}  but at 217 GHz.\label{fig:fig_ps_217} }
\end{center} 
\end{figure*}

\begin{figure*} 
\begin{center} 
        \includegraphics[width=6.2cm]{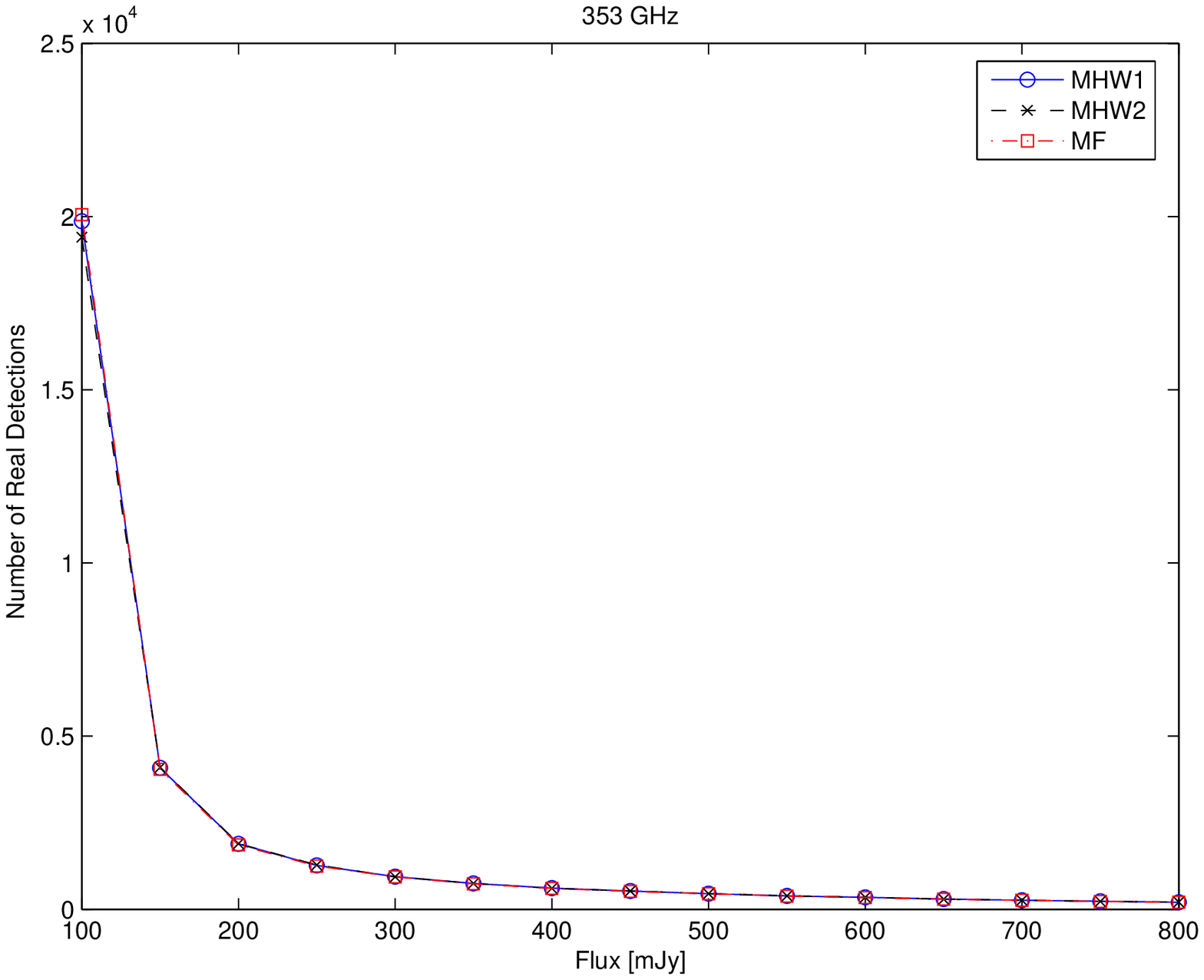} 
        \includegraphics[width=6.2cm]{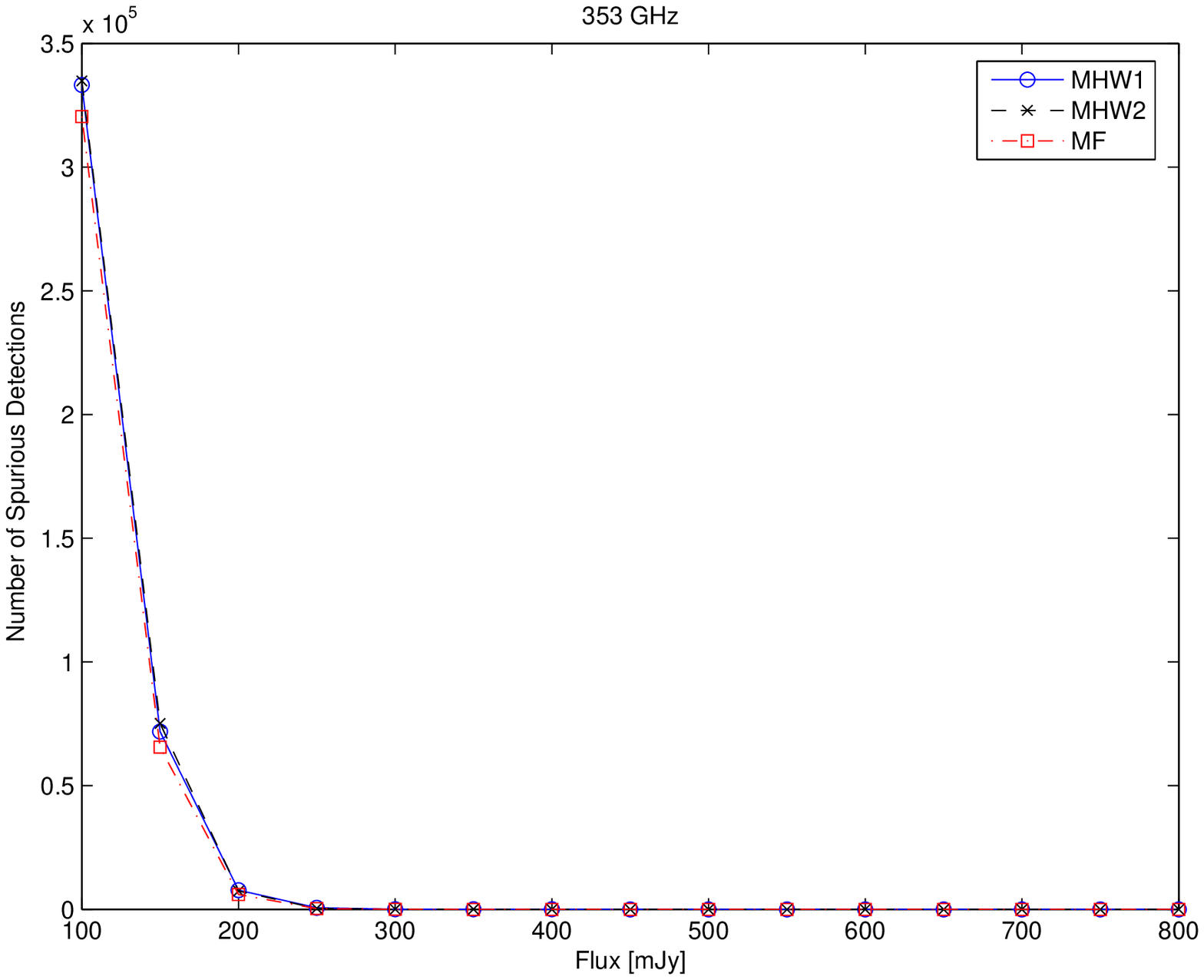} 
~~ 
        \includegraphics[width=6.2cm]{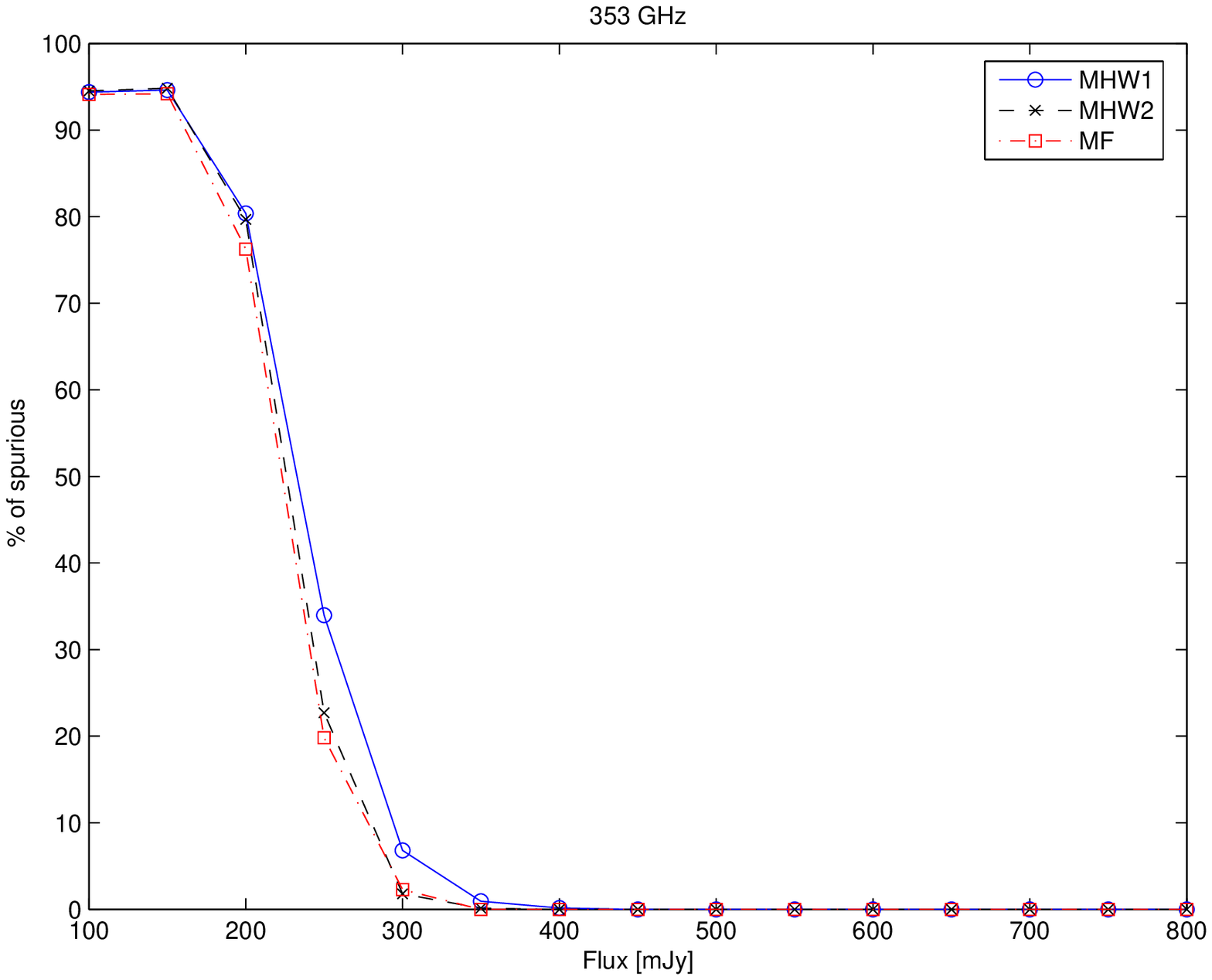} 
        \includegraphics[width=6.2cm]{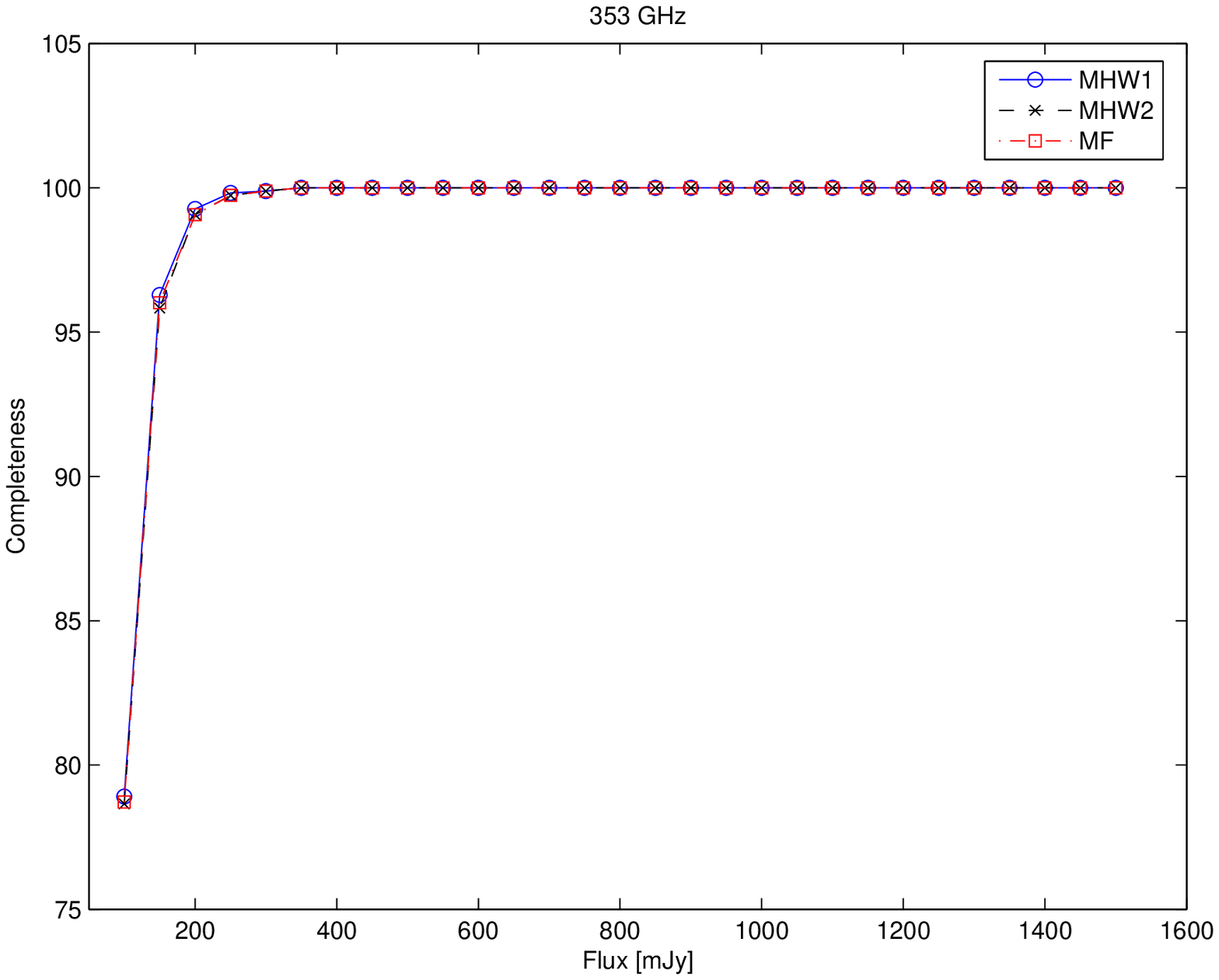} 
\caption{The same as in Figure \ref{fig:fig_ps_30} but at 353 GHz.\label{fig:fig_ps_353} }
\end{center} 
\end{figure*}

\begin{figure*} 
\begin{center} 
        \includegraphics[width=6.2cm]{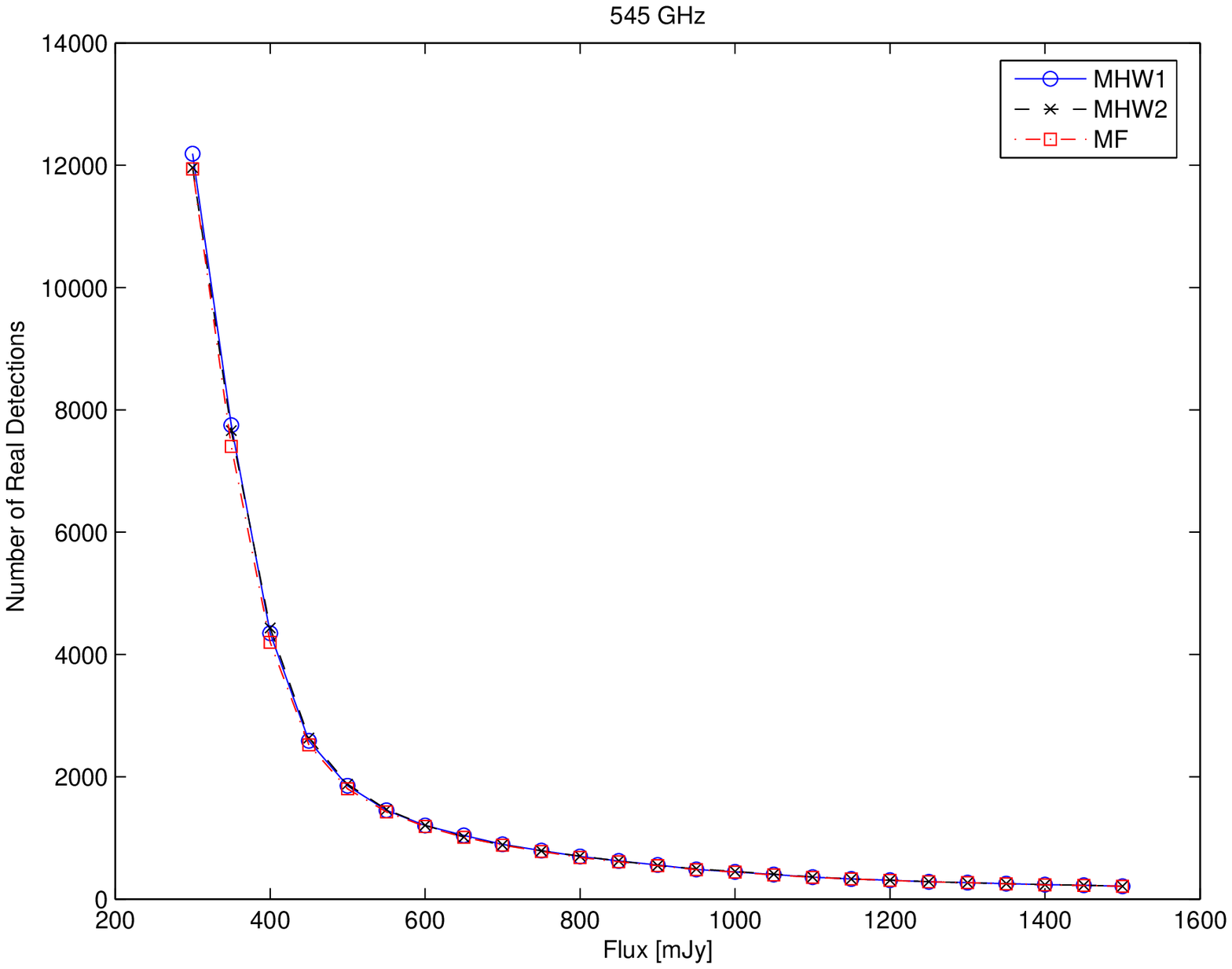} 
        \includegraphics[width=6.2cm]{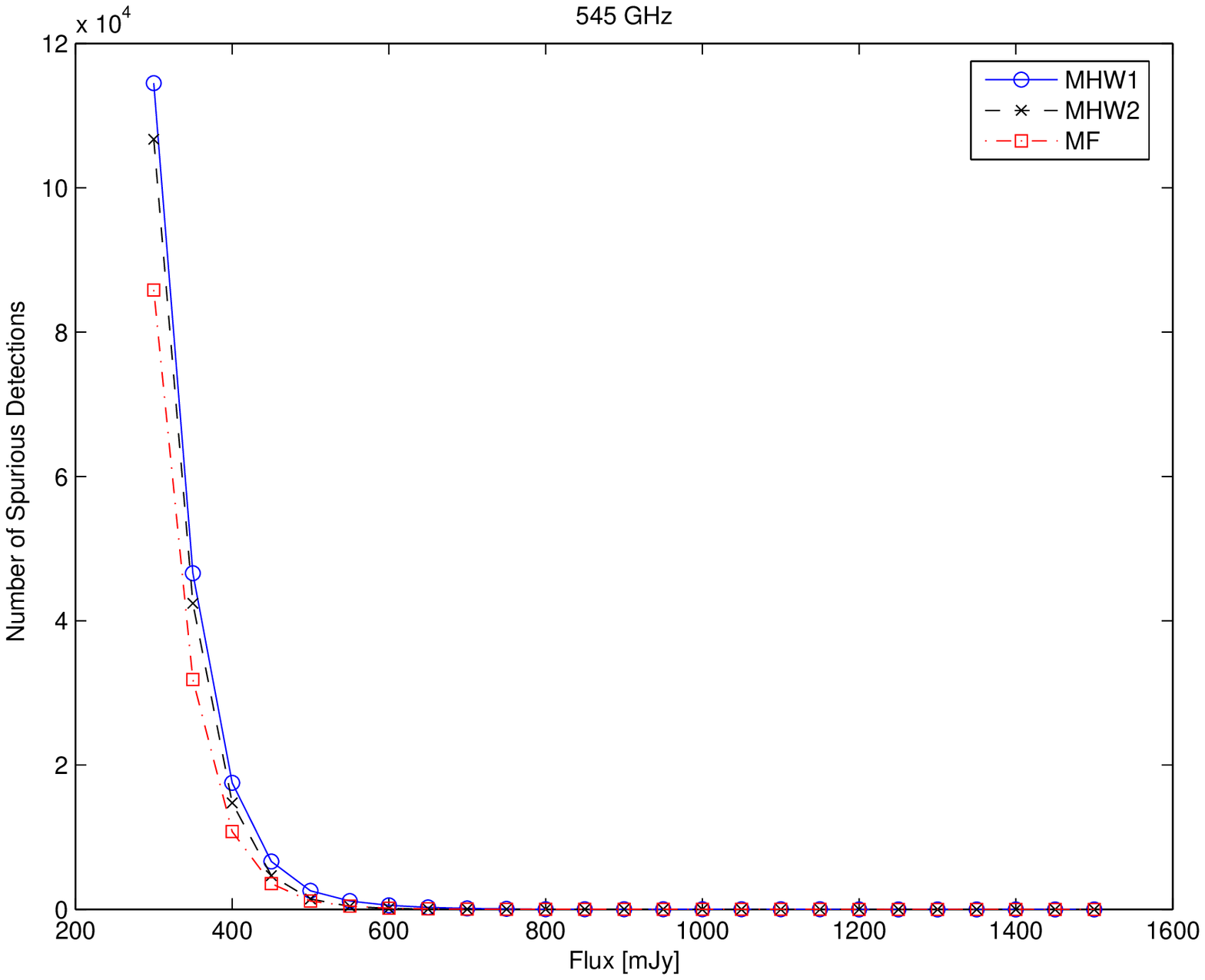} 
~~ 
        \includegraphics[width=6.2cm]{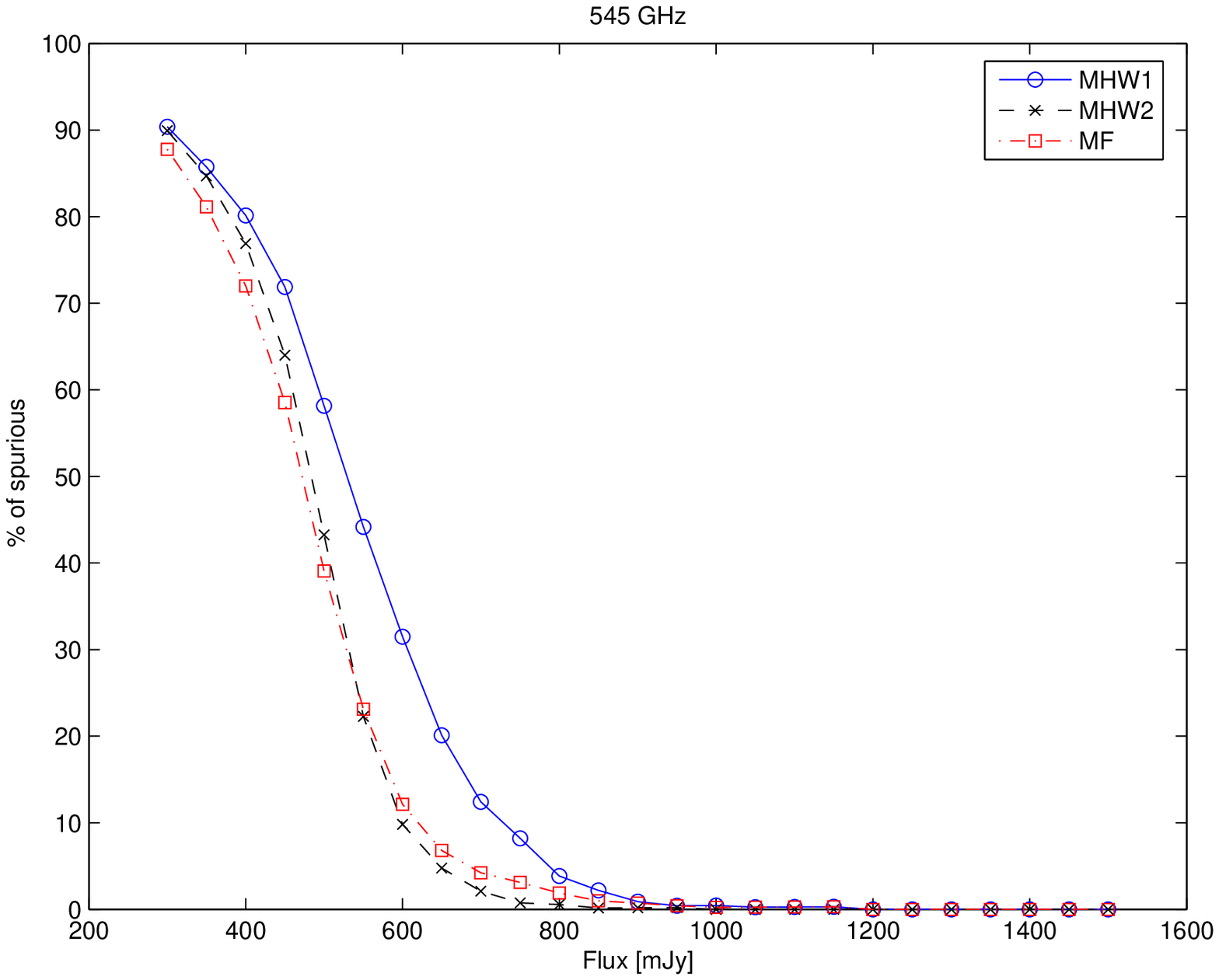} 
        \includegraphics[width=6.2cm]{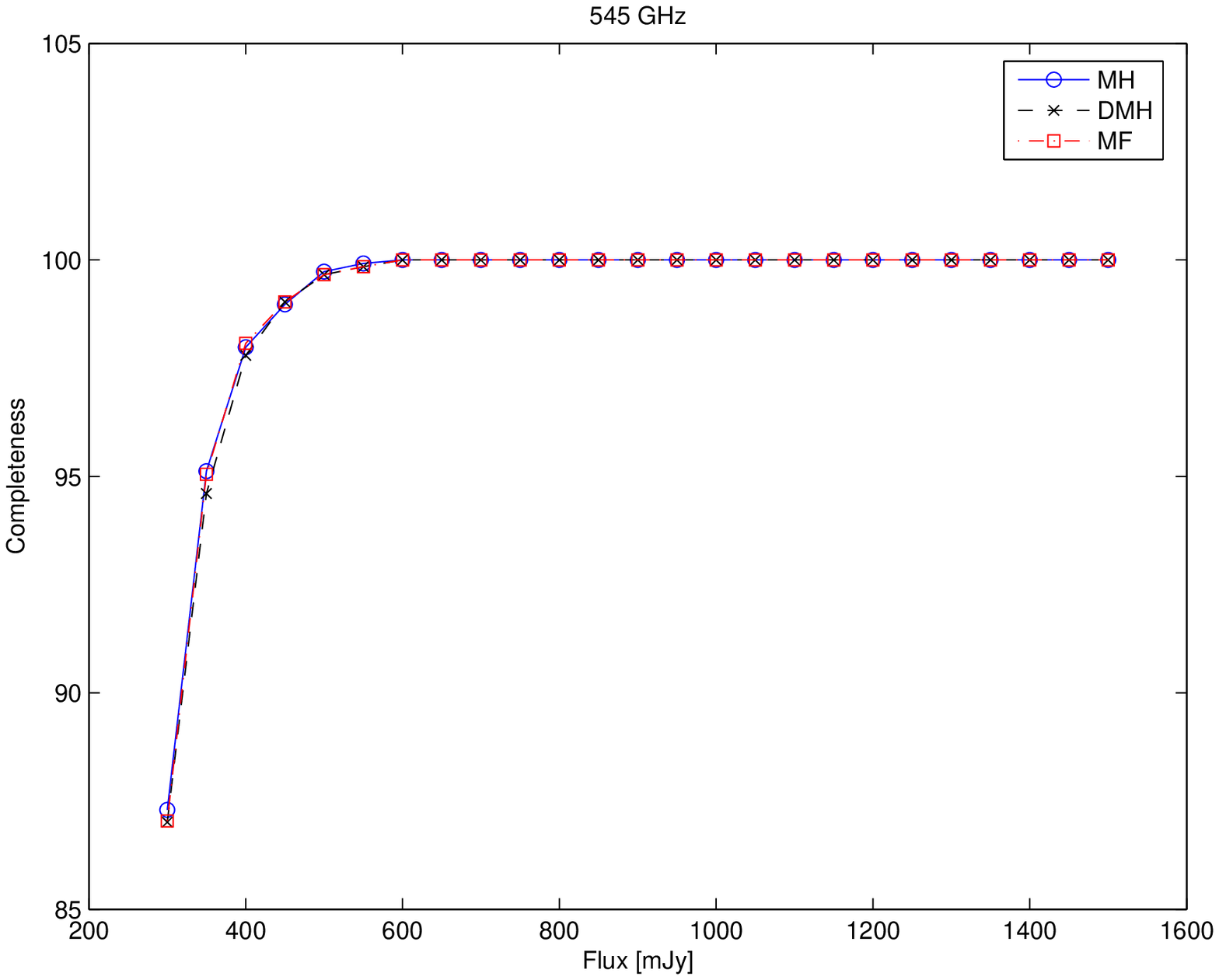} 
\caption{The same as in Figure \ref{fig:fig_ps_30} but at 545 GHz.\label{fig:fig_ps_545} }
\end{center} 
\end{figure*}

\begin{figure*} 
\begin{center} 
        \includegraphics[width=6.2cm]{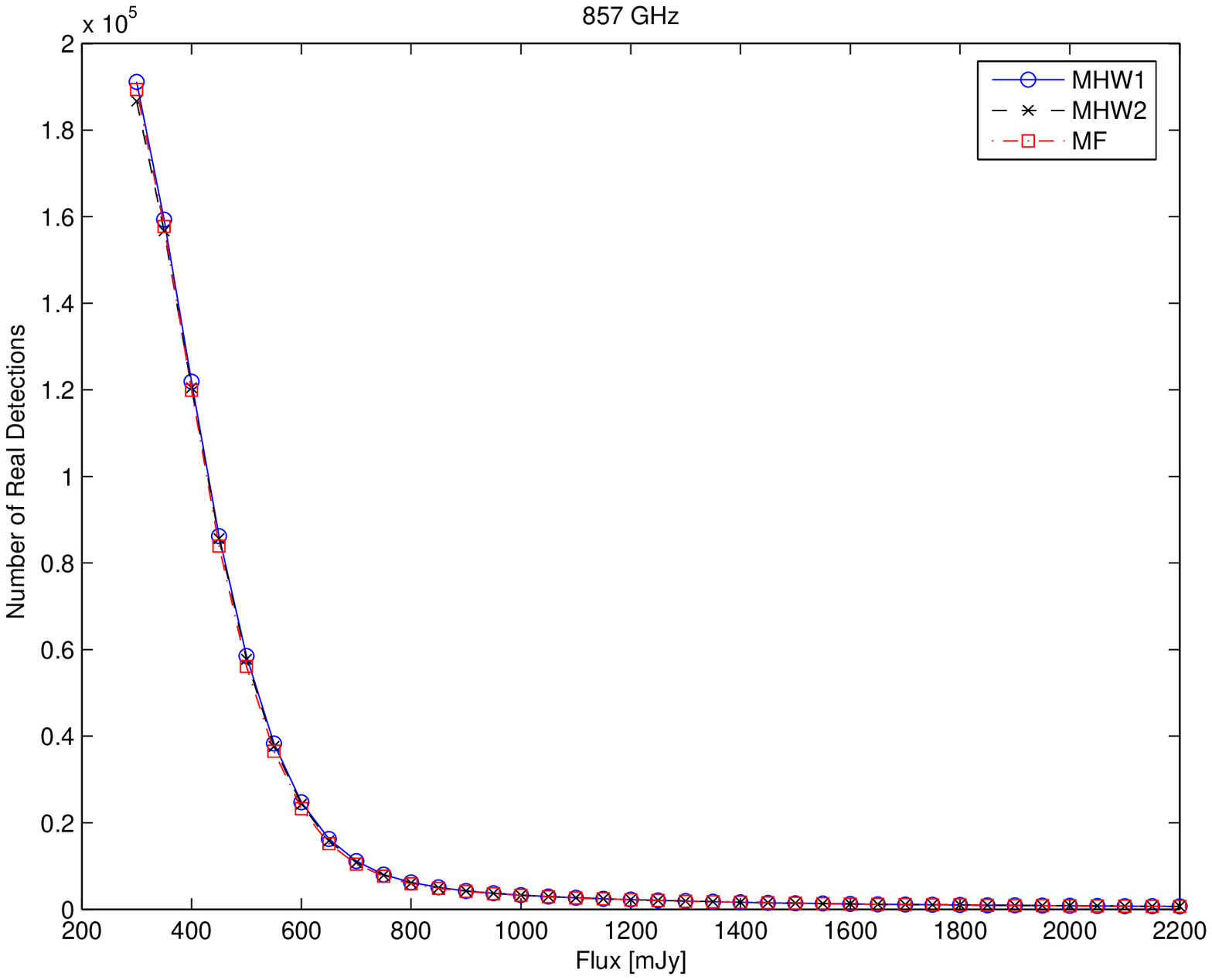} 
        \includegraphics[width=6.2cm]{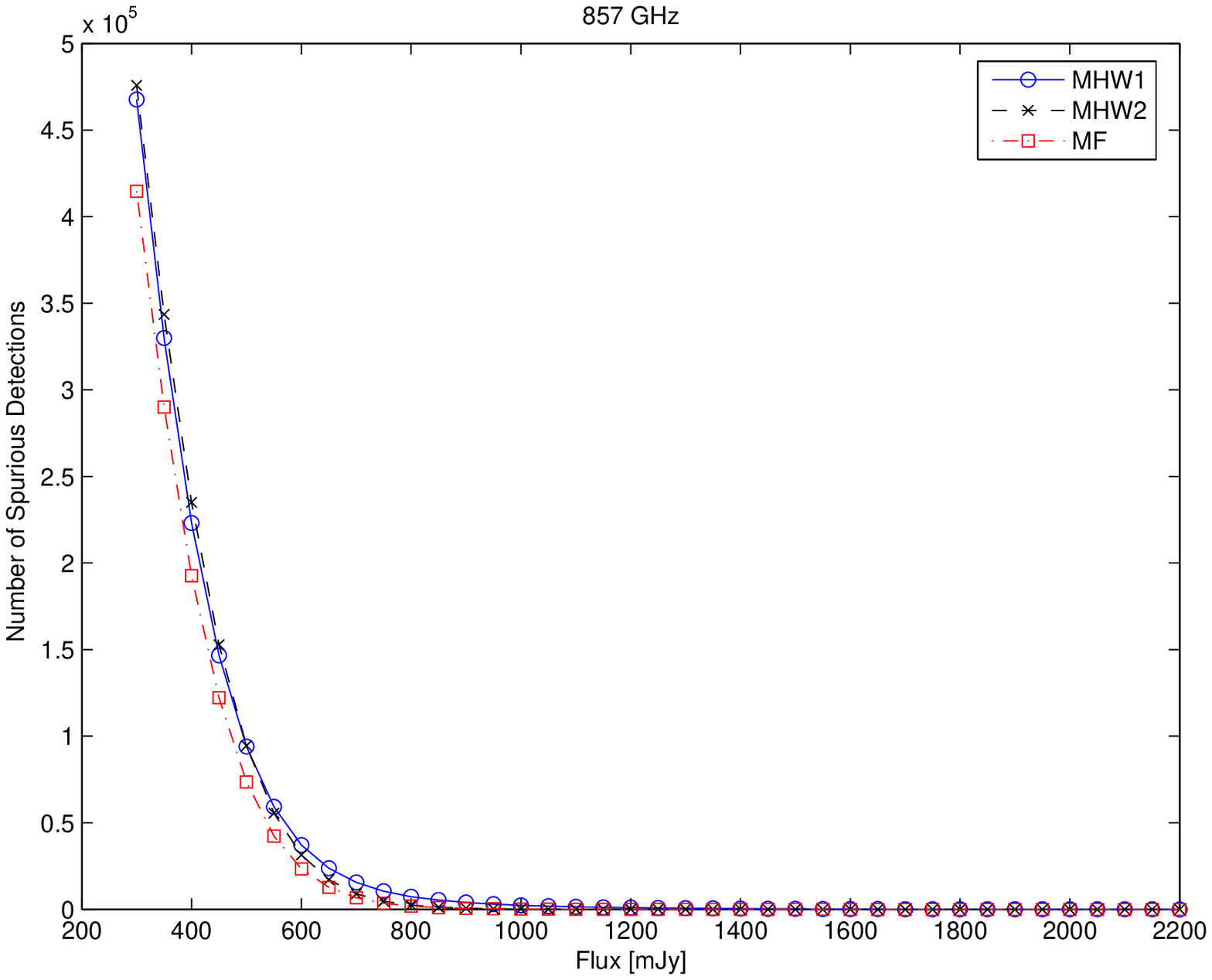} 
~~ 
        \includegraphics[width=6.2cm]{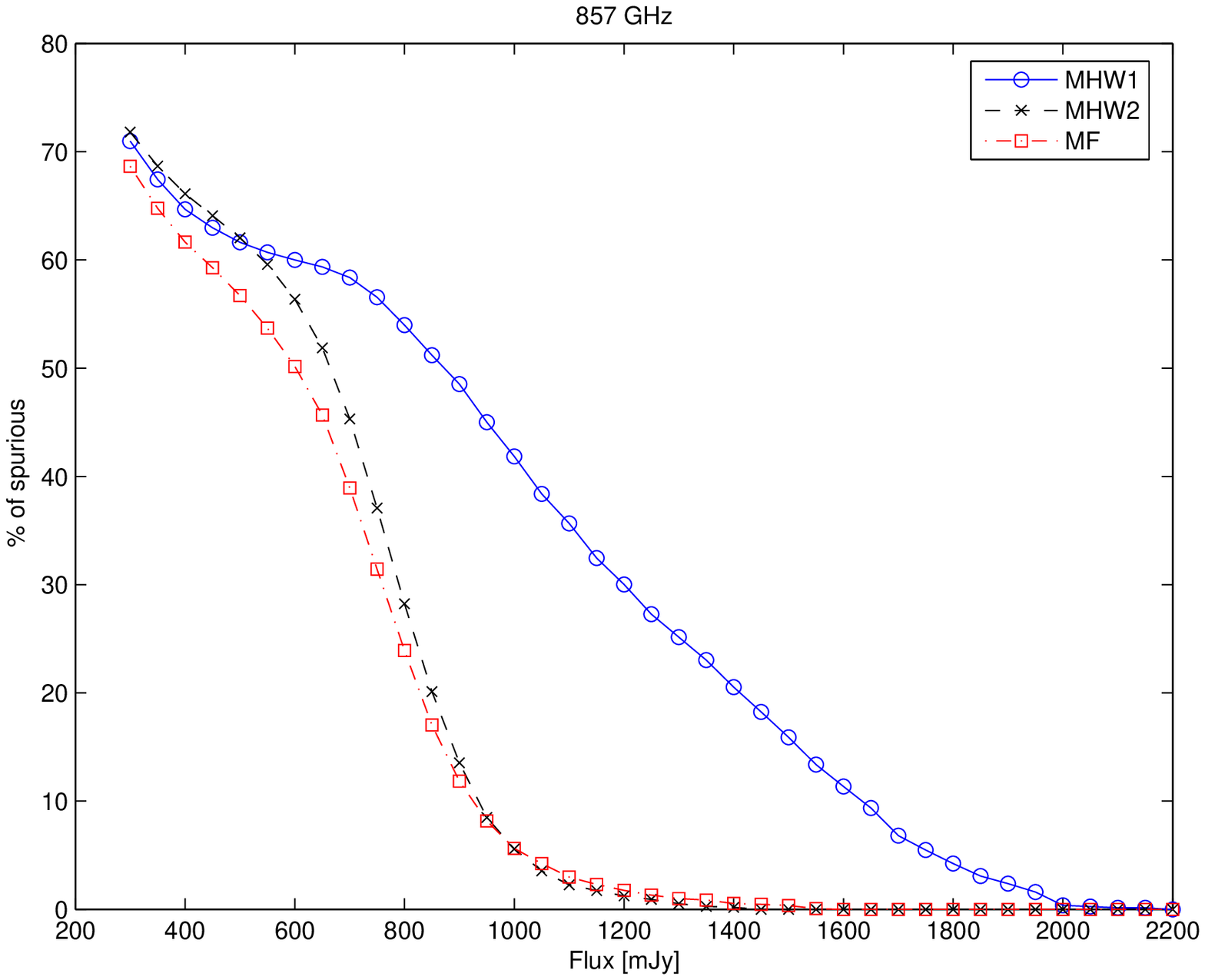} 
        \includegraphics[width=6.2cm]{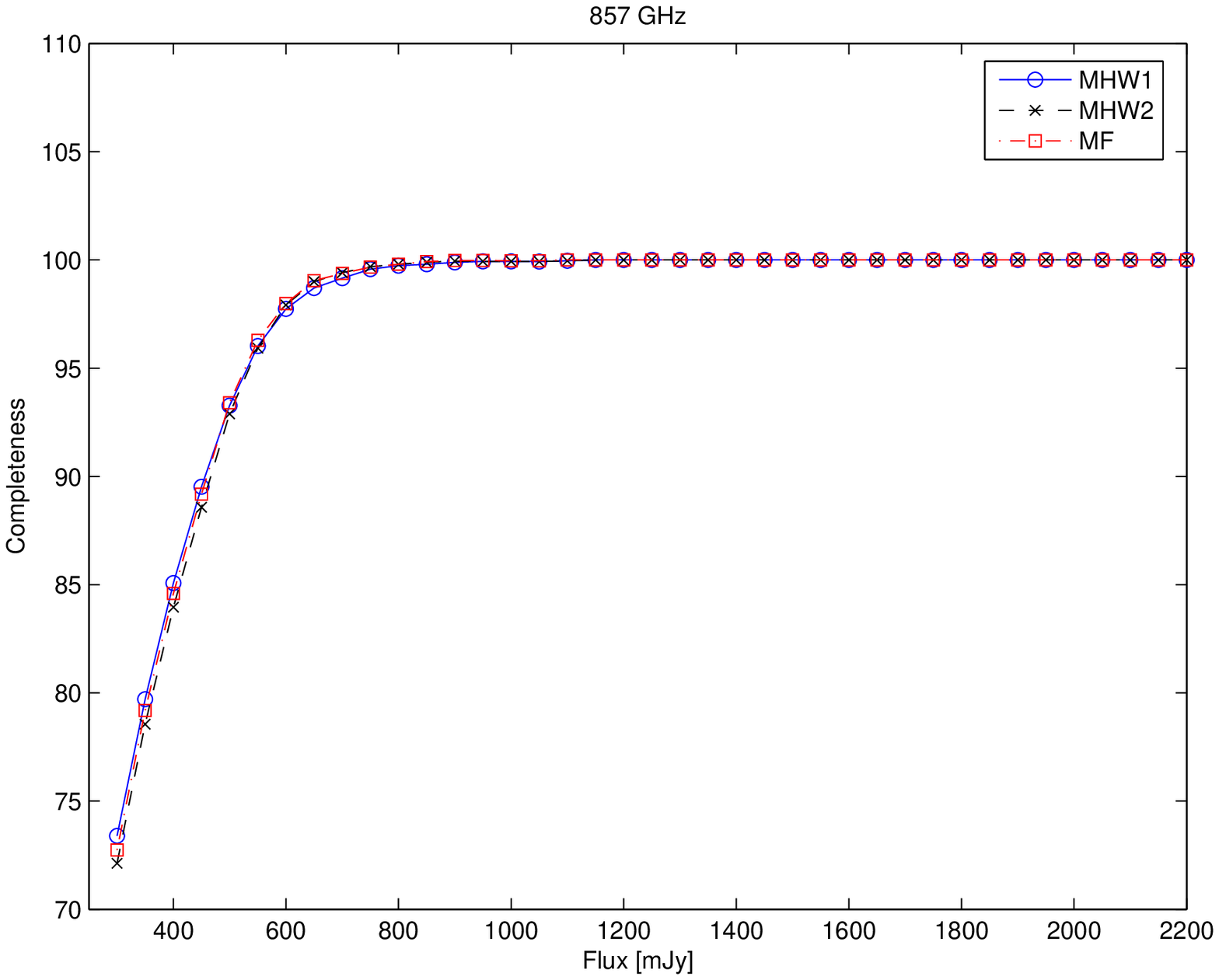} 
\caption{The same as in Figure \ref{fig:fig_ps_30} but at 857 GHz.\label{fig:fig_ps_857} }
\end{center} 
\end{figure*}

In Figures \ref{fig:fig_ps_30} to \ref{fig:fig_ps_857} we present a
number of relevant plots for all the Planck frequency channels. In the
upper panels of each Figure we show the integral (or total) number of
real and of spurious detections (left and right panels, respectively)
above different flux detection limits in half of the sky (see captions
for more details). It is easy to appreciate that the three filters
detect approximately the same number of objects in all the Planck
channels. As regards the spurious, or false, detections, the MF and
the MHW2 yield very similar results whereas the MHW1 detects a
significantly higher number of spurious sources in comparison with the
other two filters, in particular at frequencies $\leq 100$ GHz and for
faint flux detection limits.
 
In the lower left panels of Figures \ref{fig:fig_ps_30} to
\ref{fig:fig_ps_857} we show the percentage of spurious detections
with respect to the integral number of detections. These plots are
very relevant ones because they compare the relative performance of
the filters when taking into account both the number of real and of
spurious detections. By these plots we can also establish the flux
detection limit above which it is possible to reach a given percentage
of spurious, e.g. $5\%$. It can also be noted that the behavior of the
MF and the MHW2 is always comparable, whereas the MHW1 give worse
results than these two filters at higher frequencies. At the lowest
frequencies of the Planck experiment the MHW1 detects a relative
higher number of spurious sources, but only when going down to faint
flux detection thresholds.
 
In the lower right panels we also plot the completeness level of the 
catalogue above different flux detection thresholds. Each flux 
threshold represents the flux at which we have detected a given 
percentage (e.g., 95\%, 100\%) of all the simulated sources in the 
map. At low and intermediate Planck frequencies we find that our 
catalogues are $95\%$ complete (the 95\% of the simulated sources are 
detected above this flux limit) above $\simeq (300-350)$ mJy whereas 
at the highest Planck frequency, 857 GHz, the $95\%$ completeness 
level is obtained only for fluxes $\geq 550$ mJy. In Table 
\ref{tab:table2} we have summarised these results for each Planck 
frequency channel and we have also included the $100\%$ completeness 
level, the flux limit at which all the simulated point sources are 
detected. 
 
\begin{figure*} 
\begin{center} 
                \includegraphics[width=5.8cm]{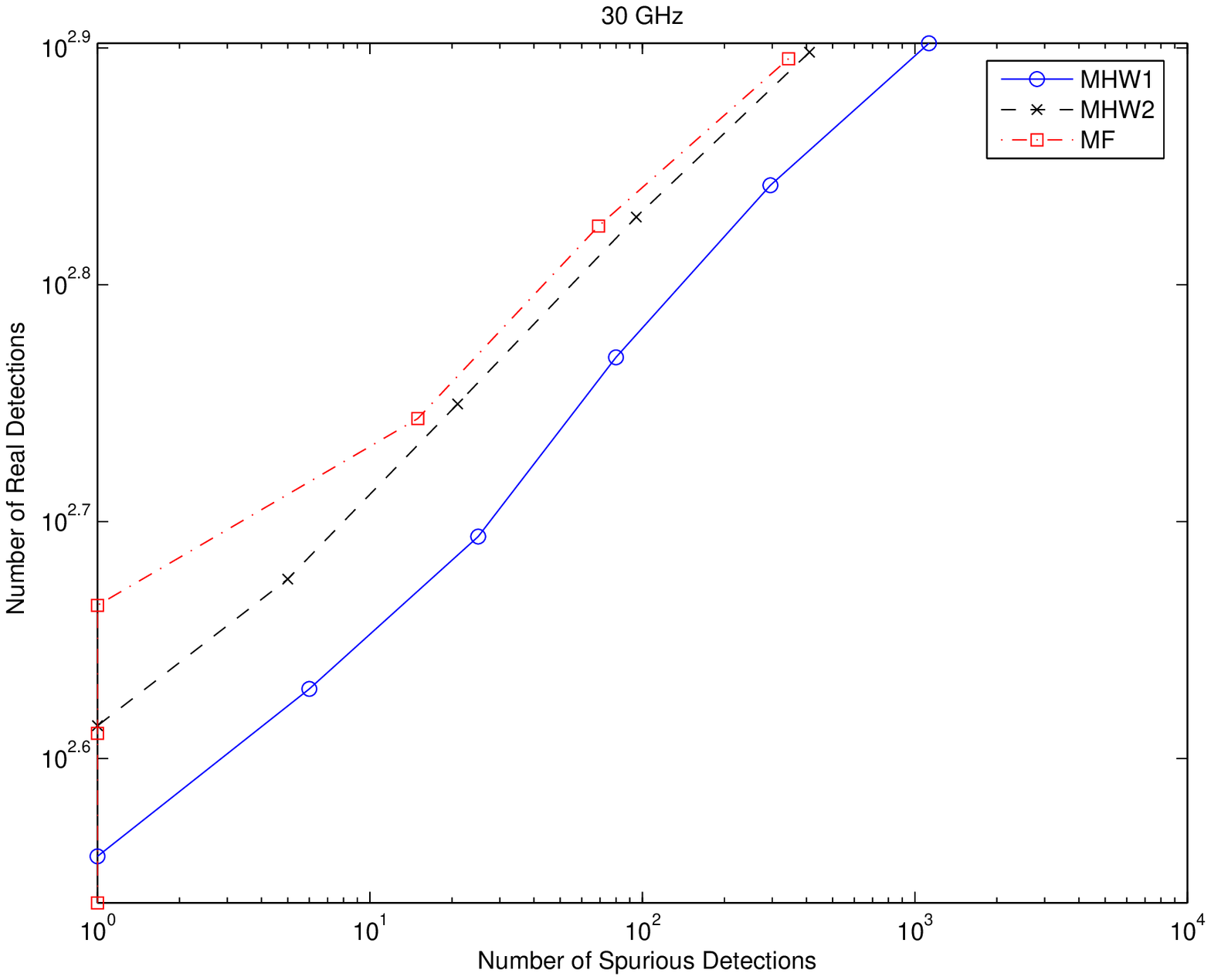} 
                \includegraphics[width=5.8cm]{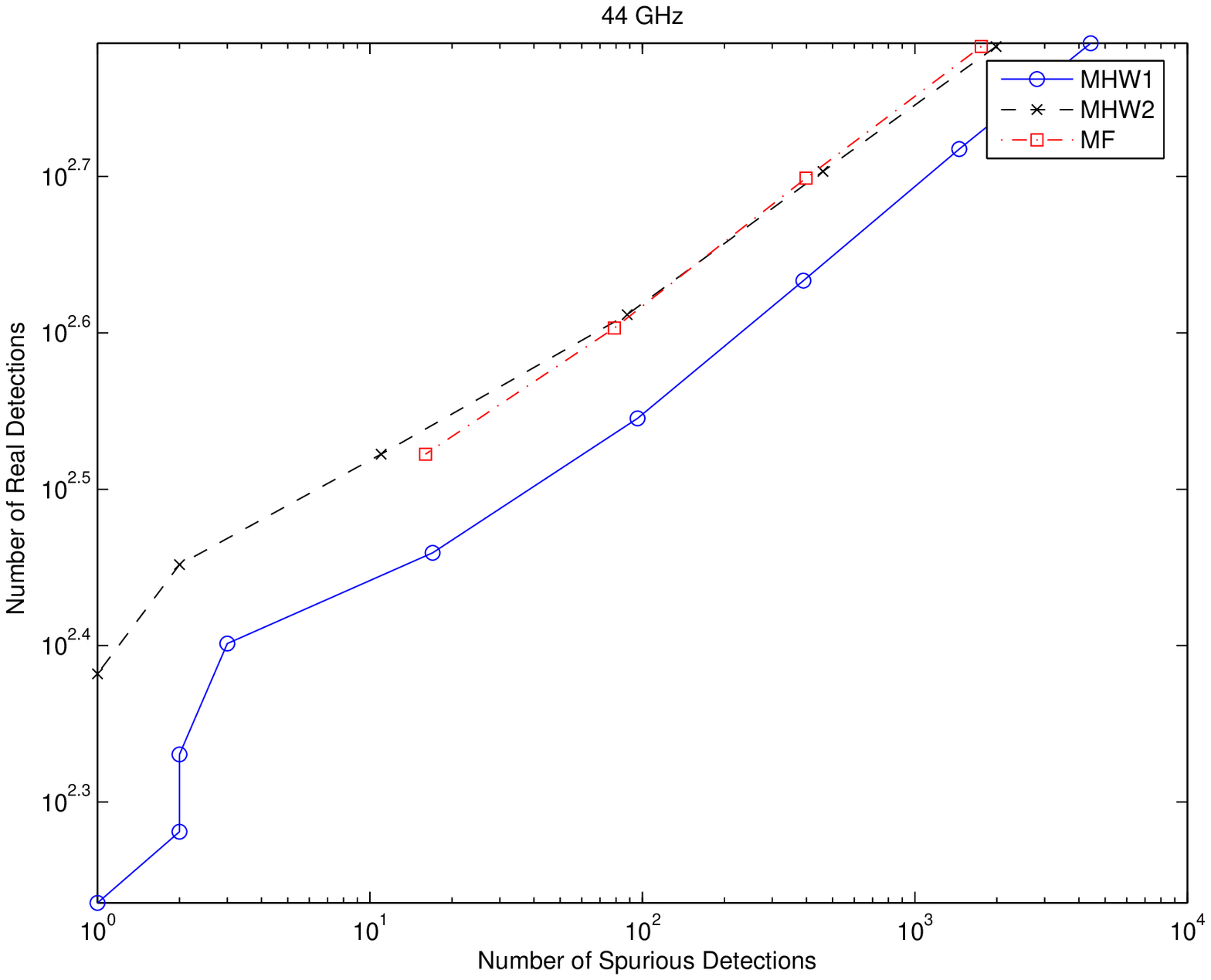} 
                \includegraphics[width=5.8cm]{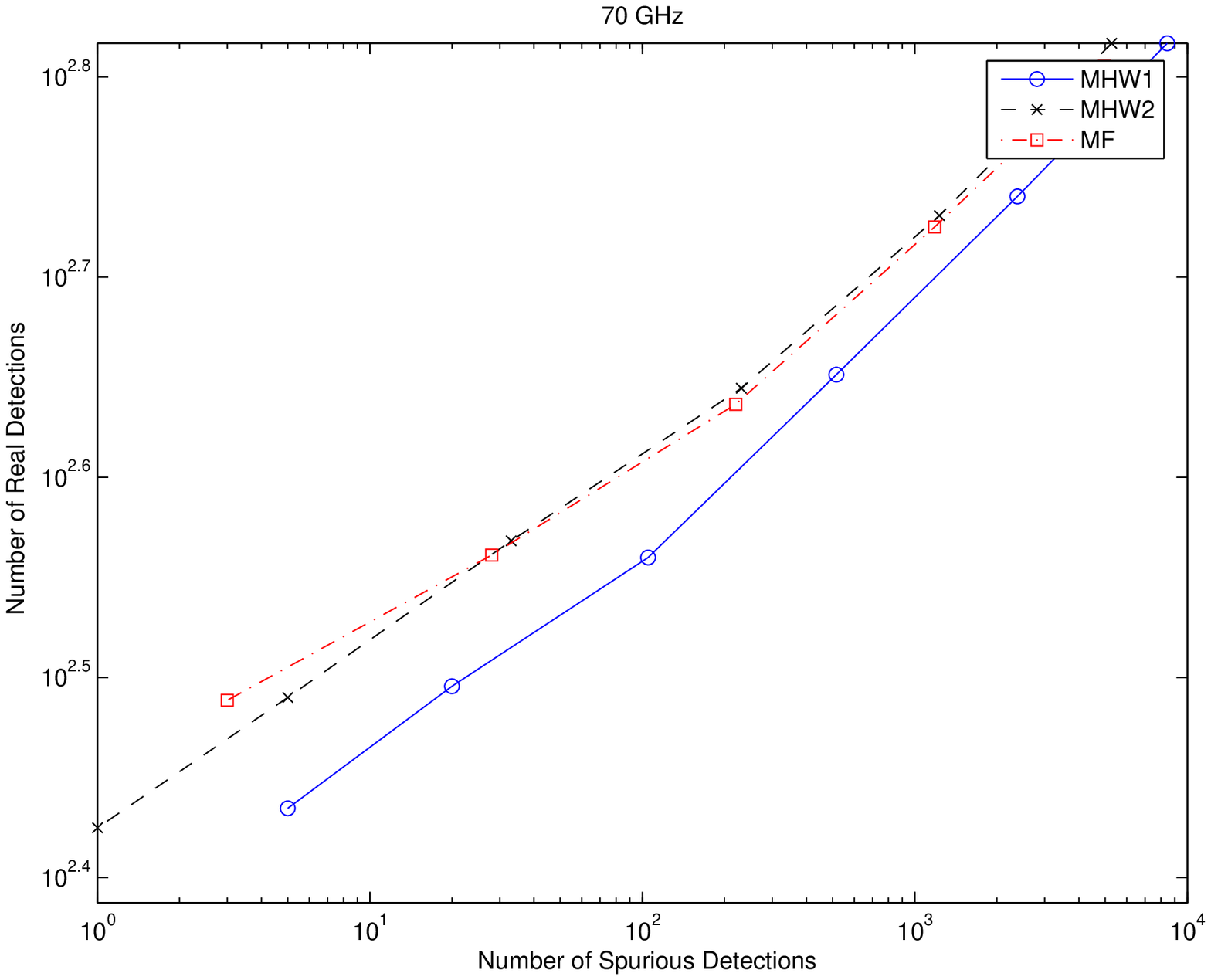} 
~~ 
                \includegraphics[width=5.8cm]{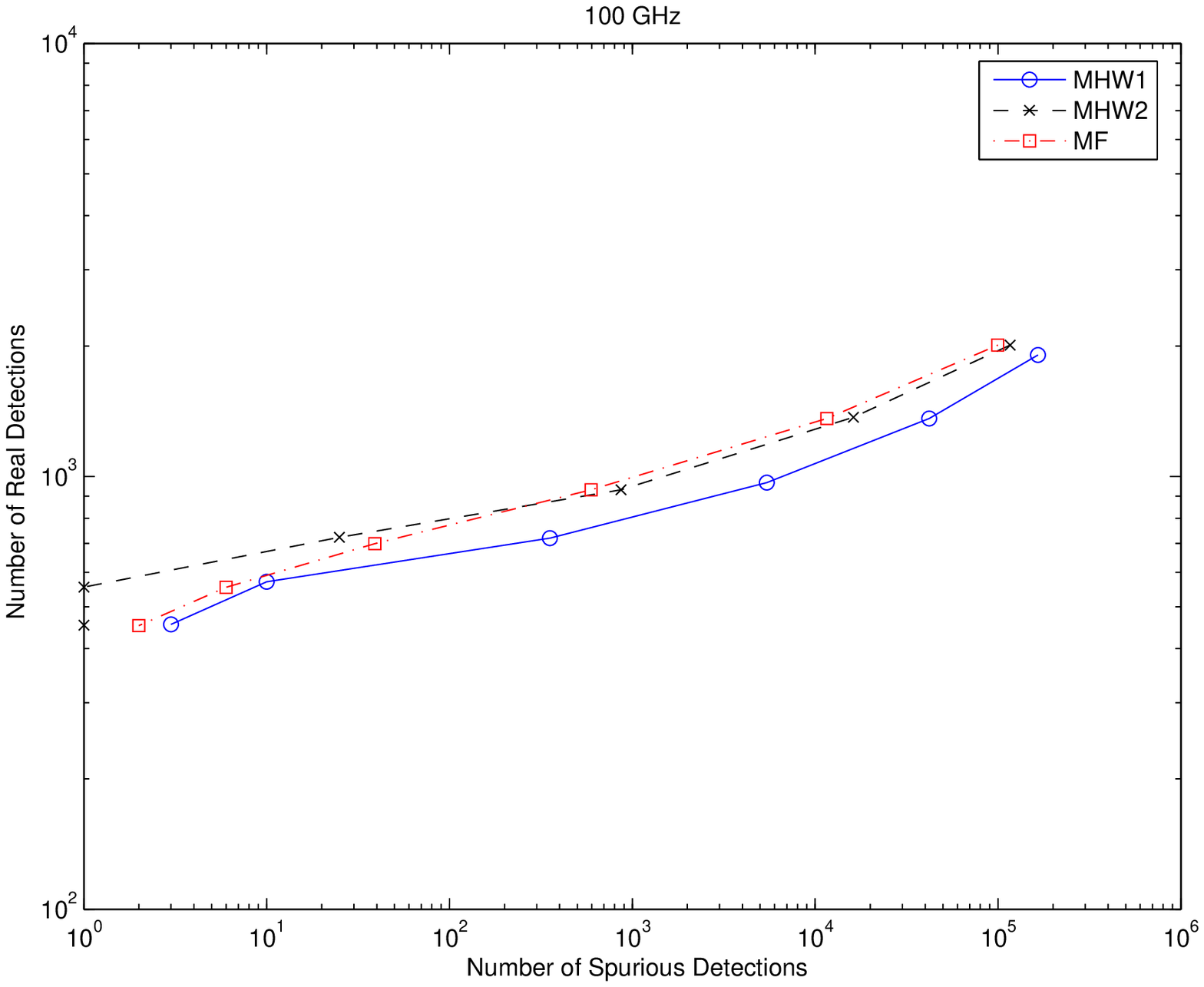} 
                \includegraphics[width=5.8cm]{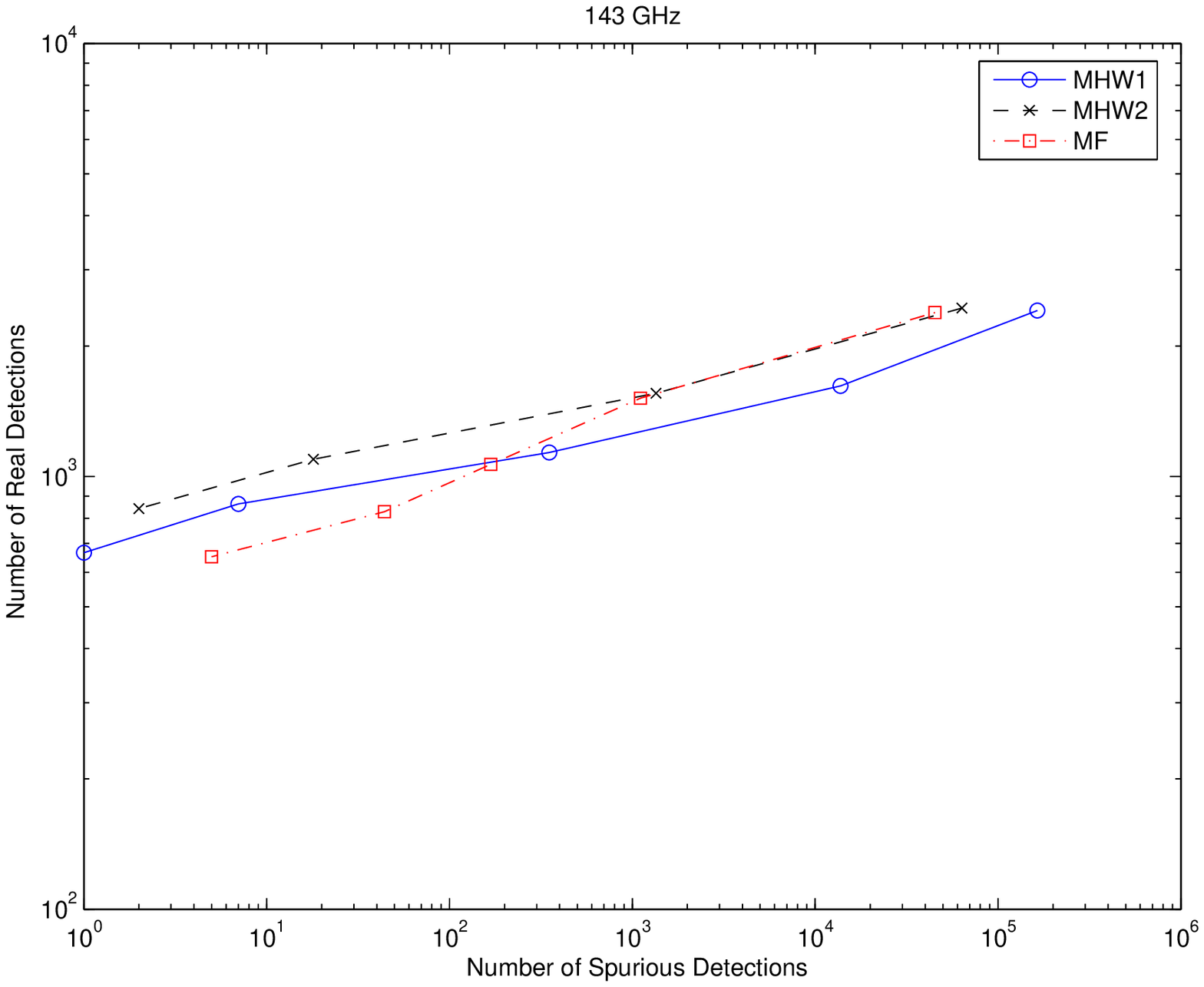} 
                \includegraphics[width=5.8cm]{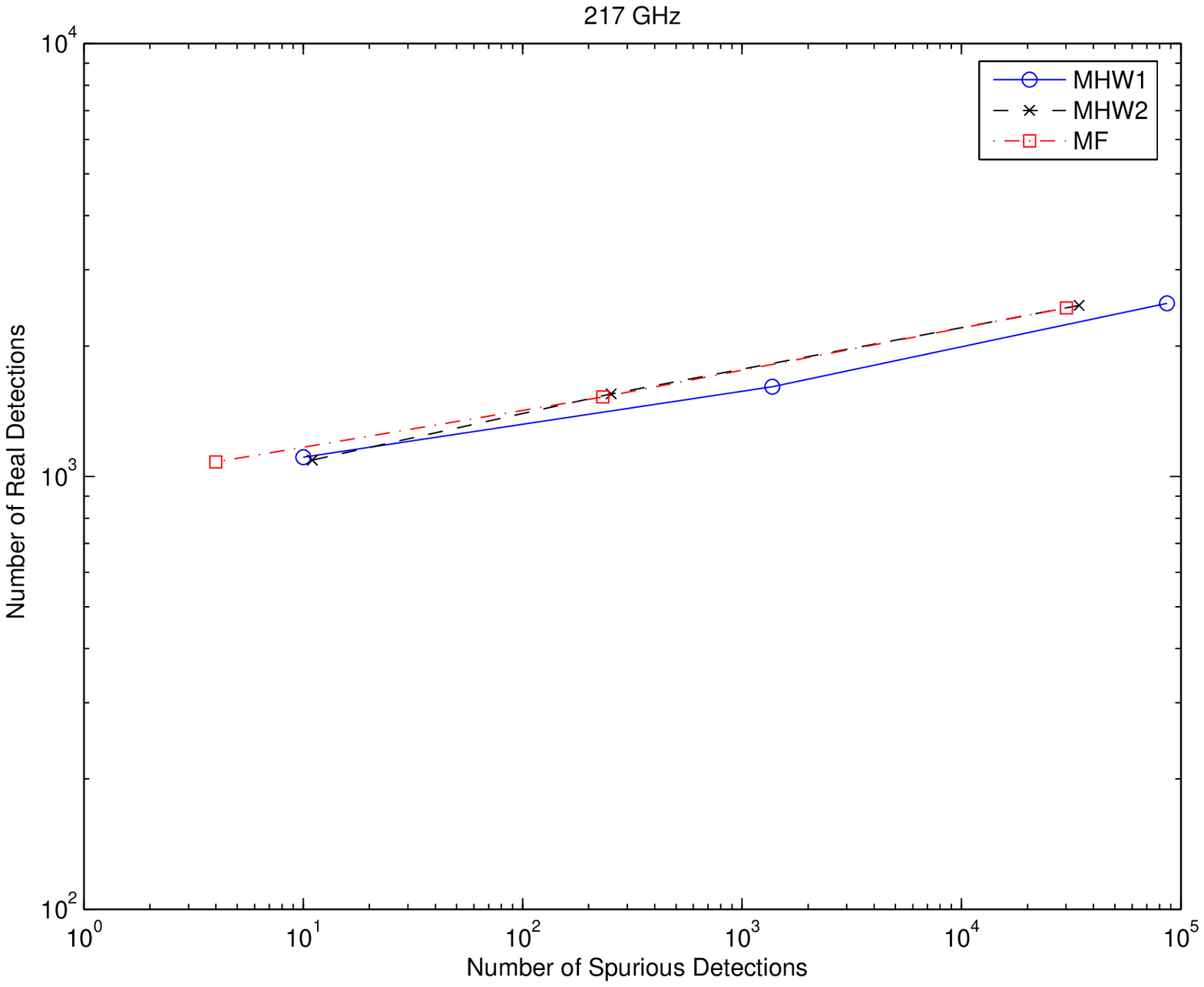} 
~~ 
                \includegraphics[width=5.8cm]{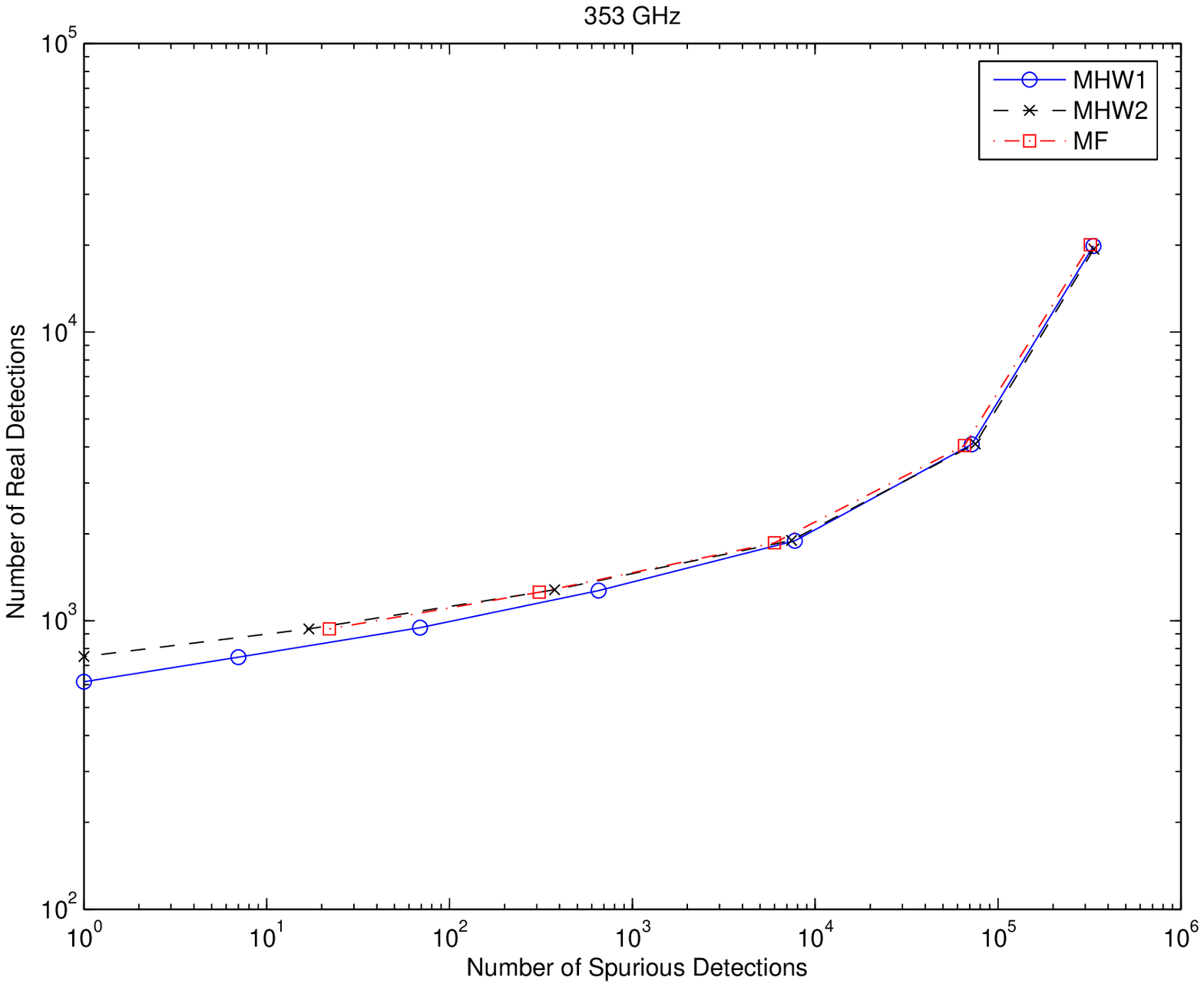} 
                \includegraphics[width=5.8cm]{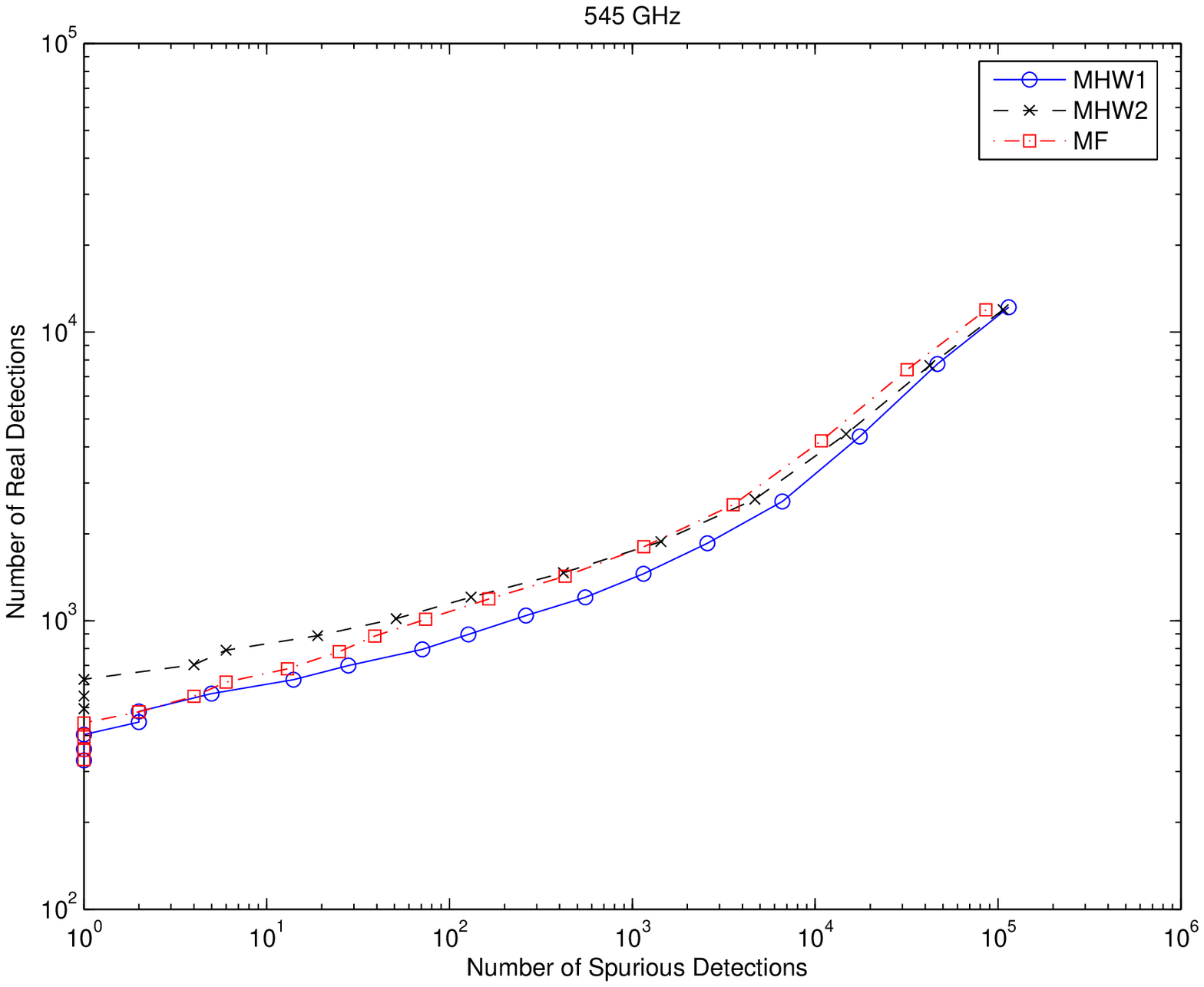} 
                \includegraphics[width=5.8cm]{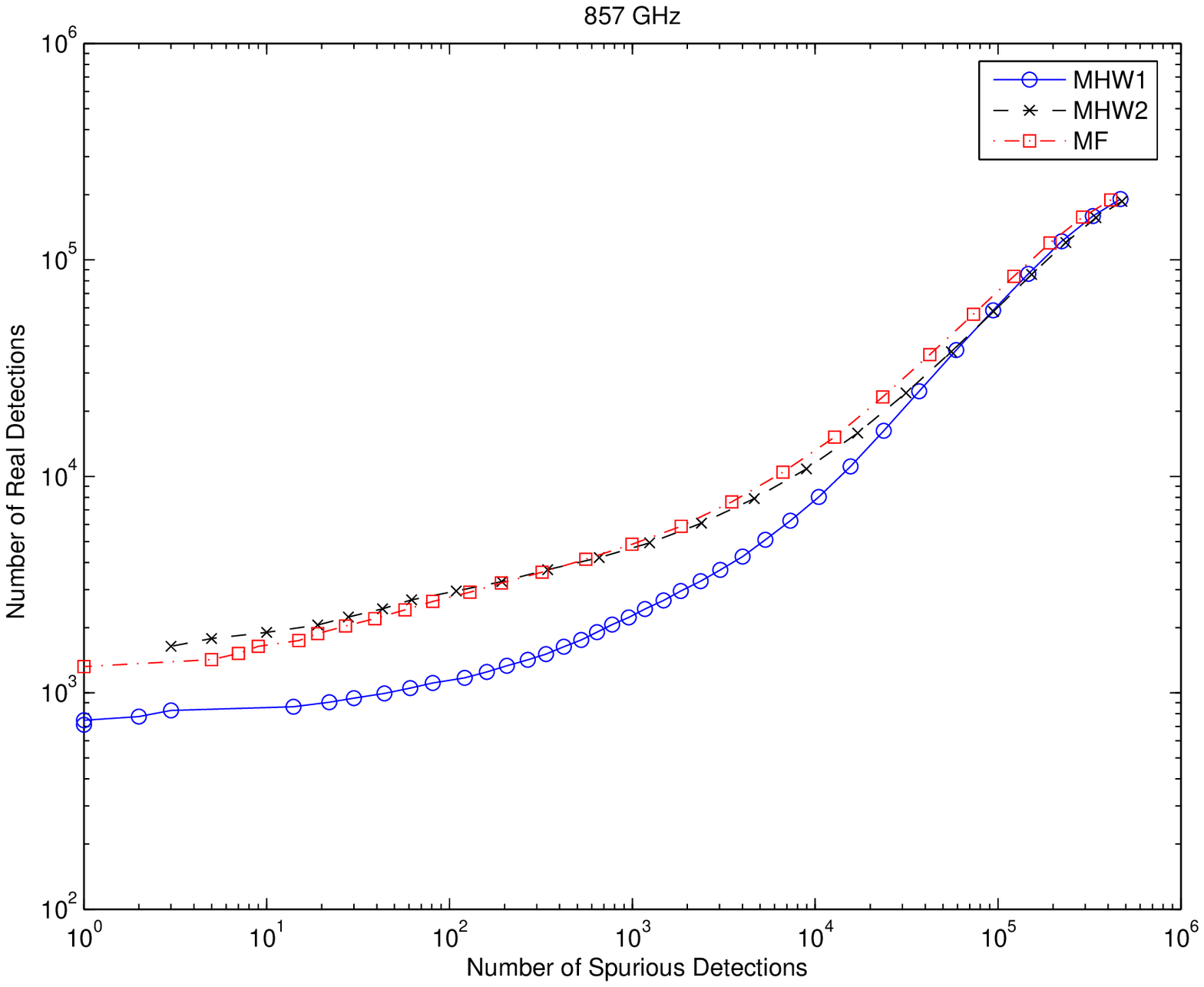} 
\caption{In each panel we plot the number of spurious detections 
vs. the number of real detections for all the Planck frequency 
channels. This plot is very useful for comparing the performance of 
the chosen filters. \label{fig:dets_spur}} 
\end{center} 
\end{figure*}

In Figure \ref{fig:dets_spur} we have plotted, for each Planck
channel, the number of spurious detections against the number of real
detections. This kind of representation, together with the plot of the
percentage of spurious detections (in the lower left panels of Figures
\ref{fig:fig_ps_30} to \ref{fig:fig_ps_857}), is very important
because it shows the global performance of the filters. It is clear
that, for a given number of spurious detections, the filter which is
able to obtain the corresponding highest number of real detected
sources should be considered the best one. In this scheme, a filter
performs better than another if its curve tends to lie above the other
ones. In the upper panels of the Figure (i.e., at low Planck
frequencies) we observe that the MF and the MHW2 behave similarly and
better than MHW1, whose curve is always below the other two. At
intermediate frequencies the three filters here discussed perform
similarly, and at 217 GHz and 353 GHz the MHW1 curve is very close to
those representing the MF and the MHW2. At 857 GHz, the curve
representative of the MHW1 is again below the other two ones, but only
at fluxes above $\simeq 650$ mJy.
 
\begin{figure*} 
\begin{center} 
        \includegraphics[width=5.8cm]{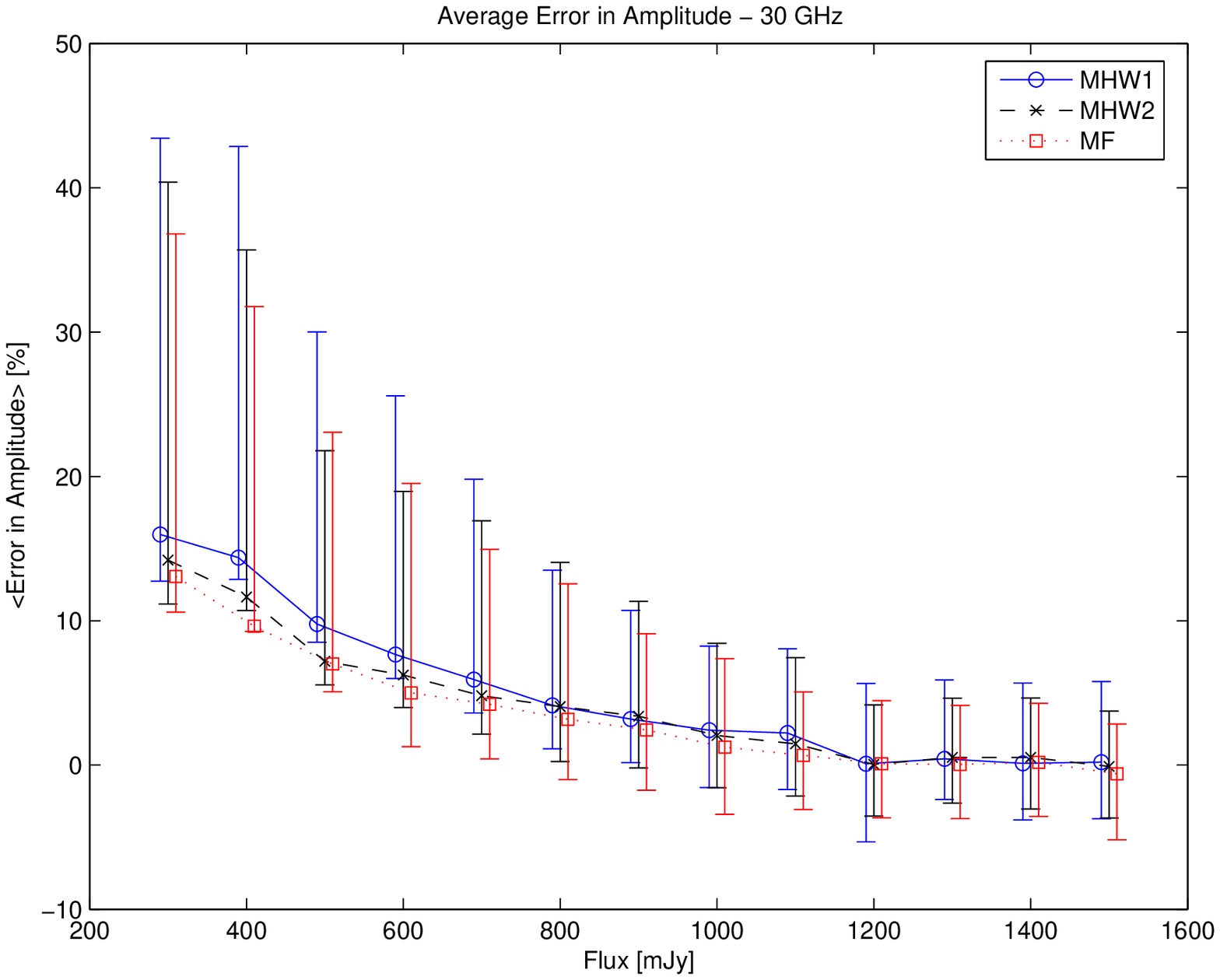} 
        \includegraphics[width=5.8cm]{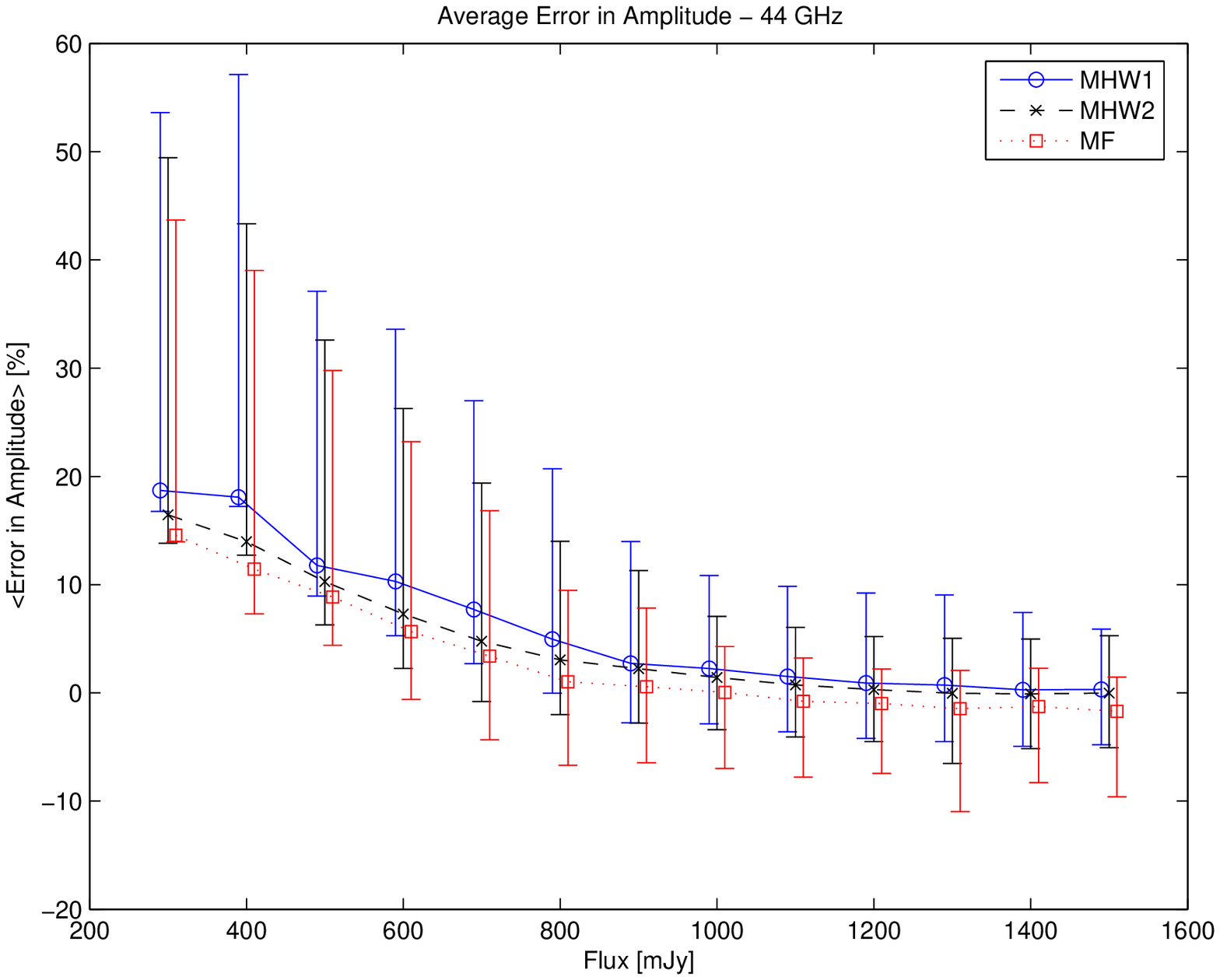} 
        \includegraphics[width=5.8cm]{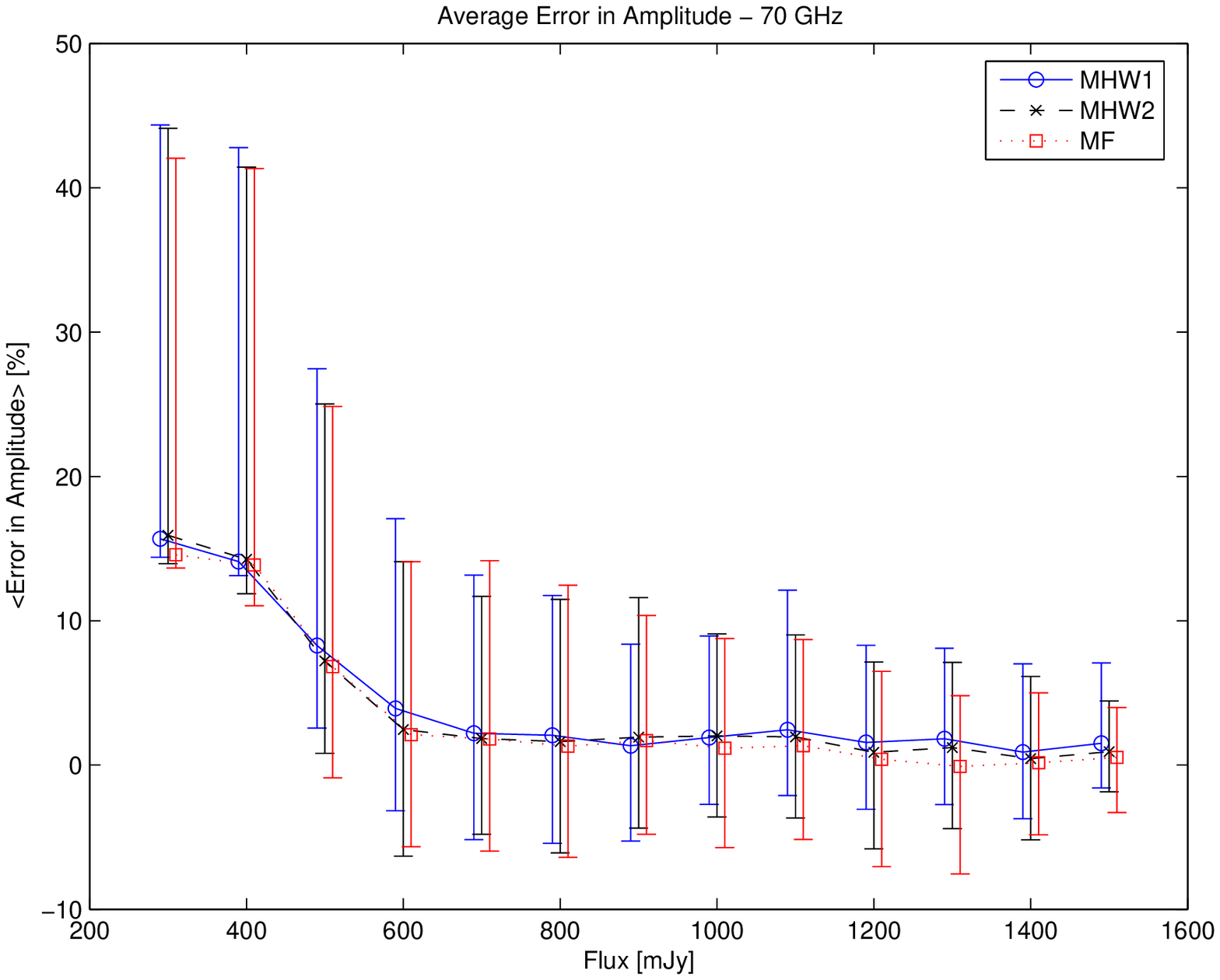} 
~~ 
        \includegraphics[width=5.8cm]{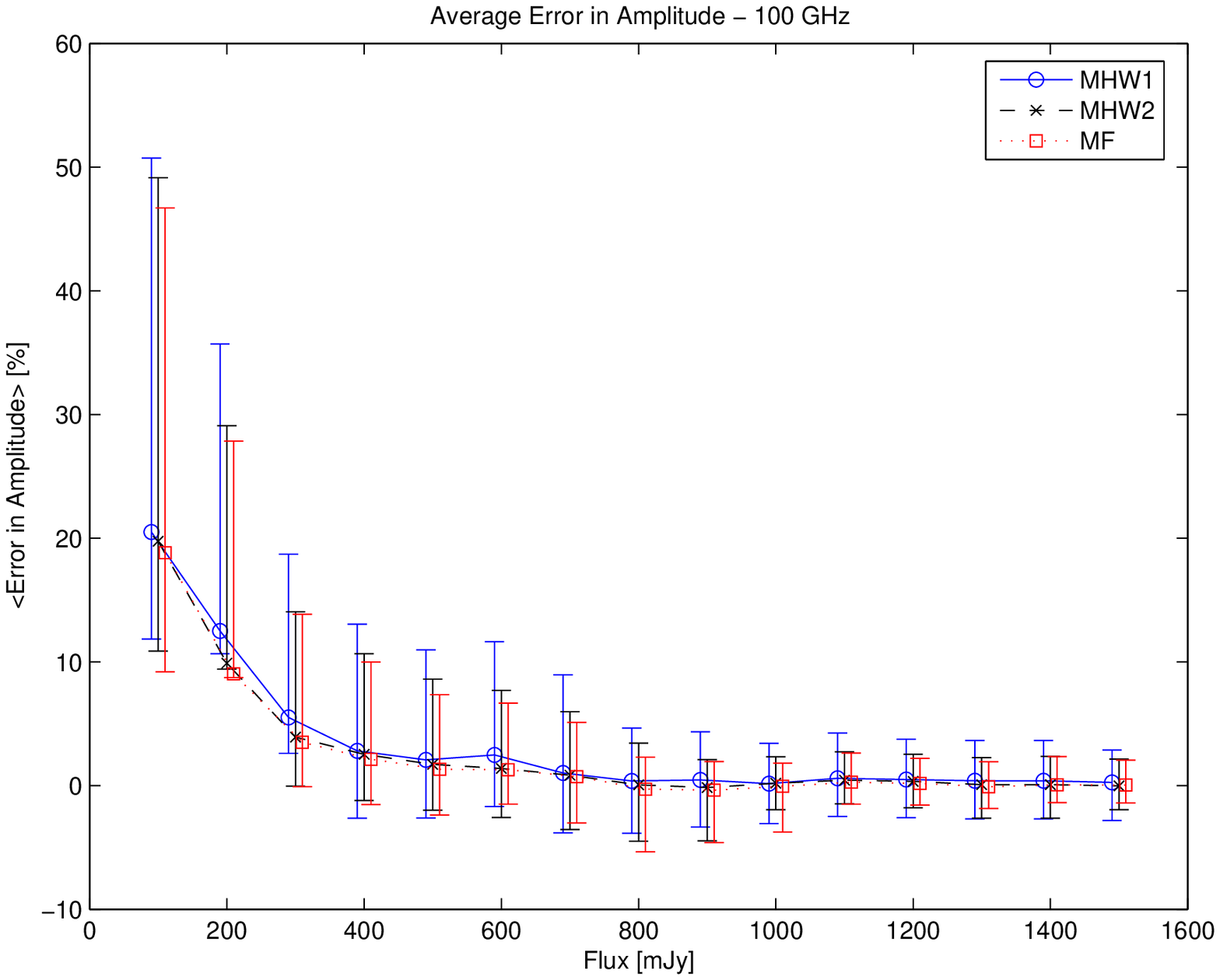} 
        \includegraphics[width=5.8cm]{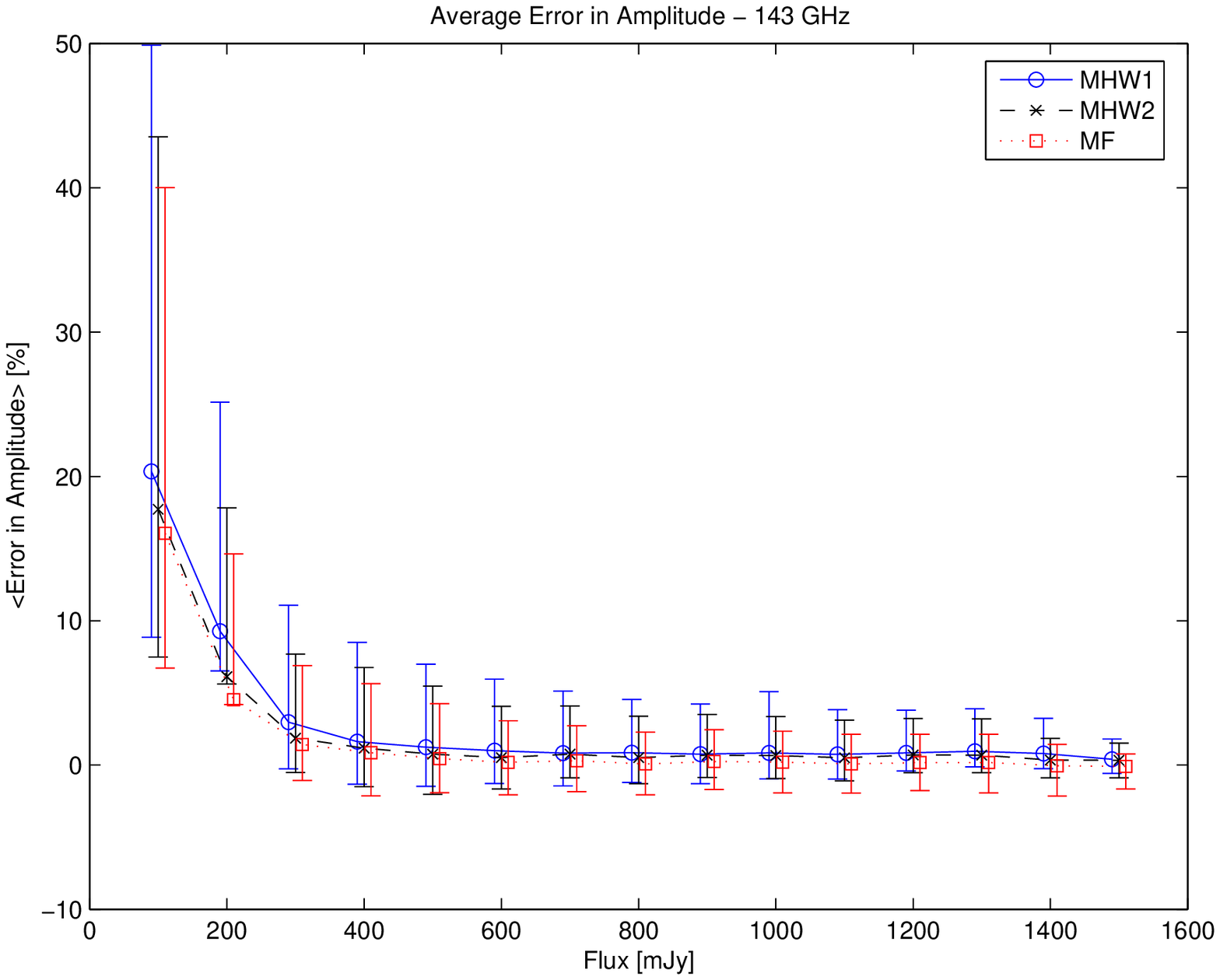} 
        \includegraphics[width=5.8cm]{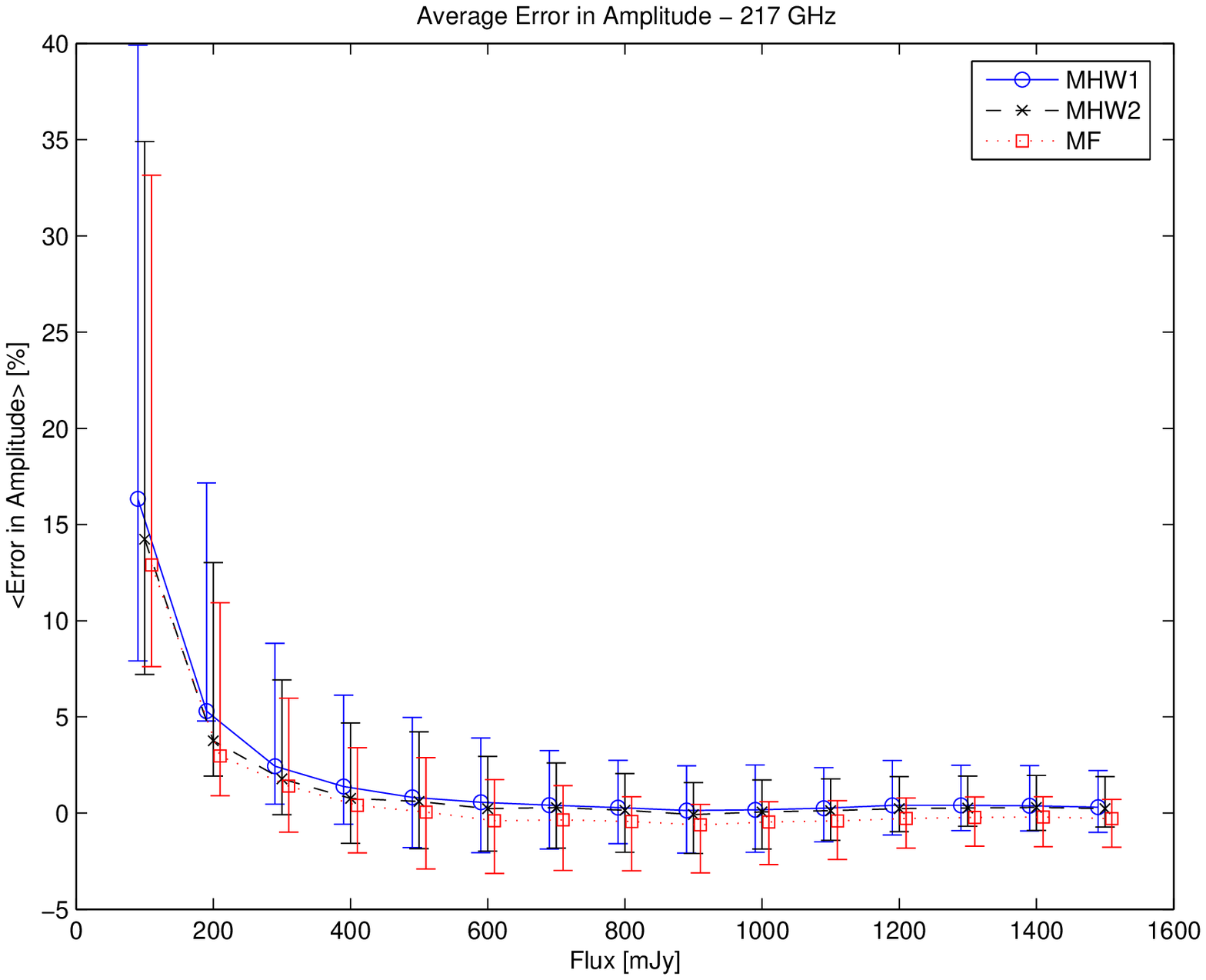} 
~~ 
        \includegraphics[width=5.8cm]{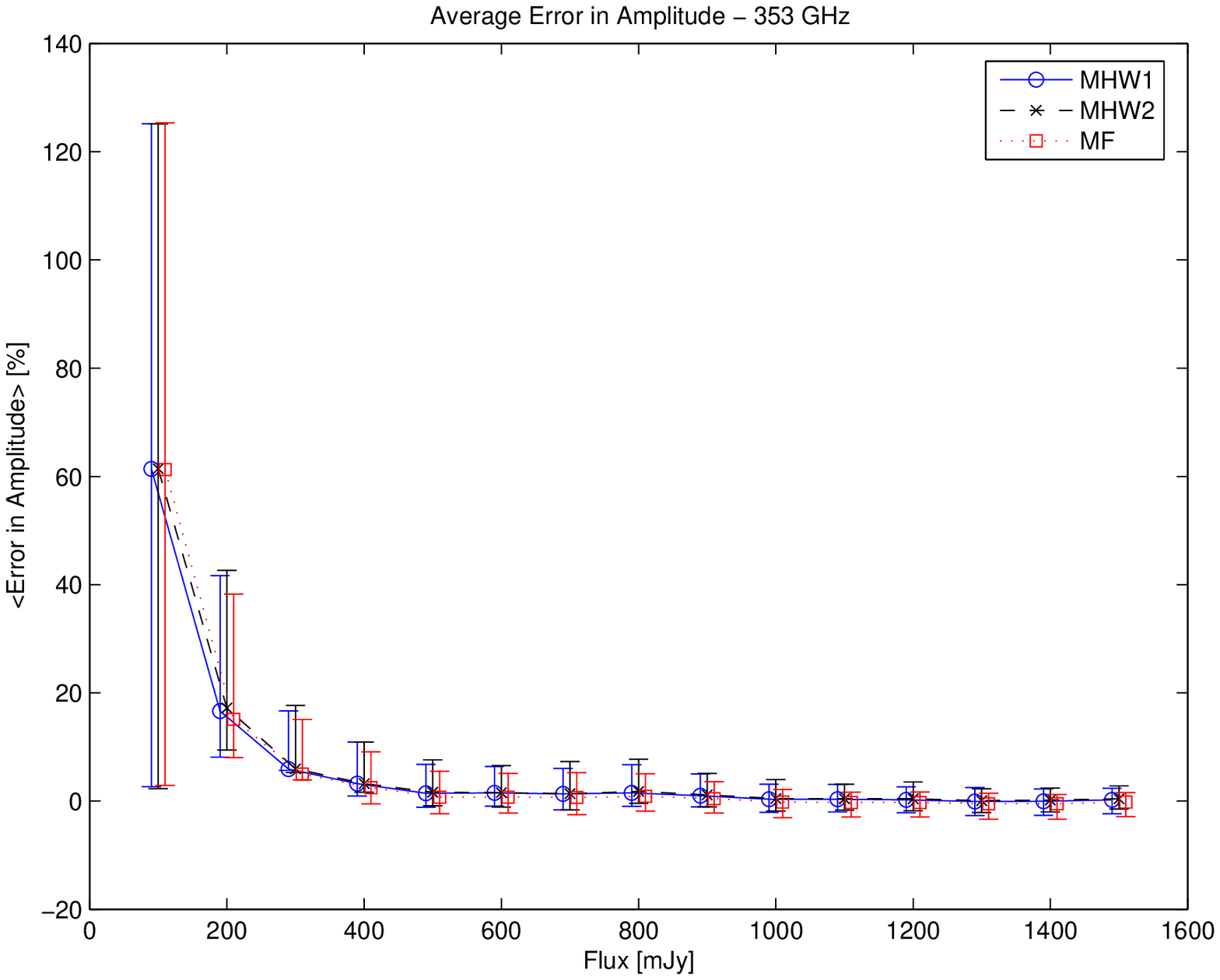} 
        \includegraphics[width=5.8cm]{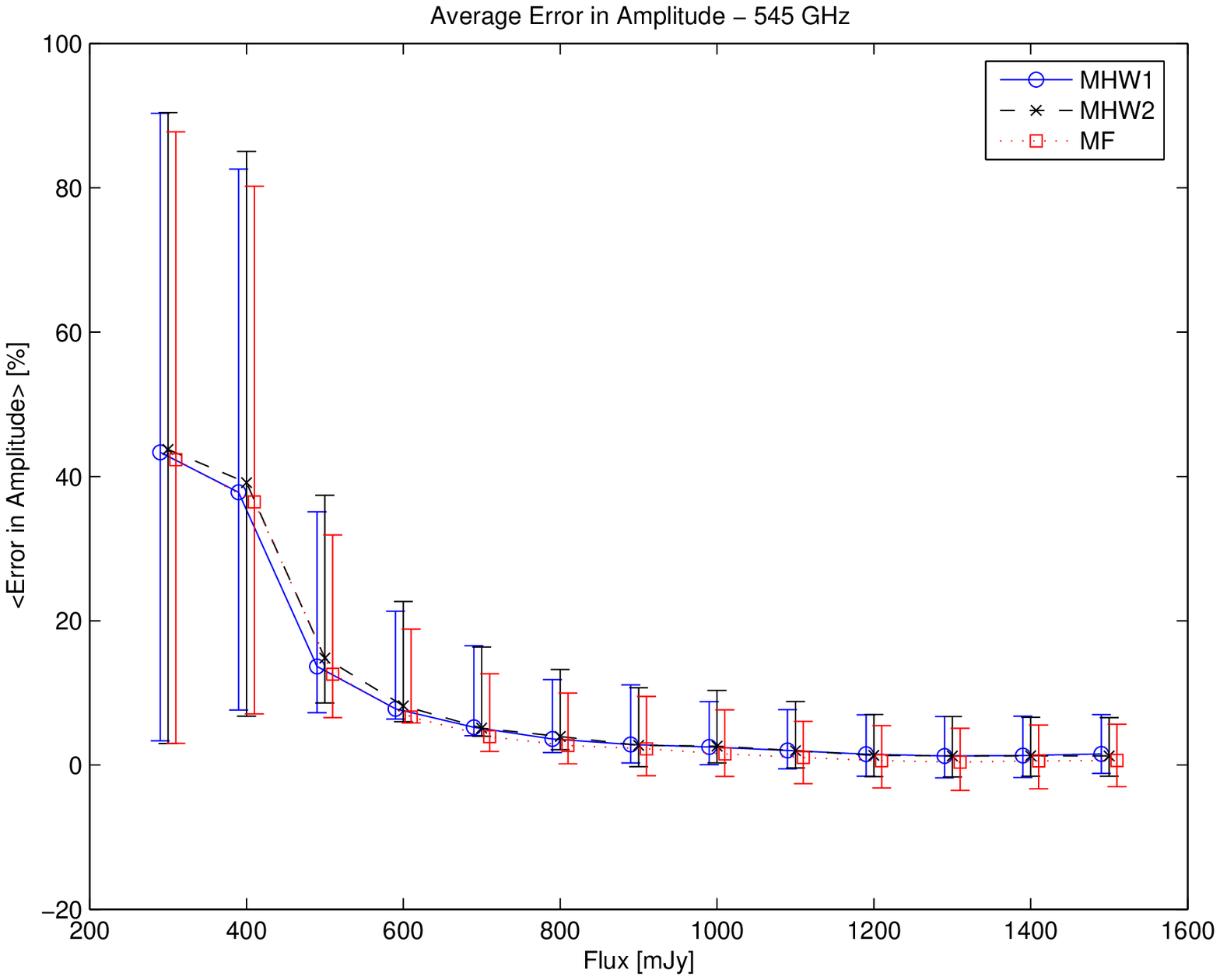} 
        \includegraphics[width=5.8cm]{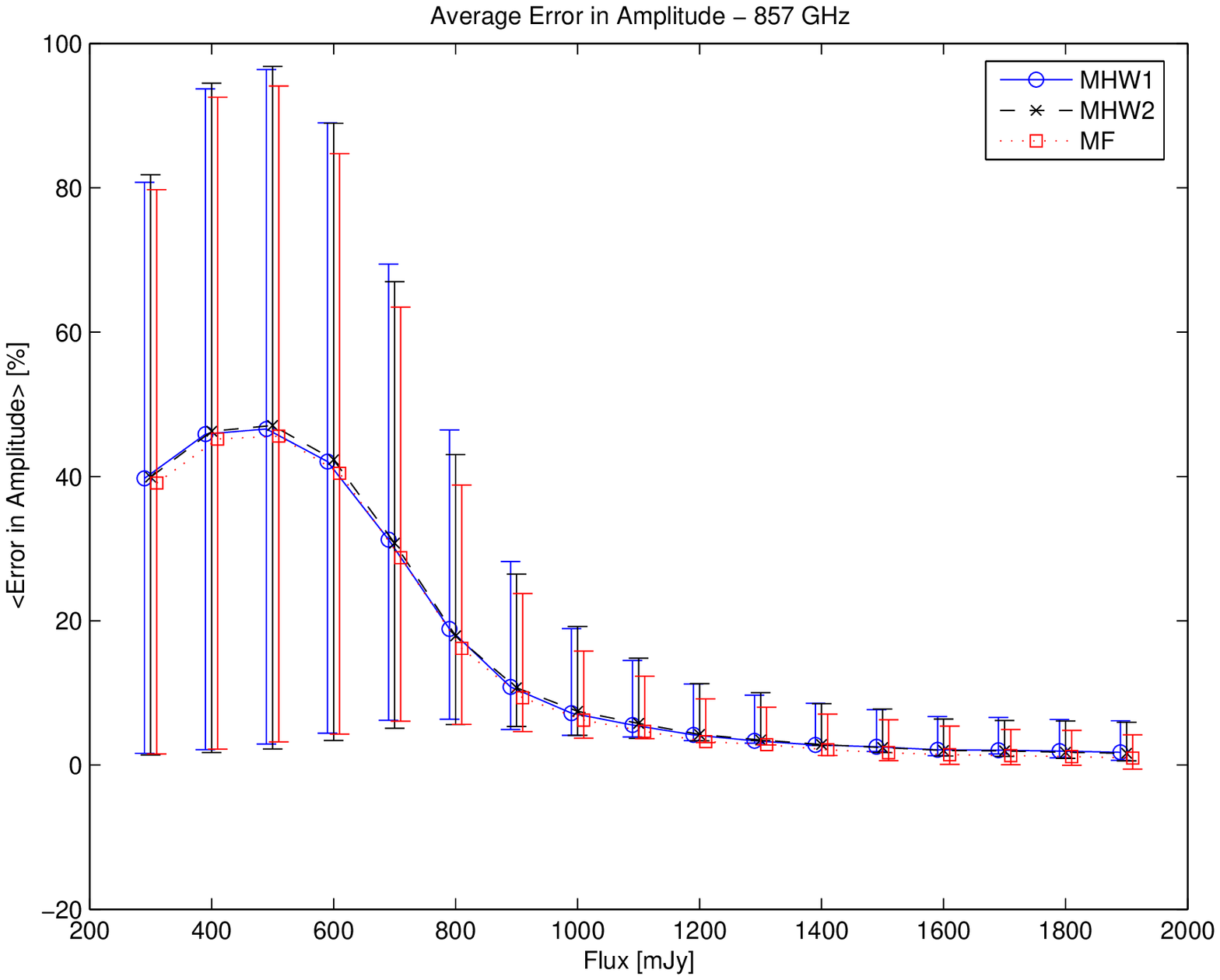} 
\caption{\label{fig:err_ampl} For each Planck channel we show the
average error (68\% error bars) in the estimated flux density for all
the sources observed above different flux detection limits. The error
is defined as ($<A_{est}-A_{real})/A_{real}>$, where $A_{est}$ is the
estimation of the flux in the simulated patch after filtering and
$A_{real}$ is the flux of the source in the catalogue. For the
sake of clarity, we have plotted the errorbars with an offset of 10
mJy in the x-axis.}
\end{center} 
\end{figure*}

In Figure \ref{fig:err_ampl} we represent the average error in the
estimate of the flux of the detected sources above different flux
detection limits, and this error is defined as $<(A_{est} -
A_{real})/A_{real}>$. This average error can take up negative values
and instead of including the standard deviation we have plotted the
true $68\%$ error bars taking into account the probability
distribution.
 
As expected, the error bars decrease towards higher fluxes. In Figure
\ref{fig:err_ampl} it can be seen that there is a positive bias in the
estimation of the flux. Such bias is negligible for high fluxes and
increases towards lower fluxes. The bias is due to the well-known
selection effect that makes more likely to detect a source when it is
located on a positive fluctuation of the background, leading to a
greater number of overestimated fluxes than the number of subestimated
fluxes. This can also be appreciated in the non-symmetric shape of the
observed error bars.
 
\begin{figure*} 
\begin{center} 
 
        \includegraphics[width=5.8cm]{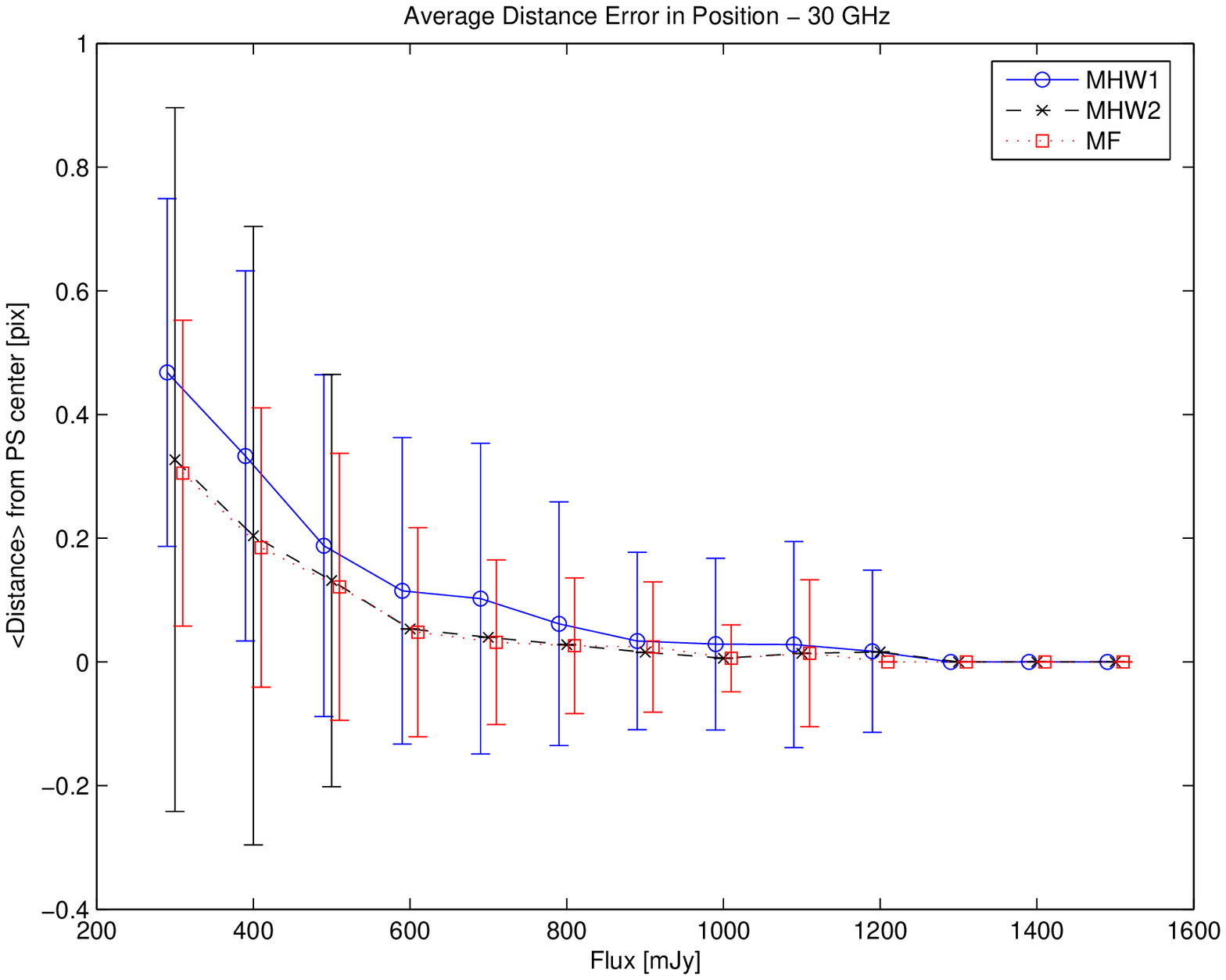} 
        \includegraphics[width=5.8cm]{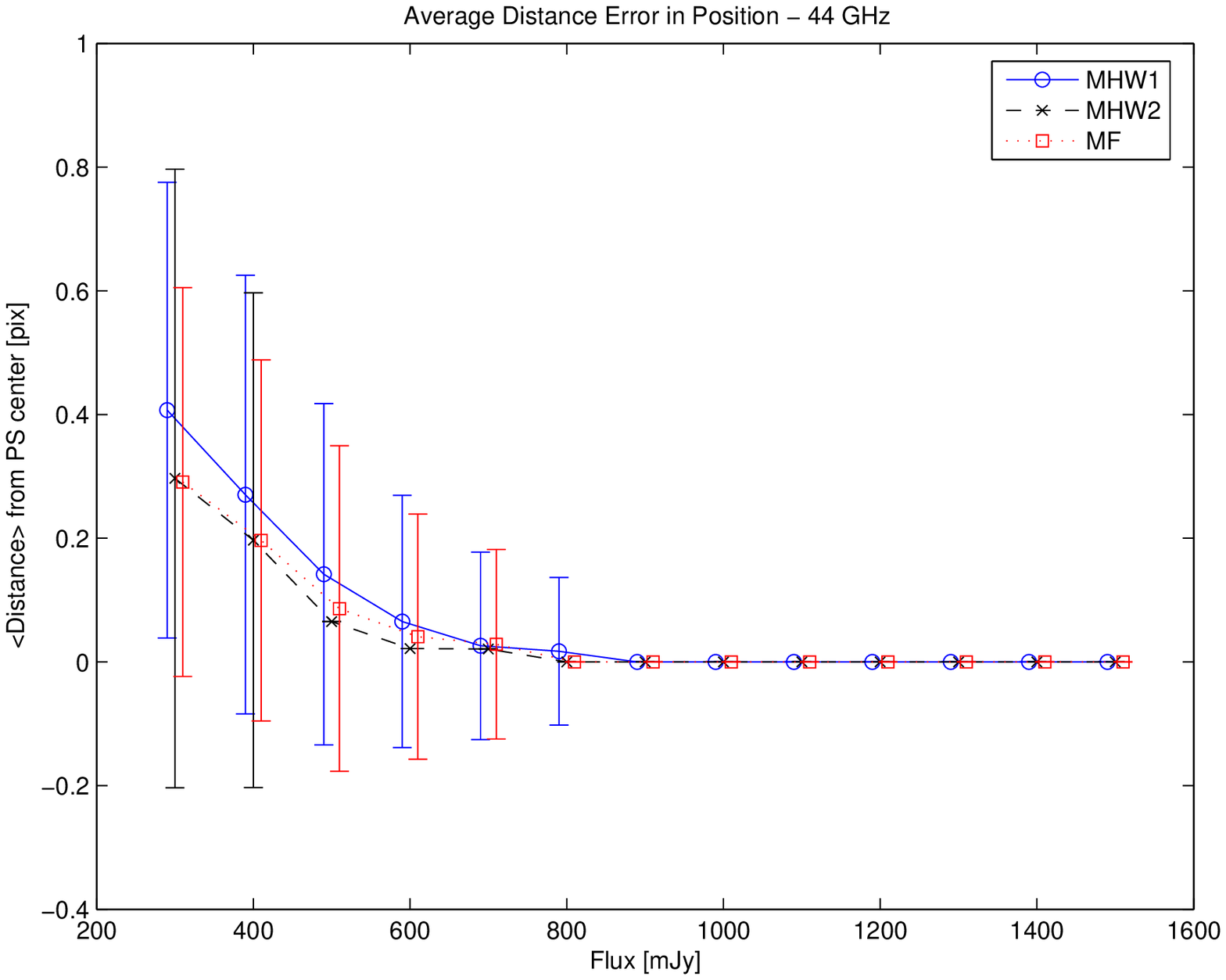} 
        \includegraphics[width=5.8cm]{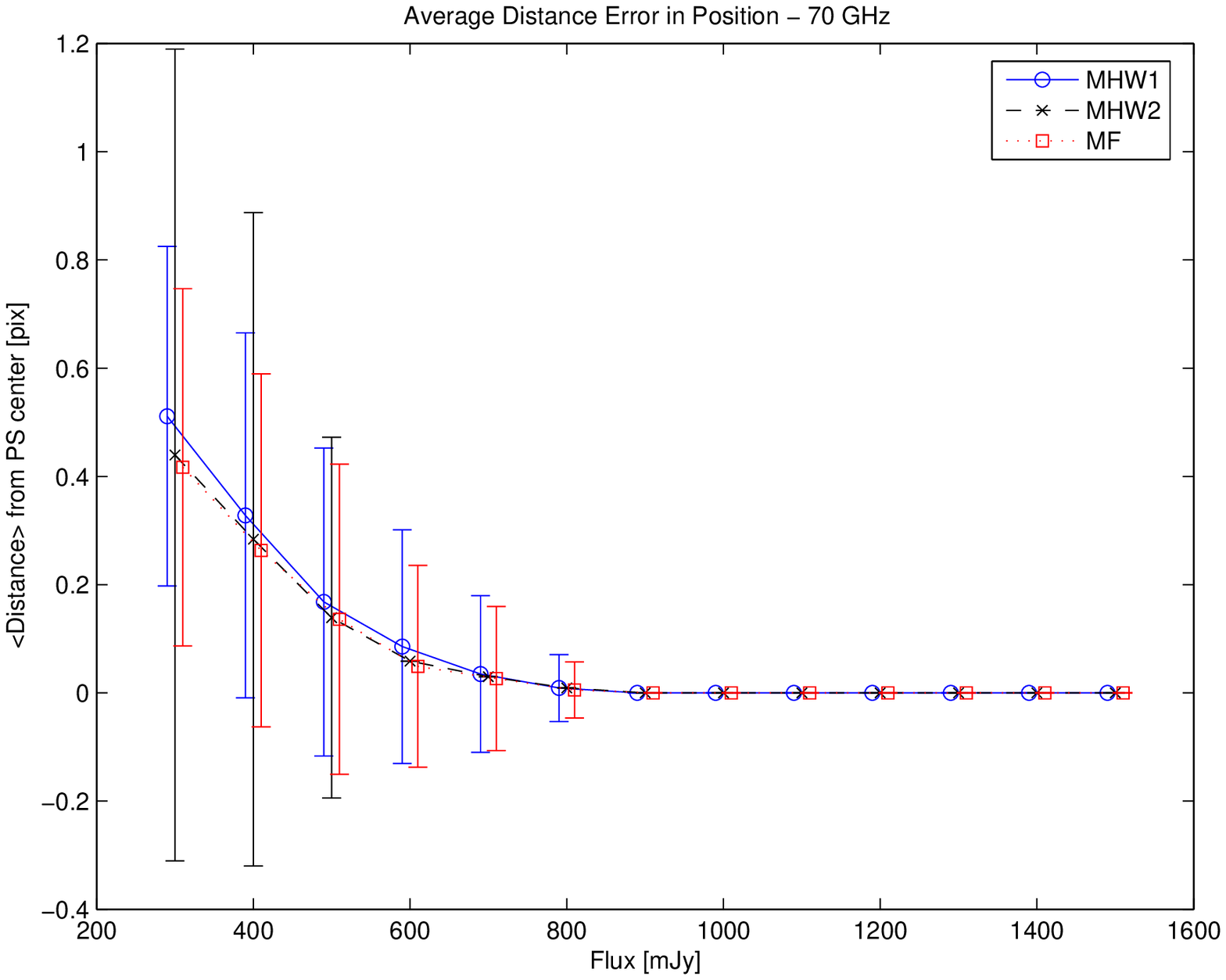} 
~~ 
        \includegraphics[width=5.8cm]{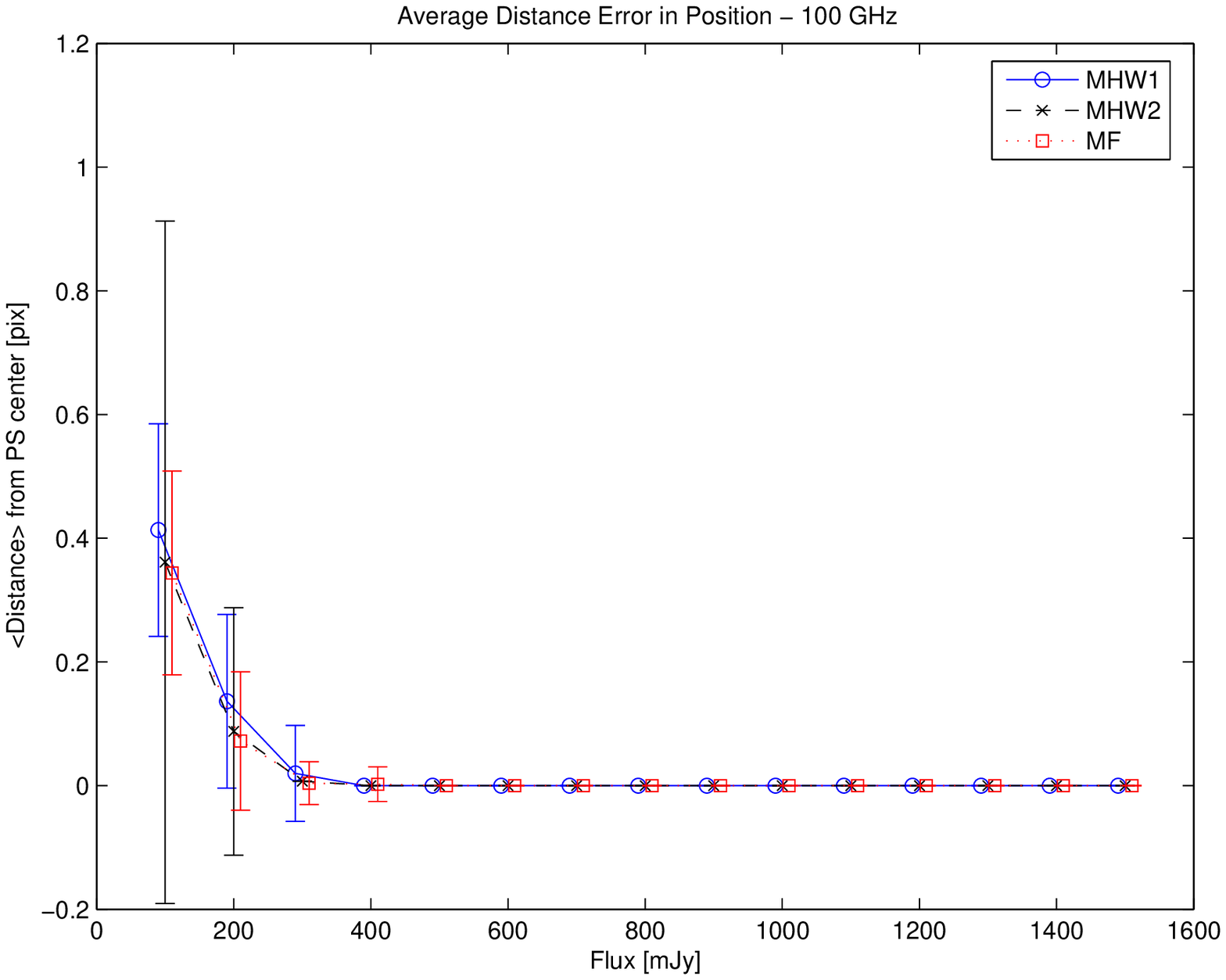} 
        \includegraphics[width=5.8cm]{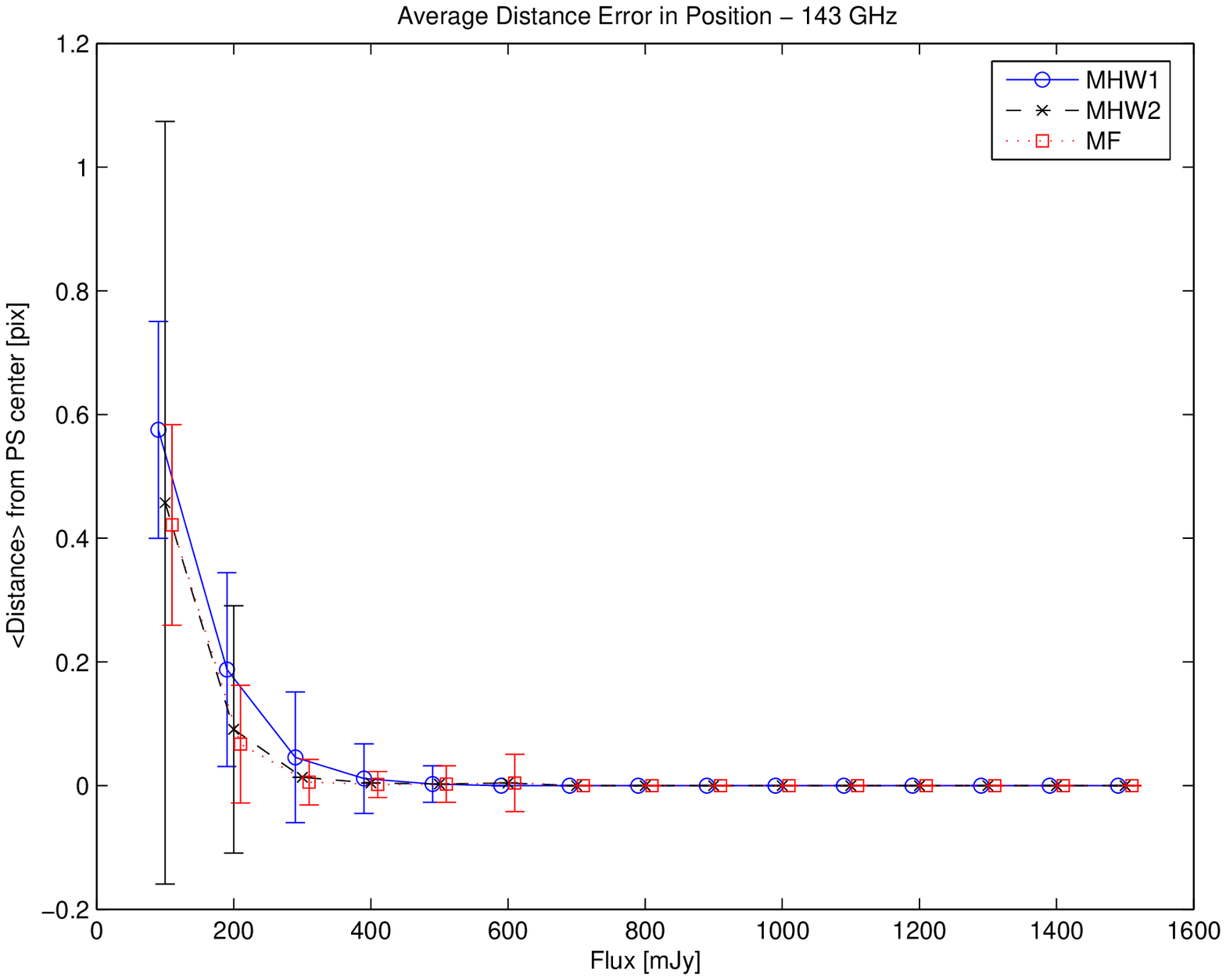} 
        \includegraphics[width=5.8cm]{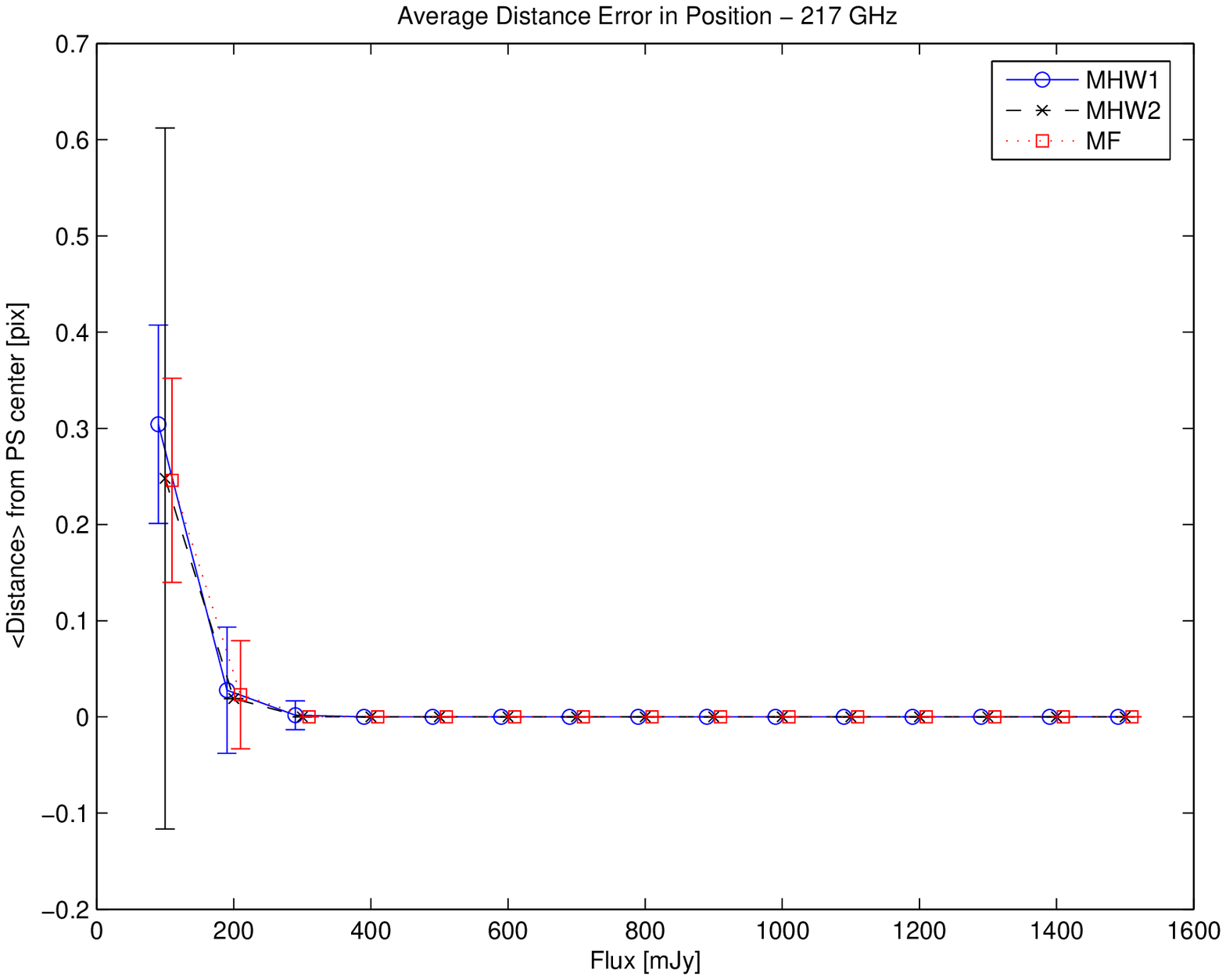} 
~~ 
        \includegraphics[width=5.8cm]{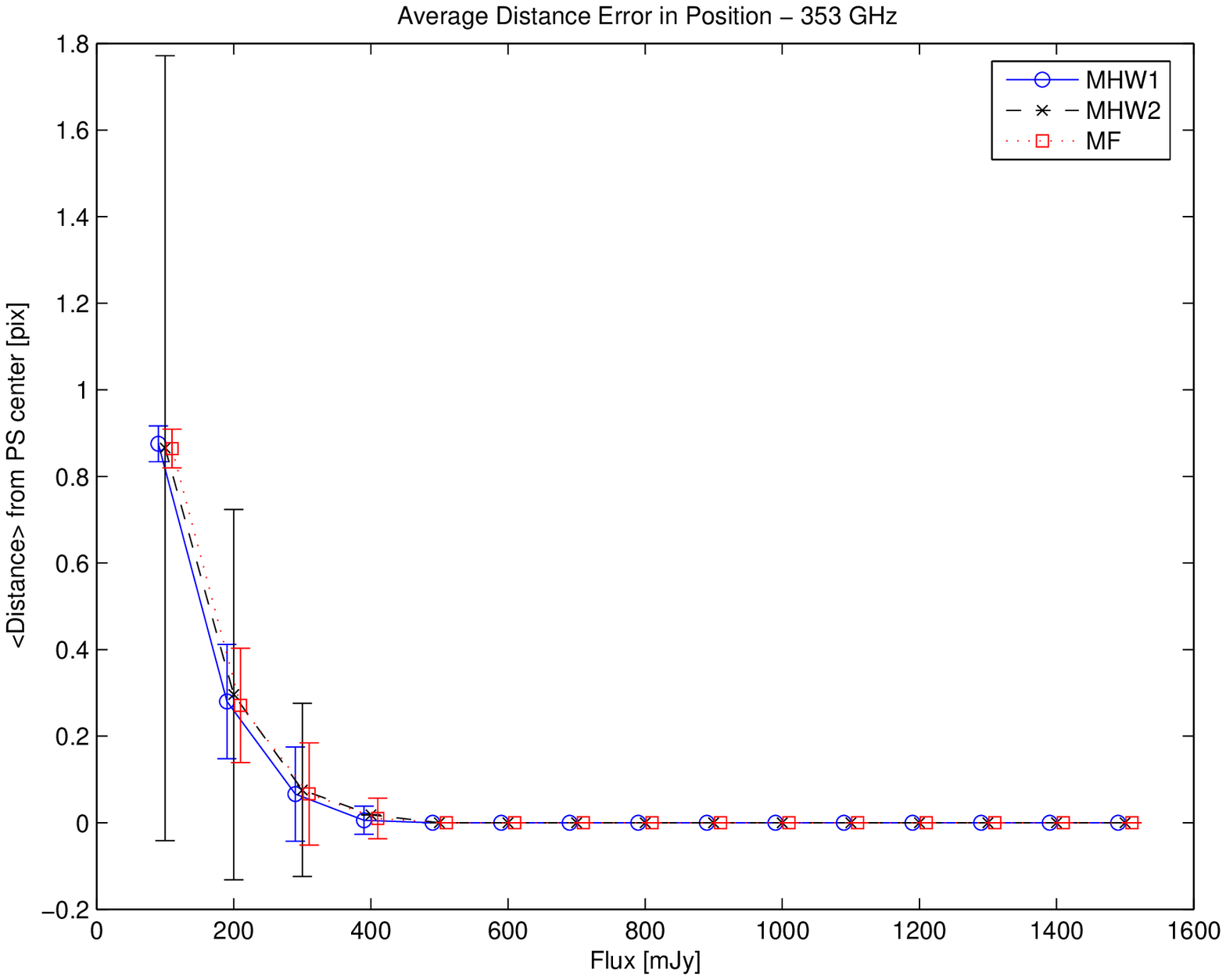} 
        \includegraphics[width=5.8cm]{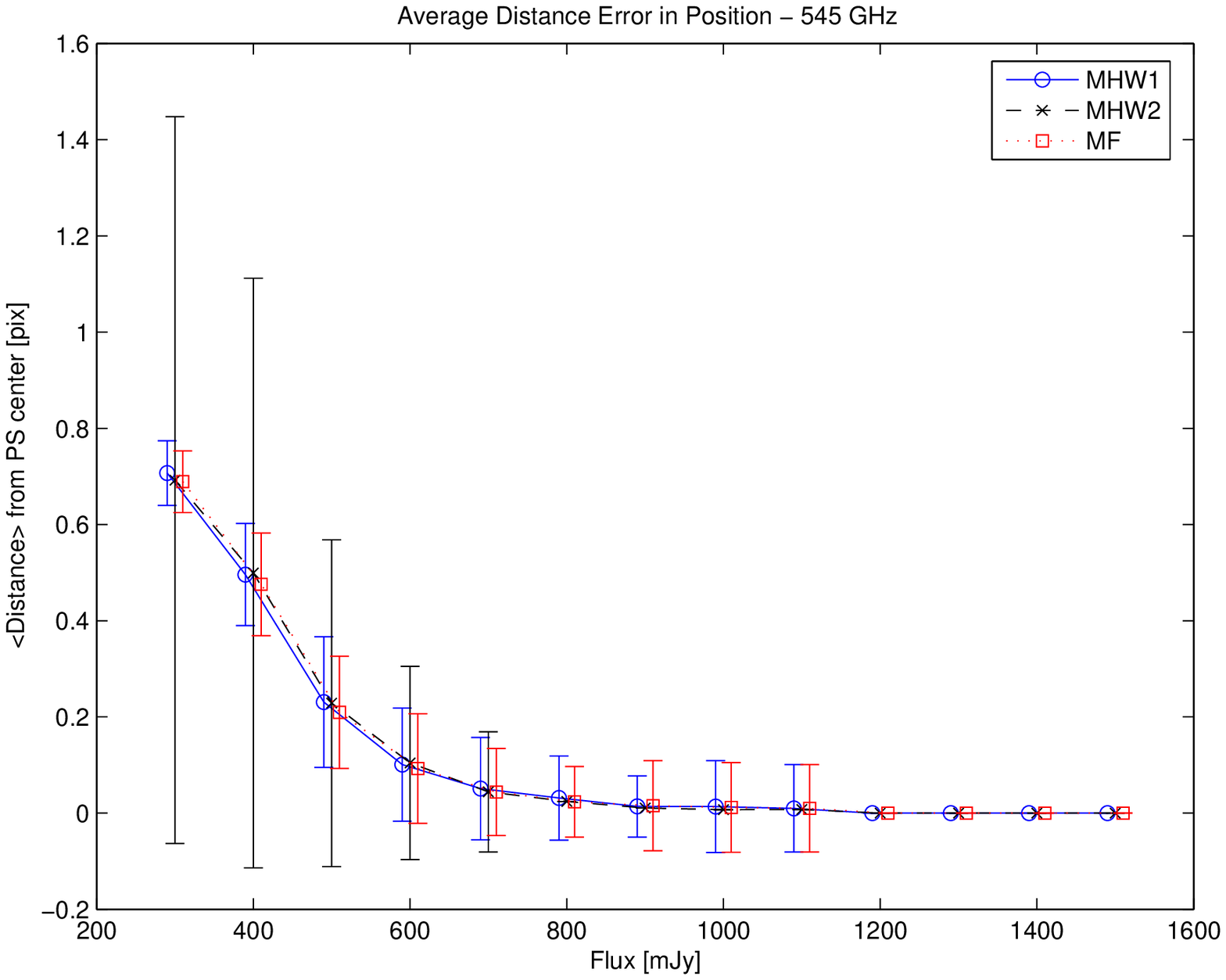} 
        \includegraphics[width=5.8cm]{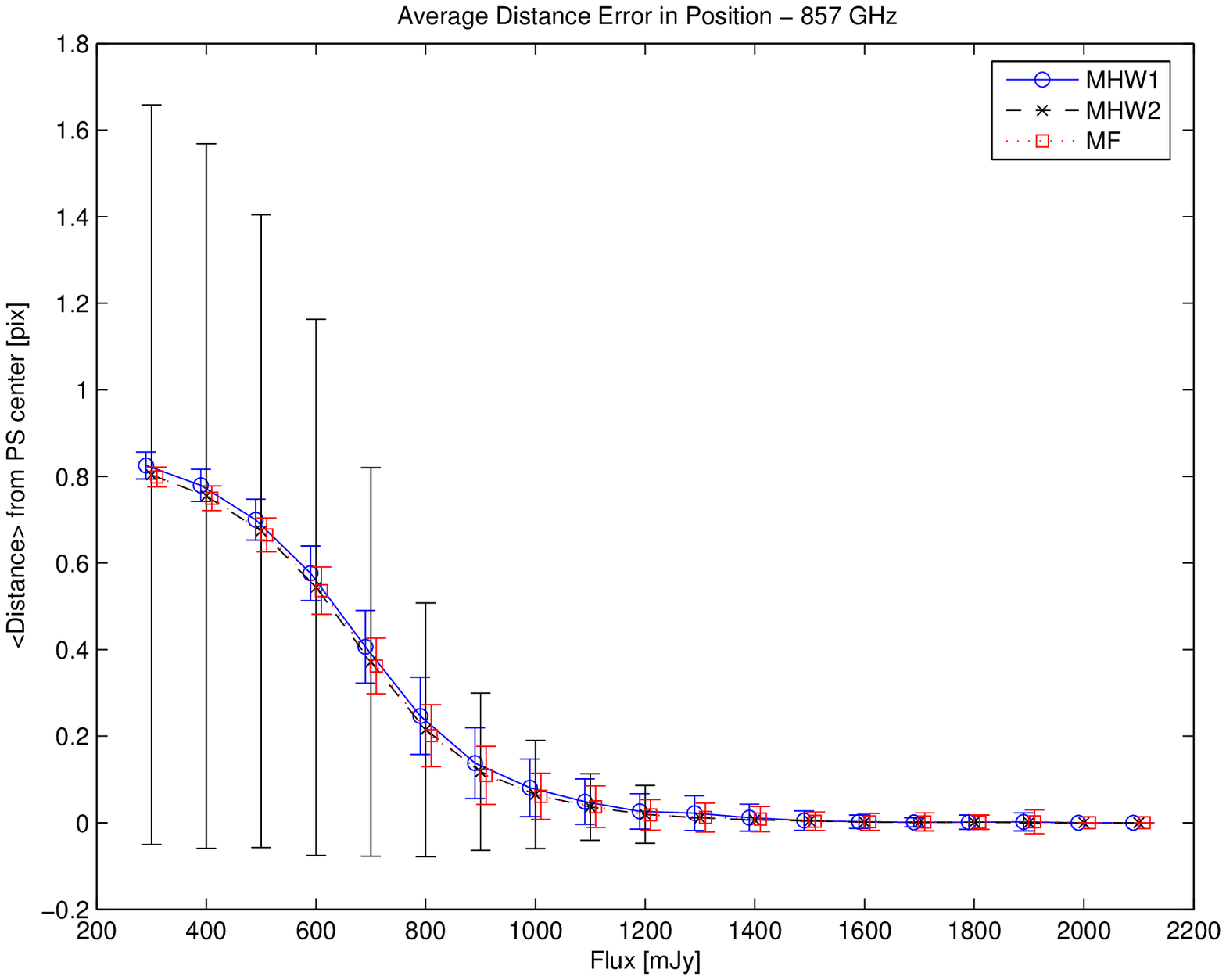} 
\caption{\label{fig:err_distance} In each panel and for different flux
detection limits, we show the average distance between the centre
coordinates of the detected source with respect to the actual
coordinates of the corresponding simulated source in the input
catalogue. For the sake of clarity, we have plotted the errorbars with
an offset of 10 mJy in the x-axis.}
\end{center} 
\end{figure*}

In Figure \ref{fig:err_distance} we have represented the average
distance between the position of the detected sources and their actual
position in the simulated map before filtering. We have experienced
that in some cases, after filtering, the maximum of a detected source
has moved to an adjacent pixel. This fact would normally not affect
our results, when matching candidate objects with the reference
catalogues, because of the way we compare the simulated (input) and
recovered (detected) sources (allowing a radius of $\leq
\mathrm{FWHM}/2$ for the proper matching). In any case, in order to
quantify this effect, we have calculated this distance for every
detected source in each map. We have found that, even in the worst
cases and only at very low Planck frequencies (i.e., 30, 44 GHz), the
average distance is always smaller than one pixel.
 
As for the MHW1, we have also checked to what extent a multi-fit 
approximation for estimating the source flux (i.e. a $\chi ^2$ fit 
of the wavelet coefficients to four scales $0.5, 1, 2$ and $3$ 
times the optimal scale) can improve the number of real 
detections, as discussed in previous works on the same subject 
\citep{patri01a,patri01b,patri03}. At frequencies above 100 GHz, 
the percentage of spurious detections with this wavelet decreases 
to the level of the MHW2, and at 857 GHz they are quite similar. 
On the other hand, at Planck frequencies below 100 GHz (LFI), 
while we still find a small improvement in comparison with the 
results previously plotted, the percentage of spurious detections 
does not reach the level of the MHW2, which performs always 
better. 
 
As a final comment, it is clear that the multi-fit approximation only 
improves the estimation of the source fluxes at the highest 
frequencies and, correspondingly, reduces the number of the spurious 
detections. However, only at 857 GHz the performance of the MHW1 -- with 
this method -- is similar to that of the MHW2. As for the error in the 
estimated flux of the detected sources, we found that this method 
applied to the MHW1 yields estimated errors similar to those 
determined for the other filters.

\begin{table*} 
  \centering 
  \begin{tabular}{ccccccccccccc} 
  \textbf{Freq(GHz)}& \textbf{Filter} & \textbf{N$_{d_5}$}& \textbf{N$_{sp_5}$} & \textbf{S$_{5}$(mJy)}&  \textbf{SNR$_{5}$} & \textbf{A$_{err}$($\%$)} &  &\textbf{N$_{d_{95}}$}& \textbf{S$_{95}$(mJy)}&  \textbf{N$_{d_{100}}$}& \textbf{S$_{100}$(mJy)}   \\ 
  \hline\hline 
   30  & \textbf{MHW1}  & 497 & 26 & 448 & 3.79   & 12 & & 583  & 329 & 327 & 545 \\ 
     & \textbf{MHW2}    & 580 & 30 & 391 & 3.75   & 12 &  & 595  & 324 & 432 & 447  \\ 
     & \textbf{MF}      & 594 & 31 & 382 & 3.63   & 10 & & 594  & 324 & 432 & 447 \\ 
  \hline 
   44  & \textbf{MHW1} & 284 & 16  & 505 & 4.10   & 12 & & 388  & 369 &  263 & 495\\ 
     & \textbf{MHW2}   & 342 & 18  & 443  & 3.99   & 12 & & 399  & 361 &  263 & 495 \\ 
     & \textbf{MF}     & 335 & 17  & 448  & 4.07   & 10 & & 418  & 349 &  303 & 447 \\ 
  \hline 
   70  & \textbf{MHW1} & 304 & 16 & 510  & 4.59   & 8  & &  445  & 339  & 276 & 495 \\ 
     & \textbf{MHW2}   & 341 & 18 & 473  & 4.54   & 9 & &  444  & 339  & 276 & 495 \\ 
     & \textbf{MF}     & 344 & 18 & 467  & 4.49   & 9 & &  443  & 339  & 276 & 495 \\ 
  \hline 
  100 & \textbf{MHW1} & 586  & 30 & 294 &  4.45  & 6  &  & 1237  & 138  & 871 & 198 \\ 
     & \textbf{MHW2}  & 730  & 38 & 248 &  4.42  & 7  &  & 1323  & 131  & 877 & 197 \\ 
     & \textbf{MF}    & 690  & 36 & 252 &  4.66  & 6  &  & 1325  & 131  & 877 & 197 \\ 
  \hline 
  143   & \textbf{MHW1} & 914  & 48  & 240 & 4.80   & 7 & & 1713 & 116 & 1048 & 195\\ 
     & \textbf{MHW2}    & 1130 & 59  & 196 & 4.73   & 7 & & 1797 & 109 & 1049 & 195\\ 
     & \textbf{MF}      & 829 & 43  & 250 & 6.25   & 2 & & 1714 & 116 & 1049 & 195 \\ 
  \hline 
   217  & \textbf{MHW1}  & 1153 & 61 & 195 & 4.87   & 6 & & 1756 & 121 &  1074 & 195\\ 
     & \textbf{MHW2}     & 1233 & 65 &  185 & 5.11   & 5 & & 1755 & 121 &  1074 & 195\\ 
     & \textbf{MF}       & 1242 & 65 & 195 & 5.57  & 3 & & 1755 & 121 &  1074 & 195\\ 
  \hline 
   353  & \textbf{MHW1}  & 887 & 47 & 314 & 5.29   & 5 &  & 2253 & 147 & 1161 & 245\\ 
     & \textbf{MHW2}     & 990 & 52 & 292 & 4.86   & 0 &  & 2243 & 147 & 1160 & 245\\ 
     & \textbf{MF}       & 986 & 52 & 292 & 4.99   & 0 &  & 2247 & 147 & 1160 & 245\\ 
  \hline 
   545  & \textbf{MHW1}  & 726 & 38 & 785 & 6.33   & 8  & & 2582   & 349 & 1495 & 495\\ 
     & \textbf{MHW2}     & 1025 & 54 & 647 & 5.09 & 9 & & 2568   & 349 & 1495 & 495 \\ 
     & \textbf{MF}       & 924 & 49 & 684 & 5.60  & 9 & & 2580   & 349 & 1495 & 495\\ 
  \hline 
  857   & \textbf{MHW1}  & 1034 & 54 & 1765  & 8.24  & 2 &  & 13885  & 532 & 5194 & 745 \\ 
     & \textbf{MHW2}     & 3183 & 167 & 1013 & 4.82 & 7 &  & 13851  & 532 & 5201 & 745 \\ 
     & \textbf{MF}       & 3093 & 162 & 1020 & 4.97 & 6 &  & 13914  & 532 & 5200 & 745 \\ 
  \hline\hline 
\end{tabular} 
\caption{In this table we summarise the $5\%$ percentage of spurious
   detections and the $95\%$/$100\%$ completeness of the catalogues at the
   Planck frequencies.  Regarding the percentage of spurious
   detections, columns three to seven show the number of detected
   sources, $N_{d_5}$, the number of spurious sources $N_{sp_5}$, the
   corresponding flux detection limit, $S_{5}$, the signal-to-noise
   ratio $SNR$ for $S_{5}$ and the average error in the estimation at
   this flux limit, respectively. We allow for a $5\%$ of spurious
   detections, where this percentage has been calculated as
   $N_{sp}/(N_{d}+N_{sp})$. Regarding the completeness level, the flux
   limits at which the catalogues are complete at the $95\%$ and
   $100\%$ are shown in columns eigth and ten, respectively, and their
   corresponding number of detections, $N_{d_{95}}$ and $N_{d_{100}}$,
   in columns nine and eleven, respectively. \label{tab:table2}}
\end{table*} 
 
\section{Further details on the detection of 
  point sources}  \label{sec:discussion} 
 
In this section we briefly discuss some additional issues related to
point source detection.
 
\subsection{Confused/blended extragalactic point sources} 
 
Every time we observe the sky with a pencil- or a synthesised beam, if
two or more point-like sources (i.e., distant galaxies or clusters)
lay inside the beam area they appear as indistinguishable since they
"blend" into a single source: a single "spot" in the map having the
same angular dimension (FWHM) of the beam. As a result, only one
``source'' -- whose flux is the integrated value of all the individual
sources in the beam -- is actually detected. However, if source counts
are not very steep \citep{dz96}, this integrated flux value is mainly
determined by the brightest source in the beam, whereas the much more
numerous faintest sources are only slightly modifying the total flux
and adding some noise \citep[]{fra89,bar92}. Thus, source
blending is unlikely to affect the counts of bright sources, which are
the ones actually detected.
 
As an example, at 30 GHz, if we reach the flux detection limit at
which we have $5\%$ of false detections, $\sim 0.4$ Jy, we have less
than 0.007 sources per beam area and, thus, the probability that two
such sources fall inside the same beam is almost negligible.
Moreover, at higher Planck frequencies the situation is still
better. At 100 GHz we find that at the $5\%$ false detections limit,
the average number of sources per beam is $\sim 0.0014$, whereas for
the 857 GHz channel that number decreases to $\sim 0.0008$ sources per
beam. On the other hand, we have to remind that our simulations do not
take into account the clustering of sources. If source positions are
correlated in the sky -- as it is well known for all the source
populations observed by Planck -- the probability of finding two bright
galaxies at the same angular separation in the sky
increases. Therefore, with a low-resolution experiment, like Planck,
we will observe the summed contribution of groups (or clumps) of
sources within the same beam area. The problem has been recently
discussed by \citet{ngr05}, who found that, in the presence of strong
clustering, source confusion can be important even above the canonical
5$\sigma$ detection limit, thus modifying source number
counts. However, this effect is likely to be important only at Planck
HFI frequencies and for the highly clustered high-redshift
proto-spheroidal galaxies (see, e.g., the thorough analysis by
\citet{jgn05}). In any case, this study goes beyond the purposes of
this paper and it shall be tackled in a forthcoming one.
 
\subsection{Other compact sources} 
 
Other compact but not completely point-like objects will be 
observed by Planck. The most important among these objects are 
galaxy clusters observed thanks to the Sunyaev-Zel'dovich (SZ) 
thermal effect. Detailed simulations of the local universe have 
shown that the SZ signal from low-redshift clusters and 
superclusters should be marginally detectable by Planck 
\citep[]{dol05,han05}. On the other hand, very few SZ clusters 
show angular scales larger than ten arcminutes. Since the scale of 
these extended clusters is significantly different from the scale 
of the point sources (i.e., the beam size), the filtering with 
either a wavelet or a matched filter -- both optimised for the 
scale of the point objects -- will tend to cancel the SZ signal 
just in the same way as it cancels the rest of the diffuse signal 
and the large scale fluctuations of the background. Therefore, 
large SZ clusters are not expected to affect significantly the 
detection process. 
 
Most of the SZ clusters, however, shall be observed as ``unresolved''
or point-like (i.e., smaller than the Planck beam).  In the latter
case, they have to be treated in the same way as point sources, except
for the fact that their fluxes are typically much fainter. Even the
brightest clusters are on the verge of the detectability, given the
sensitivities of Planck detectors. As already discussed by many
authors, it shall be necessary to combine the signal measured in
various frequency channels in order to detect clusters with Planck
\citep{yo02c}. When using a channel-by-channel detection
strategy, as the one presented in this work, the impact of SZ clusters
is very small.  In fact, at $\nu< 217$ GHz, SZ clusters appear as
``cold spots'' and therefore cannot be detected above any positive
flux threshold. At 217 GHz, the thermal SZ effect is zero and the
kinematic SZ effect is well below the level of the CMB
fluctuations. Above 217 GHz there will be some minor effect due to the
thermal SZ effect that may lead to a small number of false
detections. The detailed study of this kind of contamination shall be
addressed in a future work.
 
As for Galactic compact sources, representing different evolutionary
stages in stellar evolution -- i.e., compact pre-stellar cores, young
stellar objects and supernova remnants -- many of them shall be
observed by Planck. Current estimates predict that Planck surveys
should detect from many hundreds to thousands of these sources,
depending on the subclass (see Table 5.3 in the Planck BlueBook, "The
Scientific programme", ESA-SCI(2005)1
\footnote{http://www.rssd.esa.int/index.php?project=PLANCK}. In
particular, one of the most relevant classes of Galactic compact
sources is that of HII regions, which provide relevant information on
early stages of stellar evolution and on the Galactic spiral
structure. The thorough analysis made by \citet{pal03} has
demonstrated that the vast majority ($> 80$\%) of the 1442 HII regions
in their Synthetic Catalogue shall be detected by Planck with a very
high signal-to-noise ratio ($S/N \geq 100$). In any case, all these
sources have well known positions in the sky and, moreover, almost all
of them lay at very low Galactic latitude. Therefore, they shall be
easily identified and subtracted out. Finally, Planck should also
observe 397 asteroids and a significant fraction (50-100 objects) of
them with a high S/N ratio \citep{cre02}. However, these Solar System
bodies are moving objects very close to the Earth and, thus, they
shall show variable positions and fluxes, depending from their
heliocentric and geocentric distances, even during a single Planck
scan of 360$^\circ$.
 
\subsection{Systematics} 
 
As explained in section~\ref{sec:sims}, our simulations can be 
regarded as ``state of the art'' as for the realism of the 
simulations of the physical components (CMB, Galactic foregrounds 
and extragalactic objects) and we have used the nominal noise 
levels, beams and pixel sizes of the Planck mission. However, some 
systematics of relevance have not been included in the 
simulations. Among them, the ones that can affect our current 
results are: anisotropic noise, $1/f$ noise, thermal effects and 
non-symmetric beams. 
 
$1/f$ noise and thermal effects in the electronics of the 
satellite will lead to large scale noise fluctuations across the 
sky. Since the filters here discussed operate at small angular 
scales, these large scale fluctuations tend to be cancelled out by 
the filters and should not affect the detection of point sources. 
Anisotropic noise will appear due to the non-uniform sky coverage 
of the satellite instruments. Furthermore, mapmaking algorithms 
can introduce some degree of noise correlation over small angular 
scales. Non-uniformity of the noise will make easier to detect 
point sources in some areas of the sky (those that are better 
scanned) and more difficult to detect in others: on average, the 
integral number of detections should not change very much. Noise 
correlation, on the other hand, can be a non negligible effect and 
it must be studied in a future work. In any case,~\citet{patri03} 
have found that their results with the Mexican Hat Wavelet did not 
vary when anisotropic noise pattern and $1/f$ noise were 
considered instead of isotropic white noise. 
 
We have worked with idealised beams that are circularly symmetric and
Gaussian-shaped. In a real experiment this will not be the case. The
real response of the beam shows more complexity, with side lobes,
ellipticity and even changing the orientation in the sky of the
projected shape. As long as the beam profile is well known, it is
straightforward to construct the corresponding MF.  Even in the case
of non-symmetric beams this can be done, by increasing the
computational complexity of the problem. Therefore, when using the MF,
real beams can be in principle handled.  Regarding the standard
Mexican Hat Wavelet,~\citet{patri03} have tested the influence of
realistic asymmetric beams. Those authors have shown that, although
the MHW is an isotropic wavelet, it can also be adequate to perform
the detection of point sources that show a slight Gaussian
asymmetry. This is precisely the situation for the Planck beams (see
the Planck BlueBook, ESA-SCI(2005)1). For a more detailed discussion,
the reader is referred to~\citet{patri03}.
 
\subsection{Extension to the sphere} 
 
If we want to extend our work to the sphere, it is important to 
address the problem of very variable foregrounds from one region of 
the sky to the other. Then it is necessary to operate locally. From 
the point of view of implementation, there are two approaches to this 
problem: to work with small flat patches, as we have done in this 
paper, or to work directly on the sphere using the harmonic transform 
instead of the Fourier transform, as it was done 
in~\citet{patri03}. The first approach is valid if the patches are 
small enough (no larger than $\sim 15\times 15$ square degrees), but 
it implies to project the data from the sphere to the plane and to 
repixelise the samples (from some non-square pixels on the sphere such as 
the ones used by HEALPix to the square flat pixels of the 
patches). This leads to some deformation, especially for the outer 
parts of the flat patches and for high latitudes in the 
sphere. Therefore, the flat patch approach leads to some amount of 
``systematic'' degradation of the data. Besides, in order to cover all 
the sky it is necessary to decide how to choose the positions of the 
patches, how much overlapping among the different patches is necessary 
and how to correct the estimated flux of each source taking into 
account the deformation just mentioned. 
 
Therefore, in some cases it could be preferable to work directly in 
the sphere.  The most simple way to take into account the variability 
of the foregrounds is to use the harmonic transform to filter the 
sphere several times, using each time a filter that is adapted to the 
statistical properties of a given area of the sky. This is the 
approach adopted by~\citet{patri03}. But the number of different 
filters used should not be very large or the CPU time required would 
be unreasonably long.  Fast harmonic transform algorithms make 
possible to filter the whole sky in a relatively short time 
\citep{heal05}, but even so the process can take a significant amount 
of time for high resolution maps (for example, performing the harmonic 
transform and calculating the angular power spectrum of a NSIDE=2048 
HEALPix map takes 108 minutes in a 3 GHz Intel Pentium processor). 
Thus, it is necessary to reach a compromise between acting locally 
(that may require to filter the data many times) and saving CPU time. 
 
In the case of wavelets it has been successfully shown \citep{patri03} 
that it is possible to estimate locally the optimal scale of all the 
regions, group together different regions with similar optimal scales 
and construct a small number (around fifteen in the worst case) of 
filters on the sphere for these optimal scales. In their study it was 
possible to reduce the number of times it was necessary to filter the 
maps due to the fact that the functional form of the wavelets depend 
only on one parameter (the optimal scale) that varies slowly across 
large zones of the sky and it is easy to group the optimal scales in a 
small number of bins without loosing much efficiency. Even so, their 
algorithm takes $\sim 72$ hours for the most CPU time-demanding 
channel (857 GHz) in a Compaq HPC320 (Alpha EV68 1-GHz processor) and 
requires 4 GB of RAM memory, a non-negligible amount of computer 
resources. 
 
On the contrary, the MF functional form depends on the power spectrum 
of the data. The shape of the local power spectrum varies wildly from 
one region of the sky to other, and it is not easy to group local 
power spectra in classes to reduce the number of times the sky must be 
filtered. A taxonomic criterion to group power spectrum curves 
should be provided, introducing a new arbitrarity in the process. If 
one wants to skip this arbitrarity, it is necessary to construct one 
filter \emph{for every region of the sky and to filter the sky map a 
large number of times}, especially for high resolutions. Nowadays, 
this process requires a huge amount of computational resources. This 
gives another reason to prefer wavelets to MF when working in the 
sphere. 
 
\section{Conclusions} \label{sec:Conclusions} 
 
We have compared the performance of three filters when dealing 
with the detection of point sources in CMB flat sky maps. These 
filters are the well known MF, the Mexican Hat Wavelet (MHW1) and 
the recently introduced Mexican Hat Wavelet 2 (MHW2). The latter 
is the second member of the so-called Mexican Hat Wavelet Family 
(the MHW1 is the first), a group of wavelets obtained by applying 
the Laplacian operator to the Gaussian \citep{jgn06}. This new 
wavelet is part of an effort to improve the MHW1, a tool already 
exploited by our group that has proved very useful in the 
detection of compact sources and of non-Gaussianity in CMB maps 
\citep{patri01a,patri03,patri04}. 
 
We have tested these tools in realistic 2D simulations of the
microwave sky prepared in the framework of optimising the efficiency
in separating the various astrophysical components in the forthcoming
ESA's Planck Satellite all-sky maps. As for the Galactic foregrounds
and the S-Z effect in clusters of galaxies, we have used the available
``Planck Reference Sky Model''; we have adopted the Standard
"concordance" Model for simulating CMB anisotropy maps and, as for
extragalactic point sources, we have used simulations obtained by the
software discussed in \citet{jgn05} and with the number count models
for sources of \citet[]{dz05,gr01,gr04} and \citet[]{ngr04}. We then
applied the three considered filters to a sufficient number of flat
patches to cover half of the sky $(2 \pi \,\mathrm{sr}, b> |30^\circ|)$ (see
Section 3).
 
We have found that the MHW2 and the MF outperform the MHW1 in some
aspects, especially at the lowest Planck frequencies. The three
filters yield approximately the same number of real detections, down
to the same flux detection limit, although the MHW1 yields a
corresponding higher number of spurious detections. Moreover, the MHW2
and the MF give comparable results for almost every one of the
analysed indicators. As shown in Figure \ref{fig:mfs}, it is
remarkable that, even the estimated shape of the MF tends to that of
the MHW2, at its optimal scale, for most of the frequencies discussed
in this work.
 
In a previous work \citep{patri03}, a multi-fit approximation 
to estimate the flux density of the sources detected by the MHW1 
was shown to be able to improve the results (as compared with the 
direct approach used in this work). We have applied this technique 
to the simulated maps used in this work and we found different 
results for the LFI and the HFI Planck frequency channels. In the 
first case, LFI, the improvement is small and the final results are 
never comparable to the ones obtained with the MHW2. In the second 
case, this procedure helps to slightly reduce the number of 
spurious sources, except for the highest frequency of HFI (857 
GHz), where the decrease in the number of spurious is significant 
and the MHW1 approaches the results obtained with MHW2. 
 
These results are very important because both wavelets, the MHW1 and
the MHW2, are represented by a known analytical function. The only
parameter that needs to be obtained for each simulation is the
``optimal scale'', which is calculated locally in a very easy way.  On
the other hand, the correct definition of the MF implies a number of
steps. Firstly, it is necessary to estimate the value of the power
spectrum for all the Fourier modes present in the image, especially
for the low modes where the power spectrum is \emph{noisy}. Secondly,
the use of such a noisy power spectrum to construct the MF often
yields a filter with many discontinuities in Fourier space which, in
turn, produces ringing effects in the filtered image. Therefore some
smoothing in the spectra needs to be done before constructing the
filter, which introduces further arbitrariness. Thirdly, sometimes it
will not be possible to properly estimate some Fourier modes, for
example when using masks with missing data, and these modes will have
to be guessed.
 
Therefore, the most relevant conclusion of this analysis is that the 
MHW2 can be surely a better choice for the definition of a blind 
source catalogue -- like the ERCSC foreseen for the future Planck 
mission -- because it gives numbers of detected and of spurious 
sources comparable to the ones obtained with the MF but is easier to 
implement, more robust and has much lesser CPU requirements, 
especially in all-sphere applications. 
 
\section*{Acknowledgments} 
We acknowledge many useful suggestions from the referee, R. 
Savage, which helped us to improve the final presentation of the 
paper and the discussion of our main results. We acknowledge 
partial financial support from the Spanish Ministry of Education 
(MEC) under project ESP2004--07067--C03--01. MLC acknowledge a FPI 
fellowship of the Spanish Ministry of Education and Science (MEC). 
DH and JGN acknowledge the Spanish MEC for a ``Juan de la Cierva'' 
postdoctoral fellowship and the Italian MEC for a postdoctoral 
position at the SISSA-ISAS (Trieste), respectively. We thank G. de 
Zotti and J. Delabrouille for comments on the draft. We thank G. 
de Zotti and G.L. Granato for having kindly provided us with the 
source number counts foreseen by the de Zotti et al. (2005) and by 
the Granato et al. (2001; 2004) cosmological evolution models, 
respectively. Thanks are also due to C. Burigana for useful 
comments and remarks.

\bsp 
 
\label{lastpage}

\begin{thebibliography}{} 
 
\bibitem[\protect\citeauthoryear{Barcons}{1992}]{bar92} Barcons X., 
1992, ApJ, 396, 460 
 
\bibitem[\protect\citeauthoryear{Barreiro et 
    al.}{2003}]{bar03} Barreiro R.~B., Sanz J.~L., Herranz D., 
  Mart{\'{\i}}nez-Gonz{\'a}lez E., 2003, MNRAS, 342, 119 
 
\bibitem[\protect\citeauthoryear{Cay{\'o}n et al.}{2000}]{cay00} 
  Cay{\'o}n L., et al., 2000, MNRAS, 315, 757 
 
\bibitem[\protect\citeauthoryear{Chiang et al.}{2002}]{chi02} 
  Chiang L.-Y., J{\o}rgensen H.~E., Naselsky I.~P., Naselsky P.~D., Novikov 
  I.~D., Christensen P.~R., 2002, MNRAS, 335, 1054 
 
\bibitem[Colafrancesco et al.(1997)]{col97} Colafrancesco, S., 
Mazzotta, P., Rephaeli, Y., \& Vittorio, N.\ 1997, ApJ, 479, 1 
 
\bibitem[\protect\citeauthoryear{Cremonese et al.}{2002}]{cre02} Cremonese G., 
Marzari F., Burigana C., \& Maris M., 2002, New Ast., 7, 483 
 
\bibitem[\protect\citeauthoryear{Cruz et al.}{2006}]{cru06} Cruz M., 
Tucci M., Mart{\'\i}nez-Gonz\'alez E. \& Vielva P., 2006, MNRAS, 
submitted (astro-ph/0601427). 
 
\bibitem[\protect\citeauthoryear{De Zotti et al.}{1996}]{dz96} De Zotti G., 
Franceschini A., Toffolatti L., Mazzei P., \& Danese L., 1996, 
Astrophys. Lett. Commun., 35, 289 
 
\bibitem[\protect\citeauthoryear{De Zotti et al.}{1999}]{dz99} De 
  Zotti G., Toffolatti L., Arg{\"u}eso F., Davies R.~D., Mazzotta P., 
  Partridge R.~B., Smoot G.~F., Vittorio N., 1999, AIPC, 476, 204 
 
\bibitem[\protect\citeauthoryear{De Zotti et al.}{2005}]{dz05} De 
  Zotti G., Ricci, R., Mesa D., Silva L., Mazzotta P., Toffolatti L., 
  Gonz\'alez-Nuevo J, 2005, A\&A, 431, 893 
 
\bibitem[\protect\citeauthoryear{Dickinson, Davies, \& 
Davis}{2003}]{dick03} Dickinson C., Davies R.~D., Davis R.~J., 
2003, MNRAS, 341, 369 
 
\bibitem[\protect\citeauthoryear{Dolag et al.}{2005}]{dol05} Dolag K., 
Hansen F.K., Roncarelli M., \& Moscardini L., 2005, MNRAS, 363, 
N.1, 29 
 
\bibitem[\protect\citeauthoryear{Finkbeiner, Davis, \& 
Schlegel}{1999}]{fink99} Finkbeiner D.~P., Davis M., Schlegel 
D.~J., 1999, ApJ, 524, 867 
 
\bibitem[\protect\citeauthoryear{Franceschini et al.}{1989}]{fra89} Franceschini 
A., Toffolatti L., Danese L., \& De Zotti G., 1989, ApJ, 344, 35 
 
 
\bibitem[\protect\citeauthoryear{Gaustad et 
al.}{2001}]{gau01} Gaustad J.~E., McCullough P.~R., Rosing 
W., Van Buren D., 2001, PASP, 113, 1326 
 
\bibitem[\protect\citeauthoryear{Giardino et 
al.}{2002}]{gia02} Giardino G., Banday A.~J., G{\'o}rski 
K.~M., Bennett K., Jonas J.~L., Tauber J., 2002, A\&A, 387, 82 
 
\bibitem[\protect\citeauthoryear{Granato et al.}{2001}]{gr01} 
  Granato, G. L., Silva, L., Monaco, P., Panuzzo, P., Salucci, P., De 
  Zotti, G. Danese, L., 2001, MNRAS, 324, 757 
 
\bibitem[\protect\citeauthoryear{Granato et al.}{2004}]{gr04} Granato, 
  G. L., De Zotti, G., Silva, L., Bressan, A.  Danese, L., 2004, ApJ, 
  600, 580 
 
\bibitem[\protect\citeauthoryear{Gonz{\'a}lez-Nuevo et 
    al.}{2005}]{jgn05} Gonz{\'a}lez-Nuevo J., Toffolatti L., Arg{\"u}eso F., 
  2005, ApJ, 621, 1. 
 
\bibitem[\protect\citeauthoryear{Gonz{\'a}lez-Nuevo et 
    al.}{2006}]{jgn06} Gonz{\'a}lez-Nuevo J., Arg{\"u}eso F., 
    L{\'o}pez-Caniego M., Toffolatti L., Sanz J.~L., Vielva P., Herranz D., 
    2006, MNRAS, (astro-ph/0604376), in press.
 
\bibitem[\protect\citeauthoryear{G{\'o}rski et al.}{2005}]{heal05} 
  G{\'o}rski K.~M., Hivon E., Banday A.~J., Wandelt B.~D., Hansen F.~K., 
  Reinecke M., Bartelmann M., 2005, ApJ, 622, 759 
 
\bibitem[\protect\citeauthoryear{Haffner}{1999}]{haf99} 
Haffner L.~M., 1999, PhDT. 
 
\bibitem[\protect\citeauthoryear{Hansen et al.}{2005}]{han05} Hansen 
F.K., Branchini E., Mazzotta P., Cabella P., \& Dolag K., 2005, 
MNRAS, 361, N.3, 753 
 
\bibitem[\protect\citeauthoryear{Haslam et al.}{1982}]{has82} 
Haslam C.~G.~T., Stoffel H., Salter C.~J., Wilson W.~E., 1982, A\&AS, 47, 1 
 
\bibitem[\protect\citeauthoryear{Herranz et 
    al.}{2002a}]{yo02a} Herranz D., Sanz J.~L., Barreiro R.~B., 
  Mart{\'{\i}}nez-Gonz{\'a}lez E., 2002a, ApJ, 580, 610 
 
\bibitem[\protect\citeauthoryear{Herranz et 
    al.}{2002b}]{yo02b} Herranz D., Gallegos J., Sanz J.~L., 
  Mart{\'{\i}}nez-Gonz{\'a}lez E., 2002b, MNRAS, 334, 533 
 
\bibitem[\protect\citeauthoryear{Herranz et al.}{2002c}]{yo02c} 
  Herranz D., Sanz J.~L., Hobson M.~P., Barreiro R.~B., Diego J.~M., 
  Mart{\'{\i}}nez-Gonz{\'a}lez E., Lasenby A.~N., 2002c, MNRAS, 336, 1057 
 
\bibitem[\protect\citeauthoryear{Hobson et al.}{1999}]{hob99} Hobson 
  M.~P., Barreiro R.~B., Toffolatti L., Lasenby A.~N., Sanz J.~L., 
  Jones A.~W., Bouchet F.~R., 1999, MNRAS, 306, 232 
 
\bibitem[\protect\citeauthoryear{Hobson \& McLachlan}{2003}]{hob03} 
  Hobson M.~P., McLachlan C., 2003, MNRAS, 338, 765 
 
\bibitem[\protect\citeauthoryear{L{\'o}pez-Caniego et 
    al.}{2004}]{can04} L{\'o}pez-Caniego M., Herranz D., Barreiro R.~B., 
  Sanz J.~L., 2004, SPIE, 5299, 145L. 
 
\bibitem[\protect\citeauthoryear{L{\'o}pez-Caniego et 
    al.}{2005}]{can05} L{\'o}pez-Caniego M., Herranz D., Barreiro R.~B., 
  Sanz J.~L., 2005, MNRAS, 359, 993 
 
\bibitem[\protect\citeauthoryear{McEwen et al.}{2005}]{mac05} McEwen 
J. D., Hobson M. P., Lasenby A. N. \& Mortlock D. J., 2005, MNRAS, 
359, 1583 
 
\bibitem[\protect\citeauthoryear{Negrello et al.}{2004}] {ngr04} 
Negrello M., Magliocchetti M., Moscardini L., De Zotti G., Granato 
G.L., Silva L., 2004, MNRAS, 352, 493 
 
\bibitem[\protect\citeauthoryear{Negrello et al.}{2005}] {ngr05} 
Negrello M., Gonz\'alez-Nuevo J., Magliocchetti M., Moscardini L., 
De Zotti, G., Toffolatti, L., Danese, L. 2005, MNRAS, 358, 869 
 
\bibitem[\protect\citeauthoryear{Paladini et al.}{2003}]{pal03} Paladini R., 
Burigana C., Davies R.D., Maino D., Bersanelli M., Cappellini B., 
Platania P., \& Smoot G., 2003, A\&A, 397, 213 
 
\bibitem[\protect\citeauthoryear{Sanz, Herranz, \& 
    Mart{\'{\i}}nez-G{\'o}nzalez}{2001}]{sanz01} Sanz J.~L., 
  Herranz D., Mart{\'{\i}}nez-G{\'o}nzalez E., 2001, ApJ, 552, 484 
 
\bibitem[\protect\citeauthoryear{Spergel et al.}{2003}]{spe03} { 
        {Spergel}, D.~N. et al.}, 2003, ApJS, 148, 175 
 
\bibitem[\protect\citeauthoryear{Tauber}{2004}]{tauber} Tauber 
  J.~A., 2004, AdSpR, 34, 491 
 
\bibitem[\protect\citeauthoryear{Tegmark \& de 
    Oliveira-Costa}{1998}]{max98} Tegmark M., de Oliveira-Costa 
  A., 1998, ApJ, 500, L83 
 
\bibitem[\protect\citeauthoryear{Tegmark et al.}{2000}]{max00} Tegmark 
  M., Eisenstein D.~J., Hu W., de Oliveira-Costa A., 2000, ApJ, 530, 133 
 
\bibitem[\protect\citeauthoryear{Toffolatti et al.}{1998}]{tof98} 
    Toffolatti L., Arg\"ueso F., de Zotti G., Mazzei P., 
    Franceschini A., Danese L., Burigana C., 1998, MNRAS, 297, 117 
 
\bibitem[\protect\citeauthoryear{Vielva et al.}{2001a}]{patri01a} 
  Vielva P., Mart{\'{\i}}nez-Gonz{\'a}lez E., Cay{\'o}n L., Diego J.~M., Sanz 
  J.~L., Toffolatti L., 2001a, MNRAS, 326, 181 
 
\bibitem[\protect\citeauthoryear{Vielva et al.}{2001b}]{patri01b} 
  Vielva P., Barreiro R.~B., Hobson M.~P., 
  Mart{\'{\i}}nez-Gonz{\'a}lez E., Lasenby A.~N., Sanz J.~L., 
  Toffolatti L., 2001b, MNRAS, 328, 1 
 
\bibitem[\protect\citeauthoryear{Vielva et al.}{2003}]{patri03} 
  Vielva P., Mart{\'{\i}}nez-Gonz{\'a}lez E., Gallegos J.~E., Toffolatti L., 
  Sanz J.~L., 2003, MNRAS, 344, 89 
 
\bibitem[\protect\citeauthoryear{Vielva et al.}{2004}]{patri04} 
  Vielva P., Mart{\'{\i}}nez-Gonz{\'a}lez E., Barreiro R.B., Sanz 
  J.~L., Cay\'on L., 2004, ApJ, 609, 22 
 
\bibitem[\protect\citeauthoryear{Vio, Andreani, \& 
    Wamsteker}{2004}]{vio04} Vio R., Andreani P., Wamsteker W., 
  2004, A\&A, 414, 17 
 
\end{thebibliography}
\end{document}